\documentclass[structabstract]{aa}  
\usepackage{graphicx}
\usepackage{txfonts}
\usepackage{natbib}

\def\Msun{M_{\odot}}
\def\Msunh{M_{\odot}h^{-1}}
\def\Mst{M_{star}}
\def\lesssim{\mathrel{\hbox{\rlap{\hbox{\lower4pt\hbox{$\sim$}}}\hbox{$<$}}}}
\def\gtrsim{\mathrel{\hbox{\rlap{\hbox{\lower4pt\hbox{$\sim$}}}\hbox{$>$}}}}
\def\Ha{H$\alpha$}
\def\Hb{H$\beta$}
\def\OII{O\,{\sc ii}}
\def\kms{\mbox{km s$^{-1}$}}
\def\yr{\mbox{yr}}

\begin{document}

\title{GECO: Galaxy Evolution COde - A new semi-analytical model of
  galaxy formation}
\titlerunning{GECO: Galaxy Evolution COde}

   \author{E. Ricciardelli  \inst{1,2,3}\and A. Franceschini
          \inst{1}}

   \institute{$^{1}$Dipartimento di Astronomia, Universit\'a di Padova, 
     Vicolo Osservatorio 2, 35122 Padova, Italy \\
     $^{2}$Instituto de Astrof\'isica de Canarias, V\'ia Lactea s/n, 
     E-38200 La Laguna, Tenerife, Spain\\
     $^{3}$Departamento de Astrof\'isica, Universidad de La Laguna, 
     E-38205, Tenerife, Spain\\
             }

   \offprints{Elena Ricciardelli, \\ \email{elenaricci@iac.es}}
\date{Accepted 16/04/2010  Received ...; in original form ...}

\abstract
{}
{
We present a new semi-analytical model of galaxy formation, GECO (Galaxy Evolution COde), aimed at a better understanding of when and how the two processes of star formation and galaxy assembly have taken place, by comparison with a wide variety of recent data on the evolutionary galaxy mass functions and star-formation histories.}
{Our model is structured into a Monte Carlo algorithm based on the Extended Press-Schechter theory, for the representation of the merging hierarchy of dark matter halos, and a set of analytic algorithms for the treatment of the baryonic physics, including classical recipes for the gas cooling, the star formation time-scales, galaxy mergers and SN feedback.
Together with the galaxies, the parallel growth of BHs is followed in time and their feedback on the hosting galaxies is modelled. We set the model free parameters by matching with data on local stellar mass functions and the BH-bulge relation at $z=0$.}
{Based on such local boundary conditions, we investigate how data on the high-redshift universe constrain our understanding of the physical processes driving the evolution, focusing in particular on the assembly of stellar mass and on the star formation history. Since both processes are currently strongly constrained by cosmological near- and far-IR surveys with the Spitzer Space Telescope, the basic physics of the $\Lambda CDM$ hierarchical clustering concept of galaxy formation can be effectively tested by us by comparison with the most reliable set of observables using a minimal number of free parameters.}
{Our investigation shows that when the time-scales of the stellar formation and mass assembly are studied as a function of dark matter halo mass and the single galaxy stellar mass, the 'downsizing' fashion of star formation appears to be a natural outcome of the model, reproduced even in the absence of the AGN feedback. 
On the contrary, the stellar mass assembly history turns out to follow a more standard hierarchical pattern progressive in cosmic time, with the more massive systems assembled at late times mainly through dissipationless mergers.
}

\keywords{
galaxies: evolution -- galaxies: formation -- galaxies: halos.
}
\maketitle

\section{Introduction}

In the past decade, several observational evidences were accumulating in favour of the $\Lambda CDM$ paradigm for structure formation, now quite a successful rendition of the hierarchical clustering scenario for cosmic structure formation. In its standard form \citep{Blum:84}, it predicts that structures formed from primordial fluctuations of the density field amplified during inflation and then collapsed to form the virialized structures that we see nowadays.
The most compelling support in favour of this paradigm comes from the measurements of the temperature anisotropies of the cosmic microwave background \citep{Spergel:03, Spergel:07}. Further evidences are due to the measurements of the power spectrum of galaxy clustering from large surveys of the local universe
\citep{Perc:02, Tegm:04}, the evidence for an accelerated expansion of the universe as inferred from high-redshift type Ia supernovae observations \citep{Riess:98,Perl:99} and the baryon fraction observed
in rich clusters \citep{White:93}. 

Devised to study galaxy evolution within this cosmological framework, the semi-analytical approach favours a relatively simple handling of the main physical parameters and understanding of their possible role in driving the evolution.
This modelling has its root in the work of \citet{WR:78}, where it was proposed that galaxy formation is a two-stage process, with dark matter halos forming in a dissipationless gravitational collapse and galaxies forming inside them following the radiative cooling of baryons.
Although \citet{WR:78} and,  after,  \citet{WF:91}, based their work
only on the analytic Press-Schechter  theory \citep{PS:74}, predicting
only average  quantities, subsequently  a number  of works followed
their prescriptions using Monte Carlo (MC) merger trees.
The MC approach allows to obtain several realizations of the merging history of individual dark matter
halos. This approach was pioneered by \citet{LC:93, kauff:93a, Cole:94}, and then followed by a number of authors \citep{SK:99,  SL:99, Zen:07}.
The great advantage of the semi-analytical method (SAM), apart from being computationally very fast and flexible, is the fact that it is possible to compute merging histories with arbitrary mass resolution.
The alternative approach that can be followed in order to track the evolution of the dark matter halos, is
through the use of large cosmological N-body simulations. Their large
computational requirements are compensated by the amount of
information that can be achieved. For instance, in the Millennium Simulation \citep{Spring:01}, the evolution of substructures in massive
halos can be followed in time, with the results of a more detailed information about the galaxy dynamics and the influence  of the cosmic environment on the process \citep{DeL:04}.
In the literature, various examples of this ``hybrid'' approach, which make use of N-body simulation for the dark matter evolution and the SAM technique for the baryonic physics, have been published \citep{Hatton:03, DeL:06, Cr:06}.

In the present work we employ a MC merger tree, mainly because this allowed us to test the parameter space of the semi-analytical model with much more flexibility than using N-body simulations, and allows to compute merging histories down to arbitrary low mass resolution.

While the treatment of the evolution of dark matter structures is relatively simple, as being determined only by gravity, the physics of the baryons inside halos is much more complex to describe.
In the most classical models \citep{kauff:93b, Baugh:96, kauff:99, SP:99, Cole:00, Menci:02}, gas cooling, star formation, SN feedback and galaxy mergers are included.
In recent years, it has become clear that some other form of highly energetic feedback is needed to prevent  star formation in massive galaxies at recent epochs, where the SN feedback is ineffective. The need of such form of feedback is required in order to avoid the overcooling in massive halos, and hence the overabundance of galaxies at the bright-end of the luminosity function.
This source of feedback is commonly found in the AGN energy production. This effect, supported by the observational findings of a tight correlation between the BH and the bulge size \citep{Fer:00, HR:04}, was implemented in different ways by \citet{KH:00, Bower:06, Cr:06, Menci:06, Som:08}.

Alternative to the AGN feedback, the shutdown of star formation above a critical halo mass has been implemented as a quenching mechanism in massive galaxies \citep{Cattaneo:06, Cattaneo:08}, motivated by the prediction of stable shock heating for halos more massive than this threshold \citep{Dekel:06}.

Following the prescriptions of these models, we have built a new
semi-analytical models, the Galaxy Evolution COde (GECO), whose aim is
to identify a few key physical parameters and modify them by comparing
with several basic properties of the galaxy population at $z=0$, as
well as at high redshifts. Our main observational reference in this
paper is the redshift-dependent stellar mass function of galaxies,
which, from a suitable choice for the stellar Initial Mass Function (IMF),  
is a robust descriptor of the star formation history and the mass assembly history of galaxies.

The structure of the paper is as follows. In \S\ref{mtree} we describe
the Monte Carlo merger tree used in the model. In \S\ref{model} we
introduce the ingredients of the baryonic model. In \S\ref{param} we
explain how the free parameters are set and provide a table for them.
In  \S\ref{localUn} the basic results for the local universe are
presented, while in  \S\ref{Highz} we focus on the high redshift predictions.
We conclude in \S\ref{concl}. Throughout the rest of the paper we assume a
``concordance'' cosmological model, with $\Omega_m=0.3$,
$\Omega_{\Lambda}=0.7$, $h=0.7$, $\sigma_8=0.9$ and $n=1$ (power index of the primordial power spectrum).
However, when needed, we also show the dependence on the cosmological parameters showing the results for the new WMAP5 dataset, which are: 
$\Omega_m=0.258$, $\Omega_{\Lambda}=0.742$, $h=0.719$, $\sigma_8=0.796$ and $n=0.963$
\citep{Dunkley:09}.

\section{Merger Tree}\label{mtree}

\subsection{The formalism}

In the hierarchical theory of structure formation the density contrast
$\delta=(\rho-\rho_b)/\rho_b$, where $\rho_b$ is the mean density of
the universe and $\rho$ is the perturbation density,
is a Gaussian random field that at early times grows linearly.
When the density contrast reaches a critical value $\delta_{c}$,
given by the spherical collapse theory, the
perturbation starts to collapse and form virialized halos.
In order to associate masses to these collapsed objects,
the Press-Schechter approach considers the
fluctuations whose density contrast
smoothed on a scale $R_M$, through a spatial window function $W(r,R_M)$,
exceeds $\delta_{c}$.

The Gaussian random variable $\delta$ has zero mean and variance $S(M)=\sigma^2(M)$  which is linked to the mass through the Power-spectrum $P(k)$:
\begin{equation}
\sigma^2(M)=\frac{1}{2\pi^2}\int{P(k)W^2(kr)k^2dk}\,,
\end{equation}
where  $W^2(kr)$ is the Fourier transform of the spatial window function.
In the excursion-set approach, developed by \citet{Bond:91}, the value of $\delta$ executes a random walk as the smoothing scale $R_M$ (or $M$) is changed. We can consider the trajectories in the plane
$(S,\delta_c)$ and associate the fraction of matter in collapsed objects in the range $dM$ around $M$ at the time $t$ with the fraction of trajectories that make the first upcrossing through the threshold $\delta_c$ in the interval $S,S+dS$. This results in the well-known PS mass function:
\begin{equation}\label{ps}
f(S,\delta_c)dS=\frac{1}{\sqrt{2\pi}}\frac{\delta_c}{S^{3/2}}\exp\Big(-\frac{\delta_c^2}{2S} \Big)dS
\end{equation}
\citep{PS:74}.

In the \emph{Extended Press-Schechter} model \citep{Bond:91, LC:93} it is also possible to derive the conditional mass function, the fraction of trajectories in halos with mass $M_1$ at $z_1$ that are
in halos with mass $M_0>M_1$ at $z_0<z_1$:
\begin{eqnarray}
\lefteqn{f(S_1,\delta_{c1}|S_0,\delta_{c0})dS_1 =}
\nonumber \\
& & \frac{1}{\sqrt{2\pi}}\frac{(\delta_{c1}-\delta_{c0})}{(S_1-S_0)^{3/2}}
\exp\Big[-\frac{(\delta_{c1}-\delta_{c0})^2}{2(S_1-S_0)} \Big] dS_1\,.
\label{eps_mass}
\end{eqnarray}

Converting from mass weighting to number weighting we obtain the
average number of progenitors at $z_1$ in the mass interval $dM_1$
around $M_1$ which at redshift $z_0$ has merged to form an $M_0$ halo:
\begin{eqnarray}
\lefteqn{\frac{dP}{dM_1}(M_1,z_1|M_0,z_0)dM_1 =}
\nonumber \\
& & \frac{M_0}{M_1}f(S_1,\delta_{c1}|S_0,\delta_{c0})\Big|\frac{dS}{dM} \Big|dM_1\, .
\end{eqnarray}

\subsection{ The partition algorithm}

For generating Monte Carlo realizations of the merging history of dark matter halos we use the partition algorithm described by \citet{SL:99}, hereafter SL99, which is exact for a white-noise power-spectrum,
but needs some modifications in order to be applied to a $\Lambda CDM$ fluctuation spectrum.
Since we will describe in detail the algorithm and the test done to probe its consistency in a separate paper, we give here only a brief description of it.

The algorithm is based on the assumption that for a white-noise spectrum mutually disconnected volumes are mutually independent, as it is analytically demonstrated in the Appendix of SL99.
Let us call \emph{parent} the halo of mass $M_0$ which exists at z=0
and \emph{progenitors} the smaller halos that derive from it at higher
redshift.
Suppose to partition an halo of mass $M_0$ into progenitors by choosing
first one progenitor and then an other one from the remaining mass
and so on until the remaining mass falls
below a certain minimum mass that is our mass resolution $M_{res}$.
We have a probability of finding the first progenitor of mass $M_1$,
corresponding to $S_1$, given by
$f(S_1,\delta_{c1}|S_0,\delta_{c0})$ (equation \ref{eps_mass}) and we choose a
mass drawing a random number from this distribution.
We can consider the halo $M_0$ as a region of size $V_{M_0}=M_0b_0/\rho_b$,
where
$b_0=1/(1+\delta_0)$. If the first
progenitor has mass $M_1$ it occupies a volume $V_{M_1}=M_1b_1/\rho_b$. The
remaining mass $R=M_0-M_1$ is distributed in the volume
$V_R=Rb_R/\rho_b$. Hence, for the conservation of volumes, the density in
this region is given by
\begin{equation}
\delta_{cR}=\delta_{c1}-\frac{(\delta_{c1}-\delta_{c0})}{1-(M_1/M_0)}
\label{DD}
\end{equation}
to lowest order in $\delta$ (see equation 5 of SL99).
Now, the second progenitor must be chosen from the remaining mass $R$
and the probability that it has mass $M_2$, corresponding to $S_2$, is given by
$f(S_2,\delta_{c1}|S(R),\delta_{cR})$. Therefore we choose a random number
from this distribution. Then we iterate the process by continuing to find
progenitors until the remaining mass is below the mass resolution.

The assumption that disconnected volumes are mutually independent is right only for white noise power spectrum, but we want to build merger trees for a $\Lambda CDM$ spectrum.
SL99 show that by applying the same algorithm for scale-free spectra with $n\ne0$ leads to inconsistencies in
the excursion set mean values. Since we know that all the excursion set quantities are independent of the power spectrum when they are expressed in terms of the variance rather than the mass, we run the algorithm for the white-noise case.
Then we consider each chosen $M_{wn}$ not as a progenitor with mass
$M_{wn}$, but as a region populated by some number $\nu$ of objects
having mass $M_{\Lambda CDM}$.
In practise, we normalize the white-noise
power spectrum in such a way that the mass variance of the two spectra
is the same in $M_0$: $S_0=S_{\Lambda CDM}(M_0)$. Then, we choose
$S_1$ from the distribution of equation \ref{eps_mass}. We associate
to this variance a mass given by the white-noise spectrum
$M_{wn}=S_0M_0/S_1$  and one given by the $\Lambda CDM$ spectrum
$M_{\Lambda CDM}$. We consider a number $\nu$ of halos with this
mass, given by $\nu=M_{wn}/M_{\Lambda CDM}$, approximated to the nearest
integer. Then, we iterate the procedure on each new progenitor.

\begin{figure}
\begin{center}
\includegraphics[width=\columnwidth]{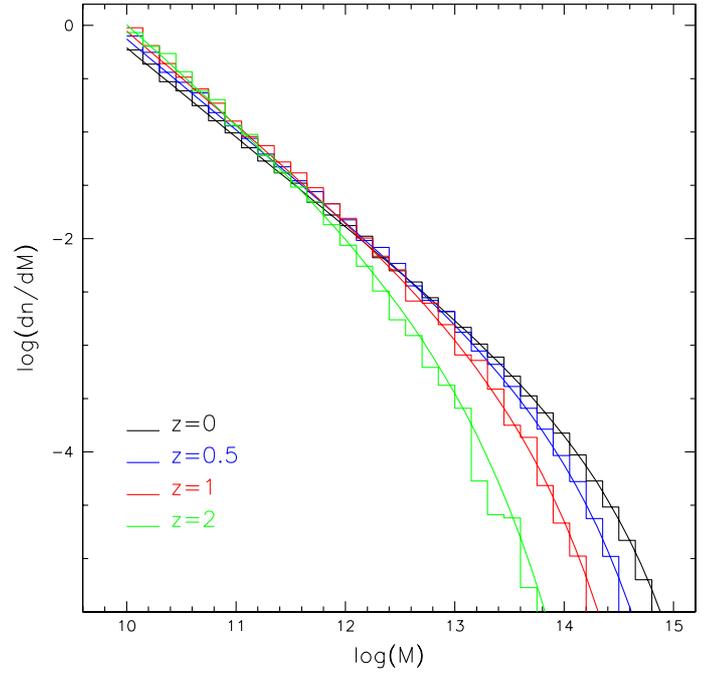}
\end{center}
\caption{Global mass function for dark matter halos at different
  redshifts. Histograms represent the mass function derived from our MC
  merger tree, while solid lines show the PS predictions.
  }
\label{umf}
\end{figure}

As a test of the consistency of our MC merger tree with the PS
predictions, in Figure \ref{umf} we show the global mass function of
dark matter halos. 
To compute it, we use a grid of 35 parent halo masses, ranging from $10^{10}\Msun$ to
$10^{15}\Msun$, logarithmically spaced, and simulate 100 realizations
for each parent halo. Then we weight each halo with
the PS number density at z=0. The results do not depend
much on the number of parent halo masses used, but they are quite
sensitive to the number of realizations for each halo. The 100
realizations used turn out to be a good compromise between
the good sampling desired and the computational-time required.
We build a grid in time of 52 intervals logarithmically spaced in redshift, ranging from
z=20 when the time-step is approximately equal to 0.05 Gyr  to z=0 with a $\sim$0.8 Gyr time-step. Because our method for building merger tree is
insensitive to the time-step used, the choice of the time-step grid is
quite arbitrary and we are not forced to use too small time intervals. 
We find very good agreement between the
MC method and the analytical predictions. A key point of this
algorithm is that it correctly reproduces the mean statistical
quantities of EPS and PS predictions for any set of
$M_1,z_1,M_0,z_0$. This means that there is no need to divide the
time interval between $z_0$ and $z_1$ in smaller time-steps. 
This is
particularly convenient because it allows us to choose the time grid for
the semi-analytical implementation in an arbitrary way.

The algorithm described is based on the Extended Press-Schechter formalism,
that means on the spherical barrier model, but it can be extended for a more
general moving barrier model, useful for the description of the ellipsoidal
collapse model \citep{Bond:96} that bears closer consistency with
N-body simulation \citep{ST:02, Giocoli:07, Zen:07, Zhang:08}.

\section{The baryonic model}\label{model}

While the theory of the hierarchical development of the dark matter
structures is relatively simple and now well established, the complex
physics of baryons presents still highly uncertain aspects, and
requires the treatment of several processes in a simplified form. The
latter is obtained with the introduction of a minimum number of free
parameters in the model, that can only be set through a comparison of
the model predictions with observational data.  These parameters flag
our lack of knowledge about the details of the physics of galaxy
formation.   Obviously, although they are set by fitting model results
with observations, only limited ranges of values are allowed: values
outside these would indicate fundamental problems in the process
treatment or lack of important physical ingredients. 
This Section is dedicated to a description of our new SAM model,
GECO. 

At variance with the majority of the models of galaxy formation, we
tested our results directly on the observables involving stellar mass
rather than luminosities, to avoid introducing further uncertainties
in the model due to the spectro-photometric synthesis, dust extinction
and the Initial Mass Function. 
The elaboration of the photometric synthesis based on the model galaxy
mass function will be deferred to a future work.

\subsection{ Linking Galaxies to Dark Matter Halos}

As described  in the  previous section, the  generation of  the merger
tree proceeds  backwards in time.  Starting from the  initial redshift
(typically $z=0$), we split each halo into  progenitors at
higher  redshifts.   On  the  contrary, galaxy  formation  is  modelled
forward in time, starting  from  the  ``leaves''   of  the  tree,
i.e. halos  at high redshift whose progenitors fall below  the mass
limit  resolution  ($M_{res}=10^{10}\Msun$  in our case)  and  their
hierarchy is  no more  followed in time.  Starting from each  of these
halos we put baryons inside  them, according to the baryonic fraction
observed by WMAP5: $f_b=\Omega_b/\Omega_m=0.17$ \citep{Dunkley:09}.

We use the grid in parent halo masses and time-steps described in 
Sect. 2.2.
To solve the differential equations for baryon evolution that we are
going to introduce in the following sections, 
we divide each step of the merger hierarchy
into 100 smaller substeps which allow us to solve the system of
equations through the use of finite increments of mass.

In order to derive statistical quantities  such as the mass function or
 the SFR  density, we simply make a weighted sum over the model
 galaxies. Since we are dealing with an EPS  merger tree, we use for
 consistency as a weight scheme the  number   density of  halos
 given  by   the  PS  mass function. 
In  some cases,  we have also checked the results using a Sheth \& Tormen
 number density \citep{ST:02},  which fits  the results  of N-body
 simulations  with higher accuracy, but we have found that it makes little
 difference. 

We will describe in the following the baryonic processes at work in
driving the evolution of galaxies, 
starting from the cooling of the gas.

\subsection{Dynamical Time-scales and Gas Cooling}

The cooling of the gas is a fundamental process for galaxy formation,
as it sets the rate at which gas becomes available for star formation.
The first to invoke the need of some dissipative process occurring
inside dark matter halos were \citet{WR:78},
in order to explain the difference between the luminosity function of
galaxies, with its characteristic mass and size, and the halo mass
function, that on galactic scales is a typical power-law without any preferred scales.
White \& Rees argued that galaxy formation is a two stage process, with dark matter halos forming through the dissipationless hierarchical clustering, and with gas cooling occurring inside them.

The gas is assumed to be heated by shocks during the violent relaxation following the collapse, up to the virial temperature of the halo, given by:
\begin{equation}\label{tvir}
T=\frac{1}{2}\frac{\mu m_H}{k}V_{c}^2\, ,
\end{equation}
where $\mu$ is the mean molecular mass of the gas, $m_H$ is the mass
of a hydrogen atom, \emph{k} is the Boltzmann's constant and $V_c$
is the circular velocity of the halo, that can be related to the mass
of the halo through the expression: $V_c=\sqrt{G M_{halo}/R_{vir}}$. The
virial radius, $R_{vir}$, is the limiting radius of a virial halo,
within which the mean density is $200 \rho_{c}$, where $\rho_{c}$
is the universal critical density, and is given by $R_{vir}=1.63
\times 10^{-2} (M_{halo}/\Msunh)^{1/3} (\Omega(z)/\Omega_m)^{1/3}(1+z)^{-1}
h^{-1} kpc$ \citep{NFW:97}.
From $V_c$ and the virial radius $R_{vir}$, an halo dynamical time can be defined:
\begin{equation}
t_{halo}=R_{vir}/V_c\, .
\label{tdyn}
\end{equation}

The rate at which the gas cools depends upon this temperature, which
determines its ionization state, and on the metallicity of the gas,
that is the chemical composition. We take the cooling function  $\Lambda(T,Z)$ 
 tabulated by \citet{SD:93}.
Note that the cooling function also depends on the metallicity of the hot gas. High metallicity values increase the cooling rate, because of line cooling from heavy elements, mainly for low-mass halos. 
We assume that the hot gas has a constant
metallicity equal to $0.3 Z_{\odot}$, close to the value for the hot gas in cluster \citep{Som:08}.
We have also checked that by adopting a solar metallicity instead of the subsolar one, the results do not change significantly.

The cooling time is defined as the thermal energy density divided by
the cooling rate per unit volume:
\begin{equation}
t_{cool}=\frac{3}{2}\frac{\mu m_H kT}{ \Lambda(T,Z) \rho_{gas}}\, ,
\end{equation}
where $\rho_{gas}$ is the gas density profile, which is assumed to be
isothermal:
\begin{equation}
\rho_{gas}(r)=\frac{M_{hot}}{4\pi R_{vir}r^2}\, ,
\end{equation}
where $M_{hot}$ is the total gas mass in the hot component, which is
assumed to extend to the virial radius.
A cooling radius, $r_{cool}$, can now be defined as the point where the local
cooling time is equal to the age of the universe at that epoch.

We can now compute the rate at which gas accretes to the centre of the
halo, becoming available for star formation.
Following the treatment of \citet{WF:91}, we distinguish two regime of cooling, depending on the value of
$r_{cool}$. The first case corresponds to $r_{cool}>R_{vir}$, the cooling radius lying outside the
virialized region of the halo.
The cooling is so rapid that the infalling gas never reaches the hydrostatic equilibrium and the supply of cold gas for star formation is limited by the infall rate rather than the cooling rate.
We assume that all the hot gas in the halo will settle to the centre in a time-scale given by the halo dynamical time:
\begin{equation}
\dot{M}_{cool}=\frac{M_{hot}V_c}{R_{vir}} .
\label{coldmode}
\end{equation}
This regime of cooling was called by \citet{WF:91} the \textit{rapid
  cooling regime}, also known as \textit{cold mode}.

The second way of infalling is the static hot halo regime, or \textit{hot mode}, that occurs when the cooling radius lies inside the virial radius.
In this case, the gas inside $r_{cool}$ will be pressure-supported and will contract quasi-statically
toward the centre.
Cooling will cause a flow of gas toward the centre in a
way exactly analogous to the cooling flow occurring in galaxy clusters.
A simple expression for the infall rate is given by:
\begin{equation}
\dot{M}_{cool}=4\pi \rho_{gas}r_{cool}^2\frac{dr_{cool}}{dt}\, .
\label{hotmode}
\end{equation}
Note that for small halos (that we find preferentially at early times), the cooling radius is greater than the virial radius, and we are in the \textit{cold mode} regime, where equation \ref{coldmode} applies.
At late times, for large values of the virial velocity, the cooling radius falls
below the virial radius and  we are in the \textit{hot mode} regime.
As a consequence, at high redshift we have a more efficient cooling.
Hydrodynamical simulations, such as that of \citet{Keres:05}, justify the use of two different modes of
cooling. Indeed, they found that a cold mode dominates for low-mass
objects, that are found at high redshift or in low-density environments
in the nearby universe, while the hot mode contributes significantly
for high-mass systems, therefore being important only at low
redshift.
Anyway it is worth to note that in our  model the cooling at high redshift might be enhanced due to the assumption of a constant metallicity. In case of a metal-dependent cooling rate we would obtain a lower cooling in the high redshift objects with primordial metallicity. We will test such effect in a future version of the code, where the chemical evolution model will be implemented.

\subsection{Galaxy Sizes and Angular Momentum}

Since several time-scales involved in the analytic treatment of galaxy formation (such as for example the star formation rate) that we are going to introduce in the following sections depend on the dynamical time of the galactic disc ($t_{dyn}=r_{disc}/v_{disc}$), it is important to have an accurate description of their sizes.
The disc size will depend on the virial radius of the halo where the galaxy was born, and on its angular momentum.
Indeed, if the halo is asymmetric and surrounded by a clumpy distribution of matter, then it can acquire an angular momentum from a net tidal torque. To quantify the angular momentum of the system, one often refers to the spin parameter, which is a dimensionless quantity, defined as:
\begin{equation}
\lambda=\frac{J|E|^{1/2}}{GM_{halo}^{5/2}}\, ,
\end{equation}
where $J, E$ and $M_{halo}$ are the total angular momentum, gravitational binding energy and mass of the halo. The distribution of the spin parameter as found in N-body simulations (\citealt{Warren:92, CL:96}) can be approximated with a log-normal distribution with $<\lambda>=0.05$ and $\sigma_{\lambda}=0.5$.
Hereafter we will assume the mean value for $\lambda$. \citet{Mo:98} related the disc radius to the spin parameter and to the virial radius assuming that the angular momentum of the disc is a fixed fraction of that of the halo:
\begin{equation}
r_{disc}=\frac{1}{\sqrt{2}}\Big(\frac{j_d}{m_d}\Big) \lambda R_{vir}\, ,
\end{equation}
where $m_d$ is the ratio between the disc mass and the halo mass, and $j_d$ is the ratio between the angular momentum of the disc and of the halo.
Higher values for the spin parameter result in larger discs, because they contract less before reaching the centrifugal equilibrium.  The above relation holds in the case of halos approximated with isothermal profile and neglecting the gravitational effects of the discs themselves, but the authors also give the correct expression in the case of NFW profiles and self-gravitating discs. As shown by \citet{Mo:98}, it is a reasonable assumption to take $j_d \simeq m_d$, which gives us:
\begin{equation}
r_{disc}=\frac{\lambda}{\sqrt{2}} R_{vir}\, ,
\end{equation}
which leads to galactic discs that are over an order of magnitude less extended than their halo hosts.
Under the assumption that we are considering negligible the self-gravity of the disc, the disc circular velocity is equal to the halo virial velocity.

The assumption that the specific angular momentum of the disc is the
same of that of the halo, as implied by $j_d \simeq m_d$, leads to
disc sizes that well match the observed spirals \citep{Mo:98}, hence it became the
standard assumption of disc modelling. Nevertheless, hydrodynamical
simulations find that the gas looses most of its angular momentum
during accretion, producing discs too small compared to that
observed \citep{NW:94}. It is still unclear if this effect is simply
due to the limited mass resolution employed, or to the absence of some
form of feedback.

\subsection{Star Formation}

Obviously, the physical processes governing the rate with which the (primordial) gas is transformed into stars play a critical role in models of galaxy formation. Unfortunately, they are still subject to major uncertainties, given the extremely complex interplay of phenomena on the largest (magnetic fields, gas angular momentum, etc.), as well as smallest (radiative cooling, thermal conduction, dust extinction, etc.) scales.
Nevertheless, observations indicate the existence of some useful laws to describe SF on global scale, larger than star-cluster scales.  Analysing a large sample of spirals and starburst galaxies, \citet{Ken:98} found two empirical correlations for the star formation per unit area,
$\Sigma_{SFR}$: (i)  $\Sigma_{SFR} \propto \Sigma_{gas}^{1.4}$ and (ii)  $\Sigma_{SFR} \propto \Sigma_{gas}/t_{dyn}$.
These empirical estimates justify the use of parametric forms for the star-formation prescription in semi-analytical models, where the star formation time-scale is given by the dynamical time.

In our model  we consider two modes of star formation. In the  first one, the \textit{quiescent mode}, the star formation rate is simply assumed to be proportional to the amount of cold gas available and inversely proportional to the time-scale of star formation, that is assumed here to be the dynamical time of the disc: $t_{dyn}=r_{disc}/v_{disc}$.
The efficiency of the star formation process is quantified by the free parameter $\alpha_{quie}$ defined by the relation:
\begin{equation}
\dot{M}_{*}=\alpha_{quie} \frac{M_{cool}}{t_{dyn}}\, .
\label{sfr_q}
\end{equation}
Since this gas is assumed to share the same high value of the angular momentum as previously discussed, stars formed in this way are added to the disc component of the galaxy.

The second mode of star formation considered in GECO is the \textit{burst mode}. Starbursts are allowed to occur after a merger between galaxies, that can be due to the cannibalism of a satellite galaxy by the central one or to a collision between satellites.
Indeed, a strong correlation between galaxy interactions and starburst activity is observed both in the present-day universe and at high redshift. This is for example the case for ULIRGs, the Ultra-Luminous Infrared Galaxies discovered locally by IRAS and then found in large numbers at high-z by ISO and Spitzer, exhibiting the most violent episodes of star formation \citep{Ken:98, Sand:96}.
The assumption of a starburst triggered by a galaxy merger is furthermore justified by several numerical
simulations \citep{MH:94, MH:96}, that show how, during the merger phase, gas sinks to the central
region of the galaxy and the increased density leads to an enhanced level of star formation (according to the Kennicutt law).
Recent simulations \citep{Cox:08} indicate that the starburst event can be triggered in the major, as well in the minor, mergers, but with an efficiency which strongly depends on the galaxy mass ratio (as implemented also in \citealt{SPF:01, Menci:04}).
Therefore, in our model, the star formation during the starburst is given by:
\begin{equation}
\dot{M }_{*}=f_{burst}\frac{M_{cold}}{t_{burst}}\, ,
\label{sfr_b}
\end{equation}
where $t_{burst}$ is the dynamical time of the largest disc, while $f_{burst}$ is the efficiency of the burst, assumed to be linearly proportional to the merger mass ratio:
\begin{equation}
f_{burst}=\beta_{burst} \frac{m_1}{m_2}\, ,
\end{equation}
where $m_1$ and $m_2$ are, respectively, the smallest and the biggest galaxy, and $\beta_{burst}$ is a free parameter. Stars formed in such way are added to the bulge component of the galaxy.

In addition to the increased rate of star formation, major mergers between spirals move stars from circular orbits to random motions, hence producing a remnant that generally resembles an elliptical galaxy.
For these reasons, we can safely assume that after that kind of mergers, the discs previously present in the colliding systems are transformed in a spheroid. We define a merger to be a major one when $\frac{m_1}{m_2}>1/3$.
Spheroid formation through disc instability is not considered in the present work.

\subsection{Feedback Processes}

Feedback processes are mechanisms which regulate the efficiency of star formation.
The need for some form of feedback in low-mass halos, which suppresses cold gas accretion, was first recognised by \citet{WR:78}, in order to flatten in some way the faint-end of the luminosity function, which turned out to be too steep compared to the observational data available at the time. SN feedback and photoionization are commonly considered to reduce the faint-end slope.

Although initially the motivation for invoking feedback was to suppress cooling, and hence star formation, in low-mass halos, in recent years the focus was shifted to attempt reproducing the bright-end of the luminosity function, which in the most classical models resulted in an excess of bright objects at low redshift compared with the observations (again a manifestation of the basic problem to shape an essentially power-law form for the DM mass function into a Schechter exponential function for the baryons).
Although the behaviour of the cooling function implies that the cooling time increases in high-mass halos, this is not enough to explain the sharp observed cut-off in the galaxy luminosity or stellar mass functions, and some mechanisms preventing the cooling rate in high-mass halos is also needed (and is commonly identified in the AGN effect, see below).

\subsubsection{SN feedback}

The first form of feedback considered in our model is that from SN explosions and high-mass stars outflows, which eject gas and energy into the surrounding Interstellar Medium (ISM).
This is supported by several evidences of Supernova driven winds in dwarf star-forming galaxies \citep{Mar:99, Strick:00, Heckman:01}, which also suggests that the gas reheating rate is proportional to the SFR.

The parametrisation of the SN feedback is based on simple energy arguments.
The rate of reheating will be proportional to the SFR, to  the number of Supernovae per solar mass of stars, given by $\eta_{SN}=4\times 10^{-3} \Msun $ for a Salpeter Initial Mass Function (IMF), to the
energy released by each SN, about $E_{SN}=10^{51} ergs$, and to the efficiency with which the SN energy is transferred into the ISM, $\epsilon_{SN}$.
This last parameter is highly uncertain, and is usually treated as a free parameter. The rate of reheating will be more efficient for galaxies living in low potential wells. In order to mark the depths of the potential well of the host halo, we use the halo circular velocity, so that the rate of reheating will be
simply proportional to $V_{c}^{-2}$. 
This implies that galaxies living in low-mass halos are affected by SN feedback effects, and their star formation will be self-regulated, while in high-mass halos this kind of feedback is ineffective. The rate of reheating will be finally given by:
\begin{equation}
\dot{ M}_{heat}=\epsilon_{SN} \frac{4}{5} \frac{\eta_{SN} E_{SN} \dot{
M}_{*}}{V_{c}^2}
\label{SN_feedb}
\end{equation}
\citep{kauff:93b, SP:99}.

The reheated gas is removed from the cold gas reservoir, and can not be used to form new stars.
We assume an \textit{ejection} model of feedback, that means that the
reheated gas is ejected from the halo and is unable to cool until it
 is reincorporated into a more massive halo
at the following step of the merging hierarchy.
Although our re-incorporation scheme assumes a fast fall-back
(analogous to the fast re-incorporation of \citealt{DeL:04}),
we found that it produces better results compared to the 
\textit{retention} model, in which the gas can subsequently
cool. In particular we obtain a better match of the faint-end of the
stellar mass function.

\subsubsection{Reionization}\label{UV}

It is now known that the hydrogen in the intergalactic medium (IGM), that
became
neutral at $z\sim 1000$, must
have been ionized at later epochs, although the redshift
at which this reionization occurred is still quite uncertain, ranging from
$z=6$, as imposed by the lack of a Gunn-Peterson trough in quasar spectra at
that redshift \citep{Fan:00}, up to $z=11$, as imposed by the constraint on the
optical depth to the last scattering surface measured from WMAP5 data. 

If a large population of galaxies and quasars exist at very high redshift, as
predicted by galaxy formation models and  confirmed up to $z \sim 6$ by
observations \citep{Fan:00}, then reionization could have occurred through
photoionization, since both young galaxies and quasars emit UV photons, able
to ionize the IGM.
This photoionizing background may also act to inhibit galaxy formation in two
ways. It can heat the IGM, 
increasing the thermal pressure of the gas and preventing it to collapse into the
dark matter halos. The
second way is through a reduction of the cooling rate of gas inside
halos, mainly reducing neutral atoms which can be collisionally excited. This
results in a strong suppression of galaxy formation in low-mass halos. 

According to previous studies \citep{TW:96, Gnedin:00, S:02}, after reionization the above mechanisms prevent gas accretion in halos with $V_c<50\,  \kms $, with the result that galaxy formation is strongly suppressed in such systems.
Hence, in our model, we mimic the effect of photoionization
by suppressing gas cooling in halos having a circular velocity lower than that  
at redshifts greater than the redshift of reionization, assumed to be $z_{re}=6$.
Although the exact value of the redshift at which reionization starts is still uncertain and other models treat it as a free parameter, we 
realized that by varying its value our results does not change appreciably and we 
choose to fix it. This is in agreement with previous works that found that  the properties of the galaxy population are almost insensitive to the assumed redshift of reionization \citep{Krav:04, Maccio:09}. Indeed, at very high redshift the leaves of the tree have very small mass and the only cooling mechanism is the $H_2$ cooling, which is very inefficient.

The simple approach used in our model for the reionization were demonstrated by \citet{Bens:03} to have very similar effects to the 
self-consistent treatment of the photoionization described, for instance, in \citet{Bens:02}.
In Appendix B we present the effect of varying the assumption on the limit of the circular velocity.

\subsubsection{AGN feedback}

There is growing evidence for a tight relationship between galaxy evolution and the growth of super-massive black holes (SMBHs) powering nuclear activity. Supporting evidences come from the tight correlations
between the BH mass and the mass of the bulge \citep{HR:04} and that between the black hole mass and the velocity dispersion of stars in the bulge \citep{Fer:00,Geb:00}.
The mutual feedback between galaxies and quasars may be the reason for such strong correlations.

We implement the accretion onto black holes in our model following the prescriptions of \citet{KH:00} and \citet{Cr:06}.
We allow two different modes of feedback, one following the starburst triggered by galaxy mergers and strong interactions, the \textit{quasar mode}, and a second one, the \textit{radio mode}, taking place at low rate  during the whole life of the galaxy.   A third natural way of growth for SMBH's is represented by the coalescence of the BHs residing in the centre of two merging galaxies. In this case, after the merger of the host galaxies, the new BH mass is simply the sum of the two progenitors. We thus simplify our treatment by ignoring the occurrence of BH binary systems and gravitational wave losses of angular momentum.

We use as a starting point a seed mass for the BH in the galaxies on the bottom of hierarchy, i.e. living in the ``leaves'' halos.  These seeds are believed to form at very high redshift, but it is quite unclear if
they formed from the direct collapse of pre-galactic gaseous discs or
if they are the remnants of massive Population III stars \citep{Volonteri:08}.  
We take a seed mass  equal to $1000 M_{\odot}$, and checked the consequence of this choice by using different mass values: the results do not depend on this choice, at least for small values of the seed.
Since we start placing the seeds at the bottom of the hierarchy, the redshift at wich they are planted depends upon the time at wich the leave haloes fall below the mass resolution of the tree and it can be higher than the redshift of the reionization. Anyway, they can not grow before the reionization starts since gas accretion, that is the main mechanism of feeding BHs, is inhibited.

The first mode of accretion is motivated by the observations of bright AGN and quasars radiating with high efficiency, close to their Eddington limit.
It is widely believed that this mechanism of accretion is triggered by galaxy mergers, since they can drive rapid inflow of gas toward the centre of the galaxy, feeding both the starburst and the BH.

As in \citet{KH:00}, we assume that during a merger of galaxies a certain fraction of cold gas is accreted on to the centre of the black hole:
\begin{equation}
\dot{M}_{BH}^{QSO}=\frac{f_{BH}M_{cold}}{1+(280\, \kms/V_c)^2} \Delta \tau^{-1}\, ,
\label{qso_mode}
\end{equation}
where $f_{BH}$ is an efficiency parameter to be chosen in order to match the relation between black-hole mass and velocity dispersion of the bulges and $\Delta \tau$ is the time interval of the integration of
differential equations. In this way QSO accretion is closely linked to the starburst activity of the galaxy, which is triggered by galaxy mergers together with BH accretion, naturally producing BH masses proportional to the mass of the spheroidal component that is formed only during the starburst.

The second way of accretion, the radio mode, is a continuous and quiescent accretion during the whole life of the galaxy, during which the black-hole accretes gas directly from the hot halo. This is supported by observations of radio galaxies (mainly in high-density environments, like groups or clusters of galaxies), which are accreting at a sub-Eddington rate, hence in a rather inefficient fashion. In spite of this inefficiency, the energy extracted from the BH accretion in the form of collimated jets is believed to contrast the cooling flow in the centre of clusters.
Assuming that in this case the BH is fuelled with a Bondi accretion rate, the resulting BH growth rate
is given by \citep{Cr:06, Som:08}:
\begin{equation}
\dot{M}_{BH}^{RADIO}=k_{AGN}\frac{M_{BH}}{10^8M_{\odot}}\frac{f_{hot}}{0.1}\Big(\frac{V_{c}}{200
\, \kms}\Big)^3\, ,
\label{radio_mode}
\end{equation}
where $f_{hot}$ is the ratio between the hot gas mass and the halo
mass and $k_{AGN}$ is a free parameter, with the dimensions of an
accretion rate ($\Msun \, \yr^{-1}$).

The most important channel building up the mass of the black hole is
the \textit{quasar} mode, in which the accretion rate can be much greater than
the Eddington luminosity. The second way of accretion, the \textit{radio} mode,
provides negligible contribution to the present-day black-hole masses,
being several order of magnitude below the Eddington rate.
Nevertheless this is an important source of feedback in high mass halos, as
it suppresses cooling flows as well as the quasar mode mechanism, and
it occurs for whole life of the galaxy, while the \textit{quasar} mode acts
only in a very short period of the galaxy life.
The feedback efficiencies of such processes, very difficult to model \textit{a-priori},
are set by trying to match various observables.
The efficiency of the \textit{quasar} mode accretion is
set by requiring the model to match the $M_{BH}-M_{bulge}$ relation, while
the \textit{radio} mode accretion efficiency is tuned to optimize the shape of
the stellar mass function, and in particular the observed knee at the
characteristic stellar masses.

The injection of energy into the ISM due to the presence of an AGN is modelled in the following way.
The mechanical heating generated by the the black hole accretion is given by
\begin{equation}\label{lbh}
L_{BH}=\eta \dot{M}_{BH}c^2\, ,
\end{equation}
where $c$ is the speed of light and $\eta$ is the global efficiency of energy production close to the event horizon in nuclear black-holes, and is set to the reference value of 0.1.
The gas cooling rate in the halo is then corrected by this energy injection according to:
\begin{equation}\label{crate_corr}
\dot{M}_{cool}^{new}=\dot{M}_{cool}^{old}-\frac{L_{BH}}{V^2_{c}/2}\, .
\end{equation}
To avoid an unphysical negative value for the cold gas accretion rate, we enforce mass conservation by assuming an amount of newly formed cold gas mass in the time-step $\Delta \tau$ the maximum between zero and $\dot{M}_{cool}^{new} \Delta \tau$, where $\dot{M}_{cool}^{new}$ is given by eq. (\ref{crate_corr}).
Hence, the cooling rate can not only be reduced, but even stopped in the case of a strong BH accretion rate.

\subsection{Galaxy Dynamics}

During the merger of two dark matter halos, the baryonic cores that they contain,
being more compact and less subject
to tidal effects, may avoid merging
with each other, and end up orbiting within the new combined halo.
An halo formed by many mergers may contain many distinct galaxies,
with the galaxy in the most massive halo becoming the central
galaxy and all the others becoming the satellites.
These satellite galaxies
gradually loose their energy and angular momentum under the action of
dynamical friction until they sink to the centre of the halo and merge
with the central galaxy. Moreover, they are subject to random gravitational collisions that
can occur with other satellites.

\subsubsection{Dynamical Friction}\label{dynfric}
The effect of dynamical friction in slowing down the orbital
velocities of galaxies moving through a sea of dark matter particles
was originally described by \citet{Ch:43} and previous SAM made use of
this approach.
Although early N-body studies \citep{Navarro:95} gave support to the
merger time-scales predicted by the classical approach, recent
numerical and hydrodynamical simulations \citep{Jiang:08, BK:08} have shown that the
Chandrasekhar's 
approach underestimates the merger time-scale for minor mergers and
overestimates it for major merger events. As an estimate of the
dynamical friction timescale we have assumed the fitting formula of
\citet{Jiang:08}, derived from hydro/N-body simulation:
\begin{equation}\label{df_epsilon}
t_{df}=\frac{0.94\epsilon^{0.6}+0.6 }{2C } \frac{M_{halo} }{m }\frac{1}{\ln(\Lambda)}\frac{R_{vir}}{V_c}\, ,
\end{equation}
where $C$ is a constant equal to 0.43, $\ln\Lambda$ is the Coulomb
logarithm, that can be approximated to $\ln\Lambda=\ln(1+M_{halo}/m)$,
$\epsilon$ is the circularity parameter of the satellite's
orbit, given by $\epsilon=J/J_c$, with $J$ the angular momentum of
the actual orbit and $J_c$ the one of the circular orbit with the same
energy. We adopt the average value found in
numerical 
simulations: $<\epsilon>=0.51$ \citep{Tormen:97, Jiang:08}. 

It is worth noting that the time-scale increases as the halo mass
increases, so that at high redshift, when the halos are less massive,
we find an high rate of merging, whereas it decreases at low redshift,
where we have an accumulation of sub-halos inside the parent halo.

An important consideration concerns the mass $m$ used to
compute the dynamical friction time-scale. As soon as a galaxy becomes a
satellite in a larger halo, its initial mass is determined by its total (dark matter and
baryonic) mass. However, as it shrinks towards the centre, a large fraction of the
dark halo is stripped away, while the mass of the gaseous core is too
dense to be stripped. Thus, the effective mass of the satellite decreases
with time and the dynamical friction time-scale should increase.
We realized that, by neglecting this effect, the
time-scales for merging were too small, leading to an overmerging,
particularly at low redshifts. 
We have taken into account the effect of tidal stripping allowing the
mass of satellite varying with time according to the time-scales predicted
by \citet{Taffoni:03}.  Hence we assume that the satellite looses their dark
halo 1.5-2 Gyr since entering the virial radius (for simplicity here we
are neglecting the tidal stripping on the baryonic core, hence we do
not consider tidally destroyed satellites).

\begin{table*}
\centering
\caption{Free parameters}
\begin{tabular}{ccccccc}
\hline \hline
Parameter & Meaning & Best-fit value & Range \\
\hline \hline
$\alpha_{quie}$  & star formation & 0.01 & 0.01-0.1 \\
$\beta_{burst}$ & burst efficiency & 0.68 & 0.5-1\\
$\epsilon_{SN}$  & SN feedback efficiency & 0.5 & 0.01-1 \\
$f_{BH} $  & AGN feedback efficiency in QSO mode & 0.006 & 0.001-0.1\\
$k_{AGN} $  &  AGN feedback efficiency in radio mode & $1\times
 10^{-6}$ & $10^{-7}-10^{-6} $ \\
\hline \hline
\end{tabular}
\label{tab_par}
\end{table*}

\subsubsection{Satellite Collisions}

In addition to the mergers driven by dynamical friction,
we include random collisions between satellites.
We consider for the satellite encounters the cross-section of \citet{Mamon:92},
which is a simplified version of that derived from N-body simulations by \citet{Roos:79}.
The resulting merger rate is given by:
\begin{equation}\label{satcoll}
k=2\sqrt{\pi}\alpha_p^2\alpha_v R_{vir}^2v_{gal}K(V_c/v_{gal})\, ,
\end{equation}
where $v_{gal}$  is the internal dispersion of the satellite galaxy, $V_c$ and $R_{vir}$ are the circular velocity and virial radius of the host halo, respectively, $\alpha_p=4$ and $\alpha_v=5.4$ are dimensionless parameters, while $K$ is a dimensionless merger rate given by:
\begin{equation}\label{mrate}
K=\Big[\frac{1}{x}+\frac{2}{x^3}-\frac{2\exp(-x^2)}{x^3}-\frac{3\sqrt{\pi}}{2}\frac{erf(x)}{x^2} \Big]
\end{equation}
\begin{equation}
x=\frac{\alpha_v v_{gal}}{2V_c} .
\end{equation}
For realistic values of the ratio $V_c/v_{gal}$ the merger rate decreases with this ratio and becomes very low for galaxy clusters.
Given the merger rate we can compute the collision time as:
\begin{equation}
t_{coll}=\frac{1}{kn}\, ,
\end{equation}
where $n$ is the number densities of galaxies in the halo. We obtain
shorter merger time-scale at high redshift, when the number density of
galaxies is higher.
The probability for having a merger for a satellite with another one,
randomly chosen, will be $P=\Delta t/t_{coll}$, where $\Delta t$ is the
lifetime of the halo.
We consider only binary mergers, and we assume that such event occurs when $P\ge1$.
During a collision between two satellites we allow a burst of
star formation in the same way as the one occurring during a merger
between a satellite and the central galaxy.
The effect of including such process, in addition to the mergers involving the central galaxy due to dynamical friction, is to decrease the fraction of low-mass galaxies in favour of intermediate-high mass objects, hence to slightly modify the shape of the galaxy mass function.

\subsubsection{Remnant}

After the merger, the total mass of the remnant (dark matter and
baryons) is the sum of the two merging galaxies, while the new
radius and circular velocities are computed from applying the conservation of
energy and the virial theorem. According to the virial theorem, the total
internal energy is given by $E_{int}=-T$. By applying the conservation of
energy:
\begin{equation}
T_{new}=T_1+T_2-E_{orb}\, ,
\end{equation}
where $T$ denotes the kinetic energy and
$E_{orb}$ is the mutual orbital energy:
\begin{equation}
E_{orb}=-f_{orb}\frac{Gm_1m_2}{r_1+r_2}
\end{equation}
($r_1$ and $r_2$ are the radius of the two progenitors and $f_{orb}$
is a parameter which weakly depends on the density profile, we assume $f_{orb}=2$). These
considerations yield to:
\begin{equation}
R_{new}=\frac{(m_1+m_2)^2}{m_1^2/r_1+m_2^2/r_2+f_{orb}m_1m_2/(r_1+r_2)}\, .
\end{equation}
In the case of major mergers involving two equal galaxies ($m_1=m_2$
and $r_1=r_2$) the size of the remnant is a factor of $4/3$ greater
than the initial size.

\begin{figure*}
\begin{center}
\includegraphics[width=\columnwidth]{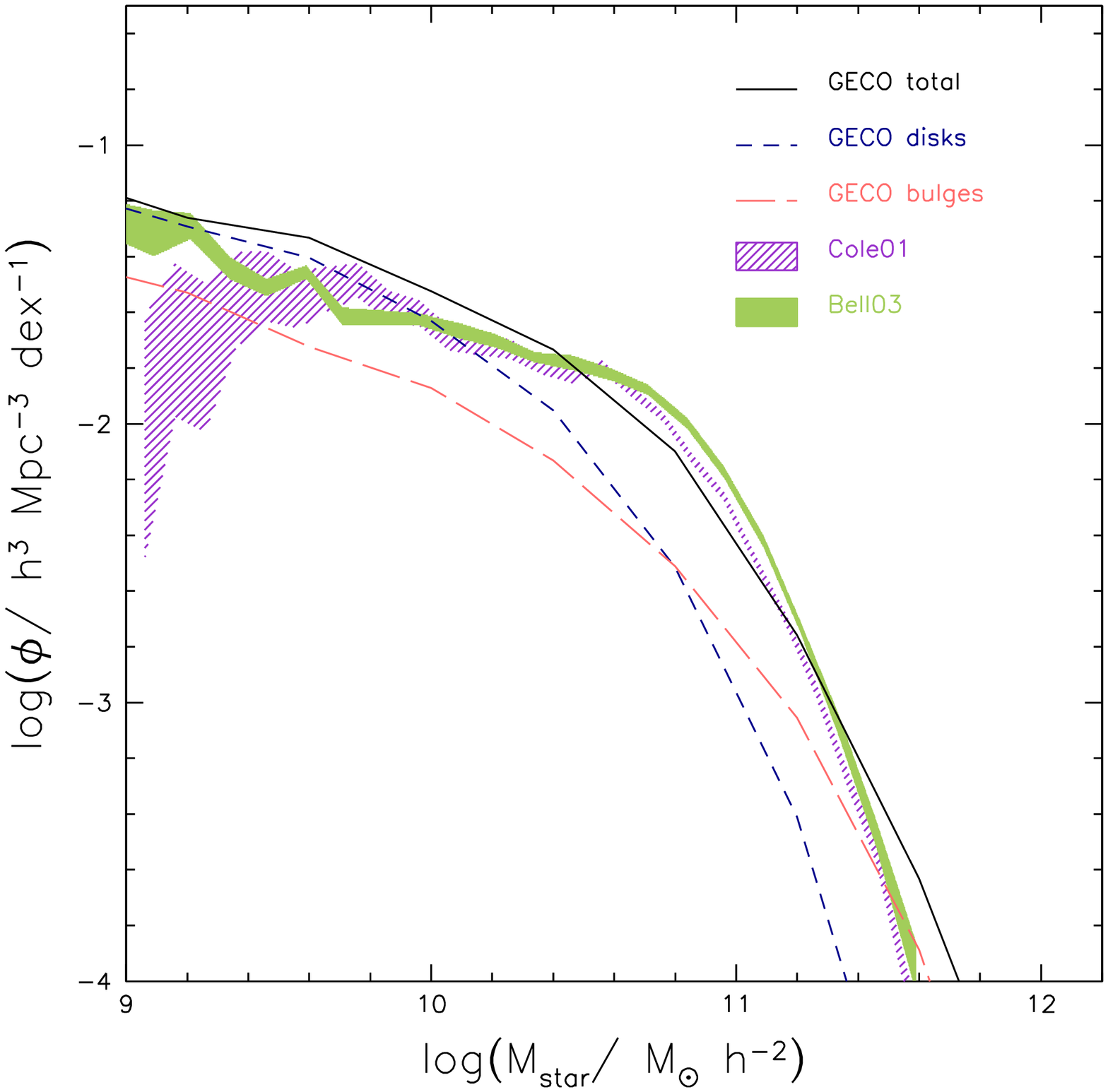}
\includegraphics[width=\columnwidth]{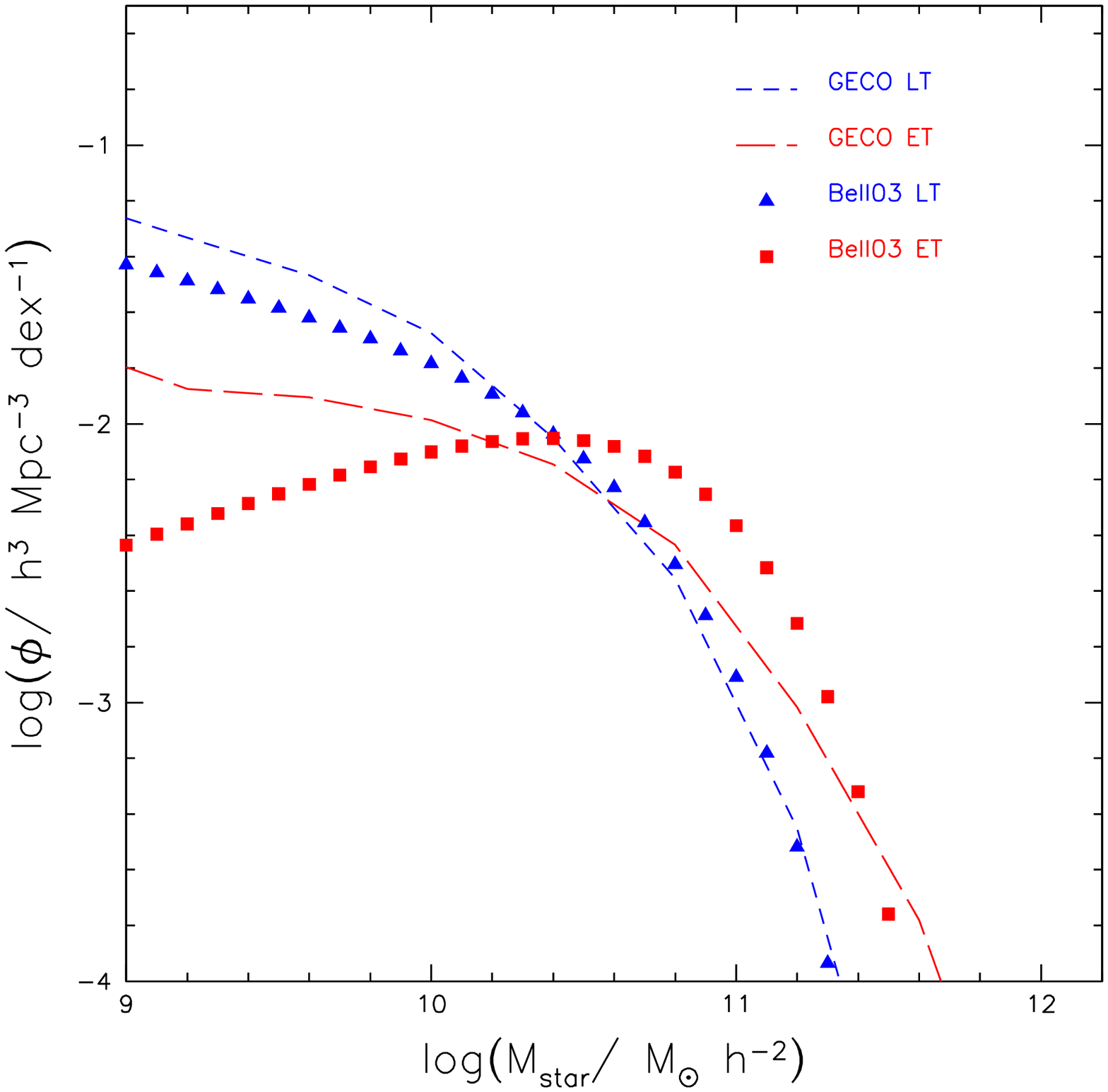}
\end{center}
\caption{Left panel:Stellar mass function at z=0. Black solid line
  represents the SMF resulting from GECO, purple shaded region stays for
  the C01 observed mass function with its error-bar, and the green shaded one
  from B03.
  Also shown is the contribution to
  the total stellar mass content from bulges (red long dashed line) and
  discs (blue short dashed line).
  Right panel: Stellar mass function at z=0 of early-type (red long
  dashed line) and late-type galaxies (blue short dashed line) compared with
  the observed one by B03: red squares for early-type
  and blue triangles for late-type. 
 }
\label{geco_smfz0}
\end{figure*}

\subsection{Galaxy Morphology}

In our model the morphology of a galaxy is determined by the
relative importance of the bulge over the disc component. The only
way of formation for discs is through the quiescent mode of star
formation, while bulges can grow in two ways: through star
formation occurring in a starburst event, hence triggered by any
event of merger, and through the disruption of discs following a major
merger. In this
case all the stars belonging to the discs undergoing a merger are added to the
bulge of the remnant galaxy. Such galaxy may eventually form a new
disc if some fraction of cold gas is still present.

We can assign a crude morphological type to each galaxy by using the ratio
between the bulge mass and the total stellar mass: $r=M_{bulge}/M_{star}$.
Using the prescription of \citet{Bertone:07}, we classify as Ellipticals
galaxies with more than $70 \%$ of their stars in bulge, as Spirals
galaxies having $0<r\le 0.7$, and as Irregulars galaxies without any bulge.

\section{Setting the free parameters}\label{param}

In the previous section we have introduced several free parameters in
the analytical prescriptions which model galaxy formation.
In summary they are:
\begin{enumerate}
\item{$\alpha_{quie}$: efficiency of star formation in quiescent mode  }
\item{$\beta_{burst}$: efficiency of star formation in starbursts }
\item{$\epsilon_{SN}$: efficiency of SN feedback   }
\item{$f_{BH}$: efficiency of AGN feedback in the \textit{quasar} mode   }
\item{$k_{AGN}$: efficiency of AGN feedback in the \textit{radio} mode   }
\end{enumerate}
We will first infer constraints on their values by comparing the model results, obtained for a large grid of
parameter values, with observations of the local universe. In
particular we focus on the build-up of the stellar mass, and we
require our model to fit the local stellar mass function, as well as
the relationship between the black-hole mass and the bulge mass.
The stellar mass function is obviously influenced by all the
parameters together.
The star formation efficiency in the quiescent mode $\alpha_{quie}$ affects the overall shape
and normalisation of the function; $\epsilon_{SN}$ affects the faint-end
(or low-mass end); $\beta_{burst}$ determines the contribution of the bulge on
the total stellar content, and we require that at the bright-end bulges
dominate over discs. The AGN efficiencies, $f_{BH}$ and  $k_{AGN}$
shape the bright-end of the mass function. Moreover, the AGN efficiency in the \textit{quasar} mode,
i.e. the one triggered by the starburst episode, determines the
relationship between the mass of the black-hole and the bulge mass,
affecting mainly the normalisation of the relation, while the
slope turns out to be very close to the one observed almost
independently on the value of this efficiency,
since in the model the growth of BHs and bulges are closely linked.
In Table \ref{tab_par} we
report the value of these parameters in our best-fit model
and the allowed range.
Unless otherwise stated, in the following we always will refer to this best-fit model.

\section{Matching data from the local universe}\label{localUn}

\begin{figure}
\begin{center}
\includegraphics[width=\columnwidth]{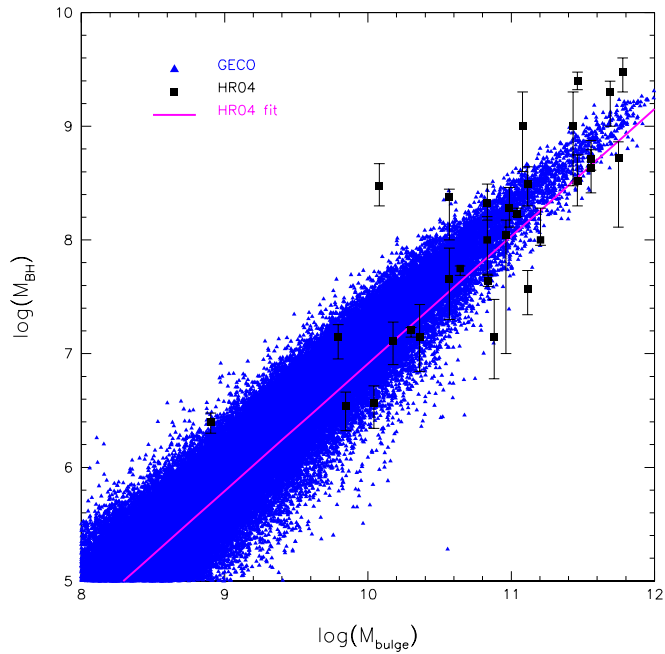}
\end{center}
\caption{Black-hole versus bulge mass relation at z=0. Blue triangles
  represent data for modelled galaxies, while black squares with
  error-bars are data from \citet{HR:04}.
  The magenta solid line is their best fit relation.
  }
\label{geco_bhbul}
\end{figure}

\begin{figure}
\begin{center}
\includegraphics[width=\columnwidth]{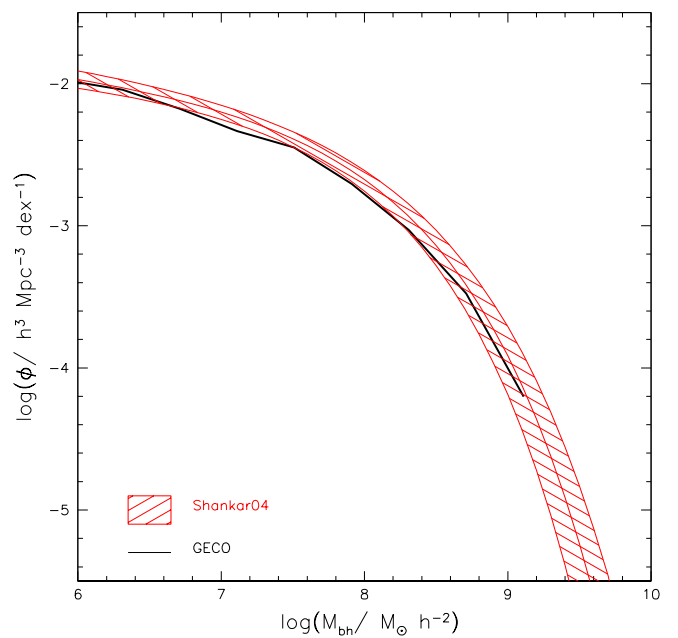}
\end{center}
\caption{Local Black-hole mass function of the model (black solid
  line) compared with the estimates of   \citet{Shankar:04} (red
  shaded region).
  }
\label{bhmf}
\end{figure}

A key observable for comparing predictions of any models of galaxy formation is the number density of galaxies as a function of their stellar mass in the local universe, the stellar mass function (SMF) at $z=0$.  We will use it to constrain our model free parameters.
We refer to two measurements of the local SMF. The first is that reported by \citet{Cole:01}, hereafter C01, obtained by combining data from the Two Micron All Sky Survey (2MASS) and the 2dF Galaxy Redshift Survey.
The second estimate, by \citet{Bell:03}, exploited a large sample from 2MASS and the Sloan
Digital Sky Survey (SDSS). We choose to show their {\it g}-selected sample (taken from their table 5).
Both estimates use near-infrared data to compute stellar masses and
hence are highly reliable. All data are transformed to a Salpeter IMF.

In the left panel of Figure \ref{geco_smfz0} we show the stellar mass function at $z=0$, resulting from the model and compared with the C01 and B03
observations. The model is able to reproduce with good accuracy the
shape and normalisation of the
SMF, especially in the bright-end ($\Mst > 10^{11}\Msunh$).
Compared to the C01 mass function, the model shows a slight excess of galaxies at the faint-end. However, the B03 estimate presents a steeper faint-end, resulting in closer agreement with our prediction.
Our model also shows an underestimation of galaxies at the knee of the mass function ($10^{10.5}-10^{11} \Msun/h^2$), an effect already present, for instance, with the Millennium galaxy catalogue \citep{Bertone:07}. Such dearth may be partly explained with the combined effect of SN and AGN feedback that are both effective in this mass range. 
Moreover, since the deficit is due to a lack of intermediate-mass early-type galaxies, (see right panel of Fig. \ref{geco_smfz0})
it is likely connected to the excess of low-mass spheroids, as satellite galaxies are not able to become massive enough in the model for the reasons described later in this section.

It is worthwhile to stress the fundamentally different shapes of the
halo mass and the luminous (stellar) mass functions presented
here. The general shape of the dark matter halo mass function (see
Figure \ref{umf}), as predicted by any theoretical models based on
hierarchical dark-matter clustering, is a pure power-law with a steep
faint-end and no knee at the high masses. A turn-off mass does exist,
but it occurs at much higher masses ($M\sim 10^{15}\ M_\odot$) than
the characteristic one observed in the stellar mass function.
In order to obtain a Schechter shape for the mass function, many effects contribute. At the faint-end side, the feedback from SN and from the photoionization background suppresses star formation in small halos, hence reducing the faint-end slope.  At the bright-end, the difference
between the exponential cut-off observed in the stellar mass function
and the power-law shape expected for the halo mass function is
explained with the inter-play of various independent effects.
On one hand, there is a marked dependence of the gas cooling on the halo mass:
the cooling time decreases with the halo mass. In addition, the two modes of gas cooling discussed in Sect. 4.3 have an halo mass dependence: in the \textit{cold mode} regime the gas accretes towards the centre in a time-scale given by the free-fall time, while in the \textit{hot mode} regime the accretion rate is governed by the cooling time, and is much less efficient than in the former case.
Since in high-mass halos the accretion mainly occurs through this last mechanism, this may explain why in this halos gas is not able to cool efficiently. Moreover, the AGN feedback acting on this scales, further reduces the cooling, leading to an approximately correct number of massive galaxies.

In Figure \ref{geco_smfz0} (left panel)  the separate contributions to the total stellar content from discs and bulges are also shown, as blue short-dashed and red long-dashed lines, respectively.
As expected, the bright-end is dominated by bulges, meaning that objects with 
very high-mass ($\Mst>10^{11}\Msunh$) do not essentially contain discs, that 
have been presumably destroyed during a major merger. On the contrary, discs 
overcome the number densities of bulges at low masses, and start to be
dominant in the mass range $\Mst \sim 10^{10.5}\Msunh$.
Note that this value is close to the transition mass, $\sim 3\times
10^{10} \Msunh$ at z=0, at which a transition in several galaxy
properties is observed \citep{Kauff:03} and where early-type galaxies start to dominate over the late-type ones in the local universe \citep{Baldry:04, Bundy:05}.

So far we have considered only the contribution to the total stellar
mass density from the disc and bulge components. In the right-end
panel of Fig. \ref{geco_smfz0} we show the mass function of early-type
and late-type galaxies. 
As defined in the previous Sect., early-type galaxies are defined as
galaxies whose bulge makes at least $70\%$ of the total stellar
mass. We include in the definition of late-type systems both spiral
and irregular galaxies. We compare the predicted mass function with
the results of B03. 
With this definition of morphologies, the agreement with the
observational data is quite good, in particular we accurately
reproduce the transition mass at which the number density of
ellipticals/S0s equals that of 
late-type galaxies locally.  
Anyway, we note quite a large excess of faint early-type galaxies with masses below $10^{10}\Msun/h^{2}$. This is a long-standing problem that afflicts all the SAMs in the recent literature \citep{Weinmann:06, Cr:06, Cattaneo:08, Som:08} and may be regarded as an 
indication of a wrong treatment of the satellite population. 
A common way among SAMs of treating the satellites is to assume that the hot gas in a system that merges into a more massive one is instantaneously stripped (strangulation)  and added to the reservoir of the new central galaxy. As a consequence, the star formation in the new satellite is quenched shortly once it has consumed its cold gas.   
This crude treatment of the strangulation is clearly an oversimplification and recent hydrodynamical simulations indicate that such a process occurs on a larger time-scale, of the order of 1-10 Gyr \citep{McCarthy:08}, allowing the star formation to continue for some time after the coalescence, hence lowering the fraction of red satellites \citep{Kang:08}.
Although even tidal disruption (see for instance \citealt{Som:08}) can help in decreasing the faint-end, it is unlikely that it can change the relative number of early and late-type satellites \citep{Kimm:09}. Other effects, such as the ram pressure on the cold gas in the galaxy \citep{Lanzoni:05} may also play a role in the formation of the satellite population and should be included in the future SAMs.
\begin{figure*}
\includegraphics[width=2\columnwidth]{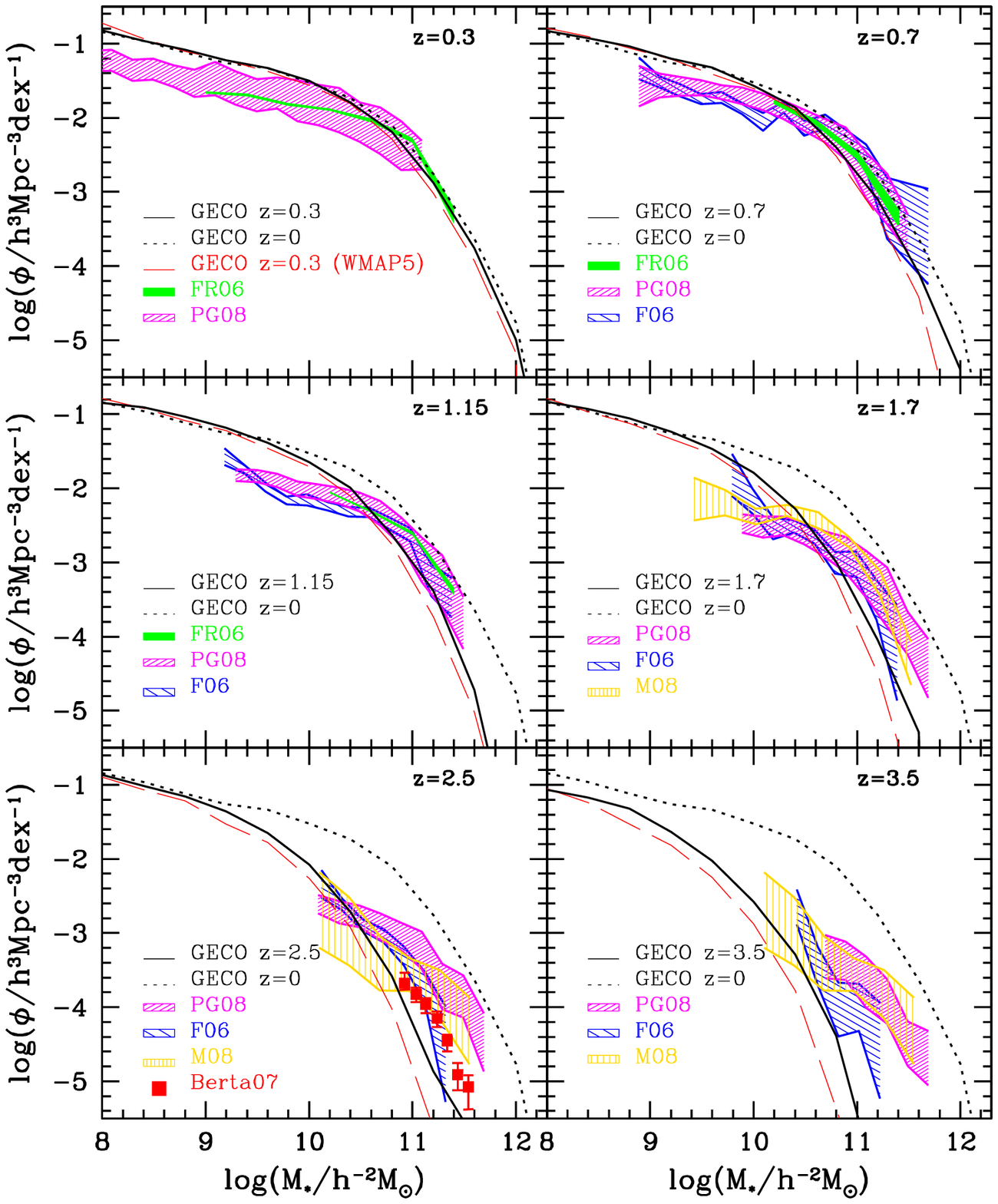}
\caption{Comparison between GECO's mass functions at high redshift,
  up to $z\sim 3.5$. The green shaded region indicates the mass
  function of \citealt{Fr:06}, the magenta one is that of
  \citealt{PG:07}, the blue region is the estimate of
  \citealt{Font:06}, the yellow shading is from
  \citealt{Marchesini:08} and the red squares at $z\sim 2$ show the results
  from \citealt{Berta:07}.
  GECO predictions are represented with a
  solid black line. The dotted line in all the panels is the GECO mass
  function at $z=0$, reported as reference and the red dashed line is the model
  prediction for the WMAP5 cosmology. 
  }
\label{geco_ev_all}
\end{figure*}

In order to set the efficiency of black-hole growth, we require our
model to match the observed relation between the black-hole mass and
the mass of the stellar bulge in the local universe. 
Since the major channel for the growth of black-hole is given by the
accretion during the \textit{quasar} mode triggered during a starbust
event, where all the stars formed are added to the bulge component, it
comes with no surprise that black-hole and bulge are closely linked. 
Nevertheless, in order to obtain an acceptable fit, where both the
slope and the normalisation of the relation are well reproduced, some
fine tuning of the $f_{BH}$ and $k_{AGN}$ parameters were required. 

In Figure \ref{geco_bhbul} we show the comparison between our derived relation
(blue triangles)
with data from Haring \& Rix (2004), represented by the black squares
with error-bars, and their best-fit relation, shown by the magenta
line. The observed relation is well reproduced by our model over the whole mass
range (note in any case that the \citet{HR:04} fit is derived only for
bulge masses $M_{bulge}>10^{10} \Msun$). 
Anyway, the scatter in the model results to be smaller than that
  observed, as the model neglects the contribution to the bulge
  formation due to disk instabilities, and the only channel to build
  spheroids is through mergers, hence directly linked to the BH
  formation.

A further test performed on the black-hole population is shown in 
 Figure \ref{bhmf}, where we show the BH mass function of the model,
 compared with the local estimate by \citet{Shankar:04}.
The remarkable agreement between the observed and predicted BH
number densities, together with the well matched BH-bulge relation is
encouraging and suggests that our method of implementing AGN feedback,
though simple, is accurate enough.

\section{Comparing GECO with high-redshift galaxy statistics}\label{Highz}

Having obtained in the previous Sect. a good match of GECO's predictions with the global properties of galaxies in the local universe, it is instructive now to attempt a first comparison of the evolutionary properties of the model with the galaxy statistics in the distant universe, with no further adjustment of
the model parameters.  Given the wide range of physical processes included in GECO, one would expect at least a rough agreement of the global properties, 
unless some key physical processes might have been overlooked.

\subsection{The Evolution of the Stellar Mass Function}

We first test the resulting mass function with observations up to
$z\sim3.5$.  A comparison of the GECO's mass functions with those
observed is reported in Figure \ref{geco_ev_all}. Here we match our
model mass functions with results from \citet{Fr:06}, hereafter FR06,
\citet{Font:06}, hereafter F06, \citet{PG:07}, PG08,
\citet{Marchesini:08}, M08 and \citet{Berta:07}, B07.
All data are transformed to a Salpeter IMF. 
We consider only mass function estimates which take into account IRAC
bands in the stellar mass derivation, since the lack of Spitzer data
in the near-IR leads to an overestimation of the stellar masses at $z>3$ 
(see F06).
Note that error-bars in FR06 take into account only Poisson
statistics, hence resulting to be much smaller than in the other
data, where uncertainties in photometric redshifts, stellar masses and
cosmic variance are considered.

Considering the spread between different estimates of the mass
function, the comoving number density of massive galaxies in our model
is roughly consistent with the observed values, although with a
tendency to underpredict the data, especially at high redshifts. 
In the highest redshift bin considered ($z \sim 3.5$), the bright-end
is marginally consistent with that of F06 but systematically lower
than M08 and PG07. Note however, that including the effects of
different assumptions in the Spectral Energy Distribution (SED) modelling leads to different MF
estimates fully consistent with each other (see M08).

A remarkable feature in the comparison of model expectations
with the data concerns the faint end of the mass function, where the
model tends to keep systematically in excess of the observed number of
low-mass objects at any redshifts: whereas there is a significant
evolution in the number density of low mass galaxies in the data, the
model predicts modest or no evolution up to redshift of one. The mismatch between data and hierarchical models at the faint-end
seems a common feature among different models
\citep{Font:06, Fontanot:09, Monaco:09}.

In the same figure \ref{geco_ev_all} we also include the predictions for the SMF adopting the WMAP5 set of parameters indicated by the red dashed line. The main difference with the Concordance cosmology is a lower value of $\sigma_8$ and a tilt in the primordial power spectrum ($n<1$), implying a delay in the formation of structures. Indeed, this is observed in the evolution of the SMF, where, at high redshift  the model predicts a later assembly of galaxies, exacerbating the underproduction of massive galaxies at higher redshifts. 
\begin{figure}
\includegraphics[width=\columnwidth]{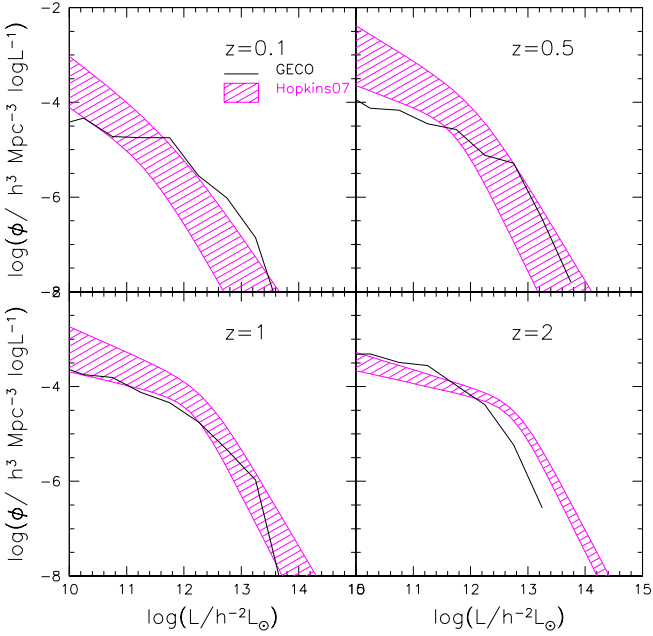}

\caption{Comparison between GECO's quasar luminosity functions at
  different redshifts and observations by \citealt{Hopk:07a}.
  The magenta shaded region indicates the quasar luminosity
  function of \citealt{Hopk:07a}, while GECO predictions are represented with a
  solid black line.}
\label{qlf}
\end{figure}

It is worth to note, however, that such a comparison must be taken with caution before deriving firm conclusions. The errors in the stellar mass estimates can strongly affect (in the sense of increasing) the counts in the bright-end where the mass function is very steep.
An improved approach would be to convolve the model stellar mass function with the error distribution in the stellar mass given by the observations \citep{Kit:06,Stringer:08, Cattaneo:08}.
Of course the uncertainties in the stellar mass estimates increase with redshift and this can influence the
amount of the observed evolution of the stellar mass function.

Finally, it is not unlikely that the derivation of the stellar mass function at the faint-end might have been seriously influenced by incompleteness effects in the source selection and have produced a spurious evolution at low masses. More extensive deep imaging by Spitzer will be carried out on large areas during the warm mission phase, which will offer an opportunity to address these issues shortly.

\subsection{Quasar luminosity function }

The gas accretion onto black holes described in Sect. 3.5.3 
triggers the AGN activity whose bolometric luminosity associated is
given by $L_{BH}$ (eq. \ref{lbh}). Although the accretion rate in our model is not limited by
the Eddington rate, we limit the quasar luminosity to lie below 
this limit:
\begin{equation}
L_{QSO}=\max(L_{BH},L_{Edd}) \,.
\end{equation}

In order to test the consistency of the
feedback from AGN activity on the galaxy population we compare the
quasar luminosity function predicted by GECO with observational estimates. 
To avoid to introduce further uncertainties due the
obscuration factor in the optical bands, we compare our predictions with
bolometric luminosity functions. 
We refer to the compilation  from \citet{Hopk:07a}, obtained by
collecting a large multi-wavelength dataset, from the optical, soft and
hard X-ray, near- and mid-IR. The authors give obscuration-corrected
luminosities, aimed to represent the intrinsic quasar population.
In Figure \ref{qlf} we show our predicted quasar luminosity function
up to z=2. The model gives an acceptable description of the evolving
quasar population at low and intermediate redshift, but at higher
redshift (panel at z=2) it
underpredicts the number of luminous quasars. 
Varying the efficiency of the accretion onto black hole ($f_{BH}$ and
$k_{AGN}$) does not solve the problem, because the match at high
redshift would be obtained at the expense of the agreement of the
local black-hole mass function. 
This may be an indication that the parametrisation of the AGN activity
used in the present work is not completely adequate, as already
suggested by \citet{Marulli:08}, and an accretion efficiency increasing
with redshift may be required (as suggested even by recent hydrodynamical
simulations, see \citealt{Viola:08}), to fit simultaneously the low and the
high redshift luminosity function \citep{Marulli:08, Lapi:06}.
We also suggest that higher accretion efficiencies at high redshift may be reached also with
an enhanced cooling efficiency at early times. This alternative 
may also provide a way to form galaxies more efficiently at high redshift, hence,
improving the match of the stellar mass function evolution.

\begin{figure*}
\begin{center}
\includegraphics[width=0.7\columnwidth]{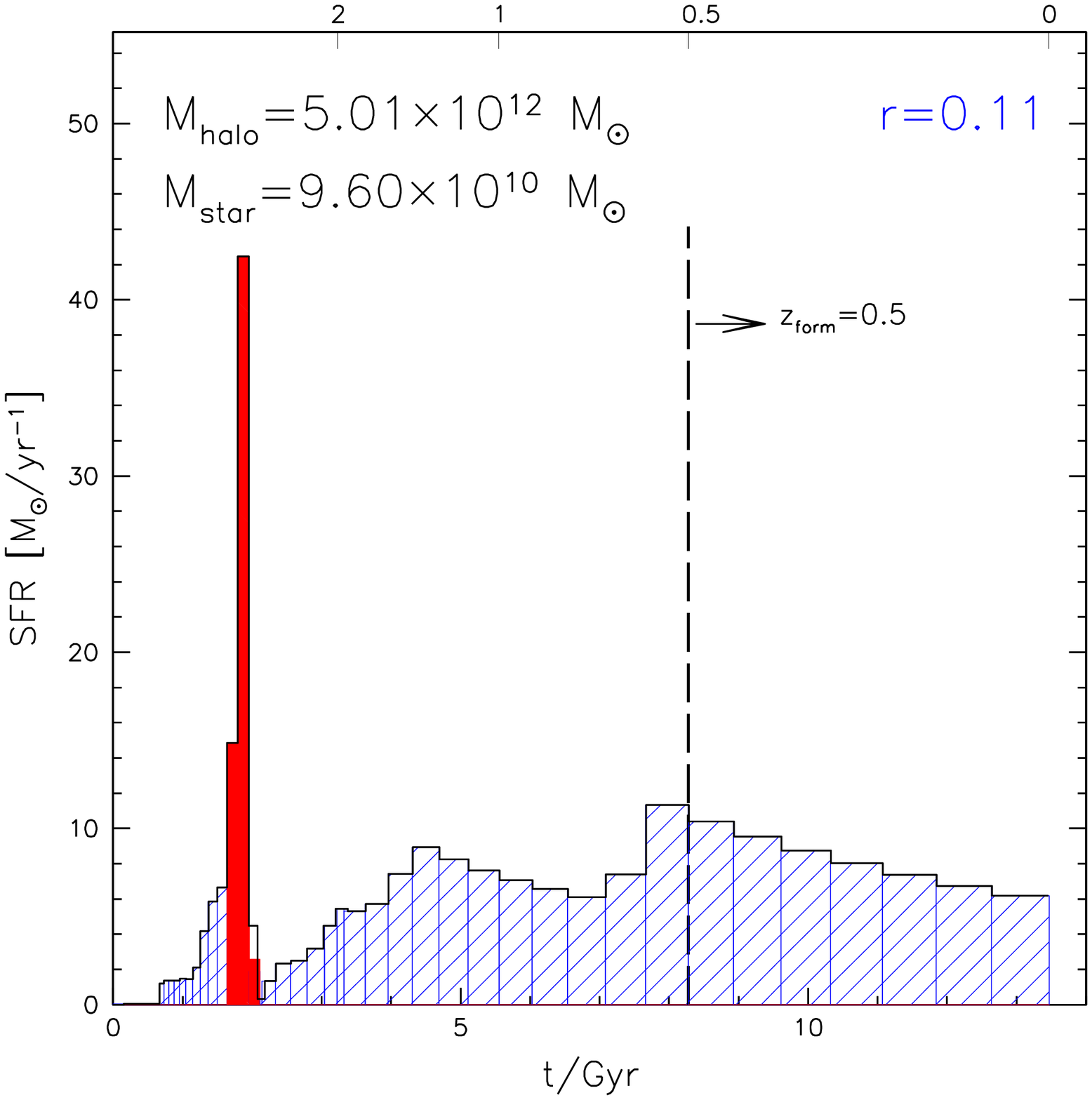}
\includegraphics[width=0.7\columnwidth]{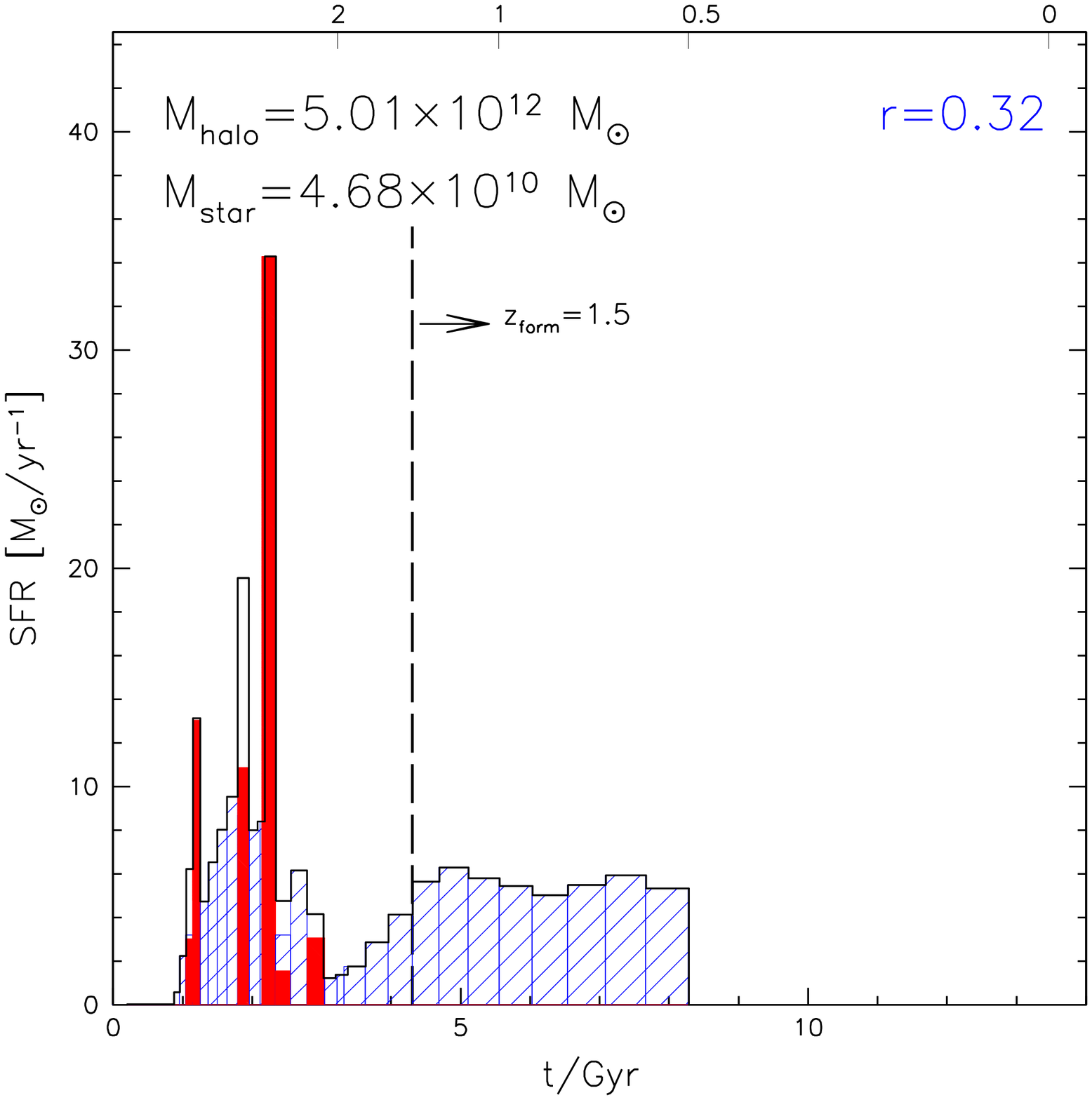}
\includegraphics[width=0.7\columnwidth]{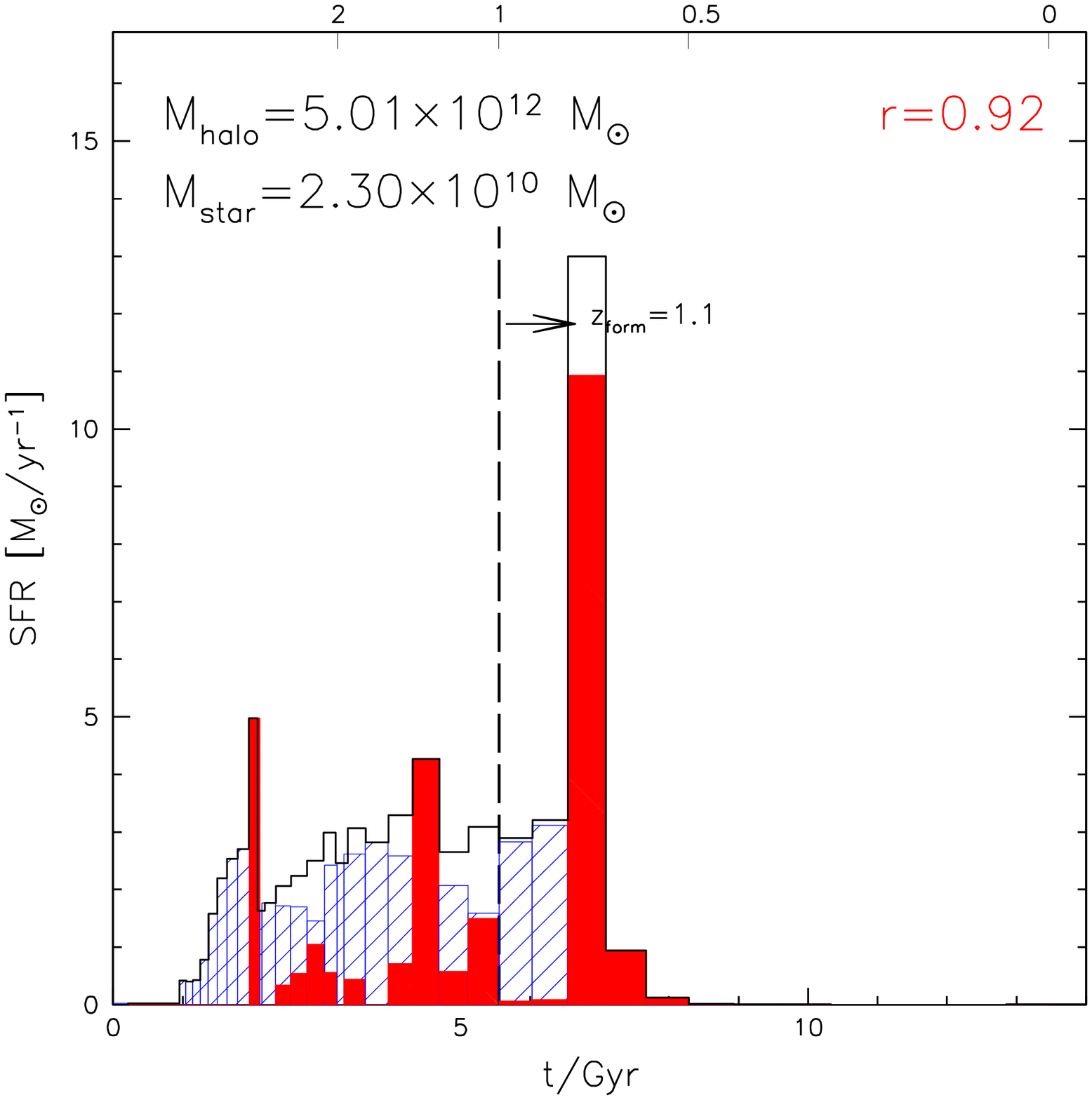}
\includegraphics[width=0.7\columnwidth]{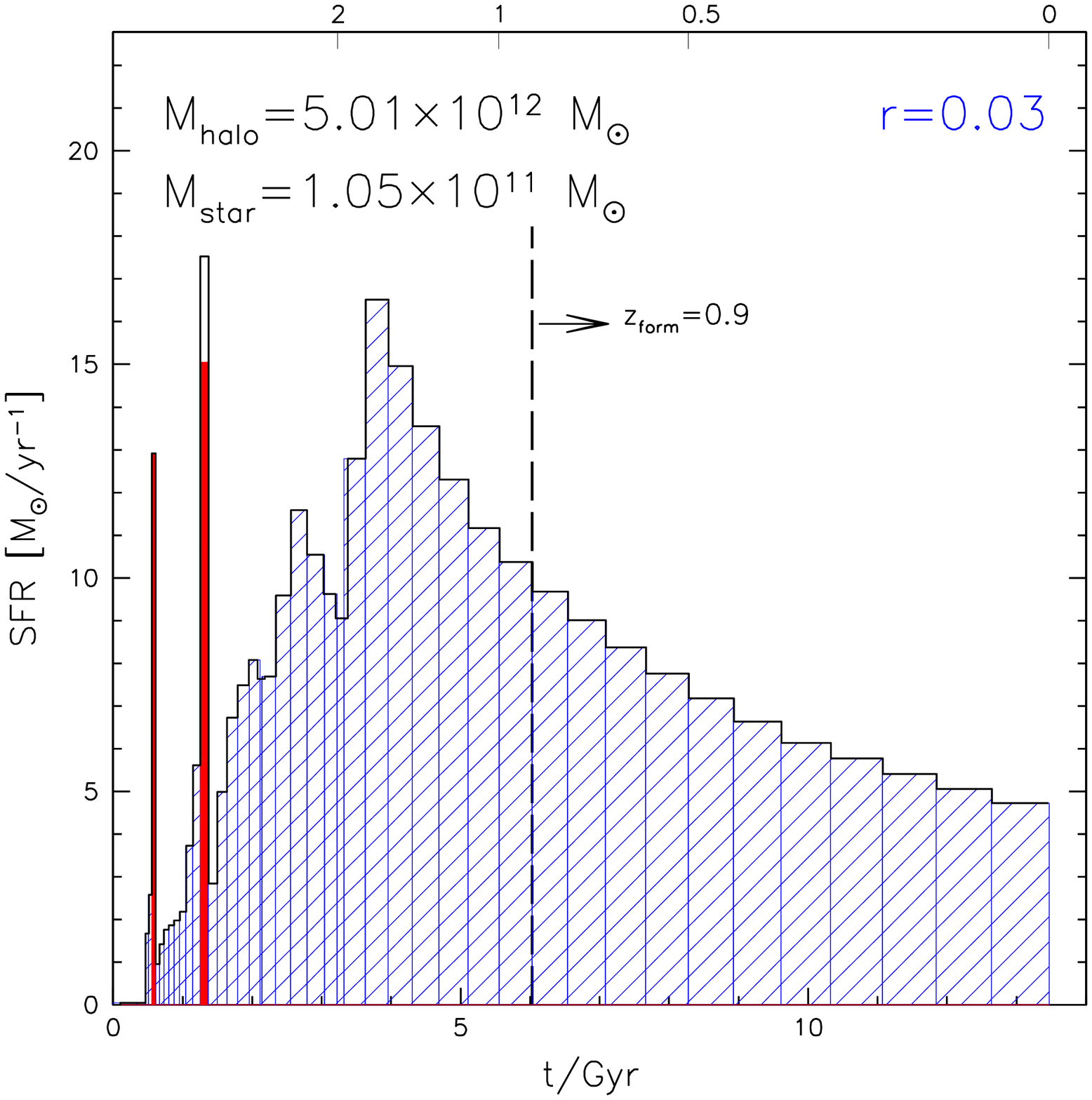}
\end{center}
\caption{SFHs of stellar populations for different realizations of the central galaxy of a Milky-Way like halo, having masses equal to $5 \times 10^{12} \Msun$. These SFHs sum up the contributions to the final galaxy stellar mass by all the stars present in the galaxy at z=0.
Blue shaded histograms show the star formation occurring in the quiescent mode, the red ones represent the bursty mode, while the black envelope is the total SFR. The vertical dashed line indicates the formation redshift of the galaxy.
The final stellar masses in each realizations are indicated. }
\label{geco_sfh181}
\end{figure*}

\begin{figure*}
\begin{center}
\includegraphics[width=0.7\columnwidth]{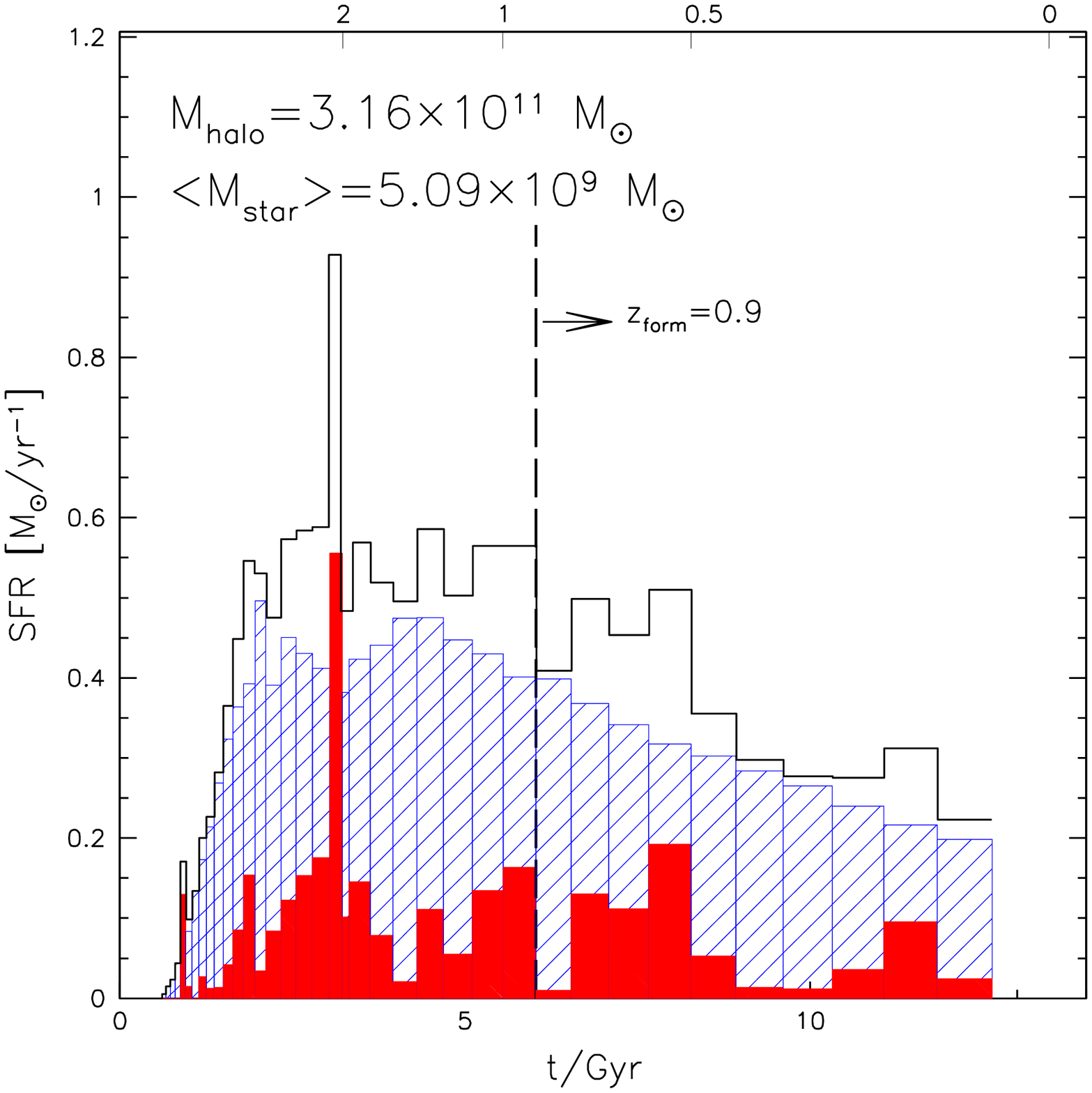}
\includegraphics[width=0.7\columnwidth]{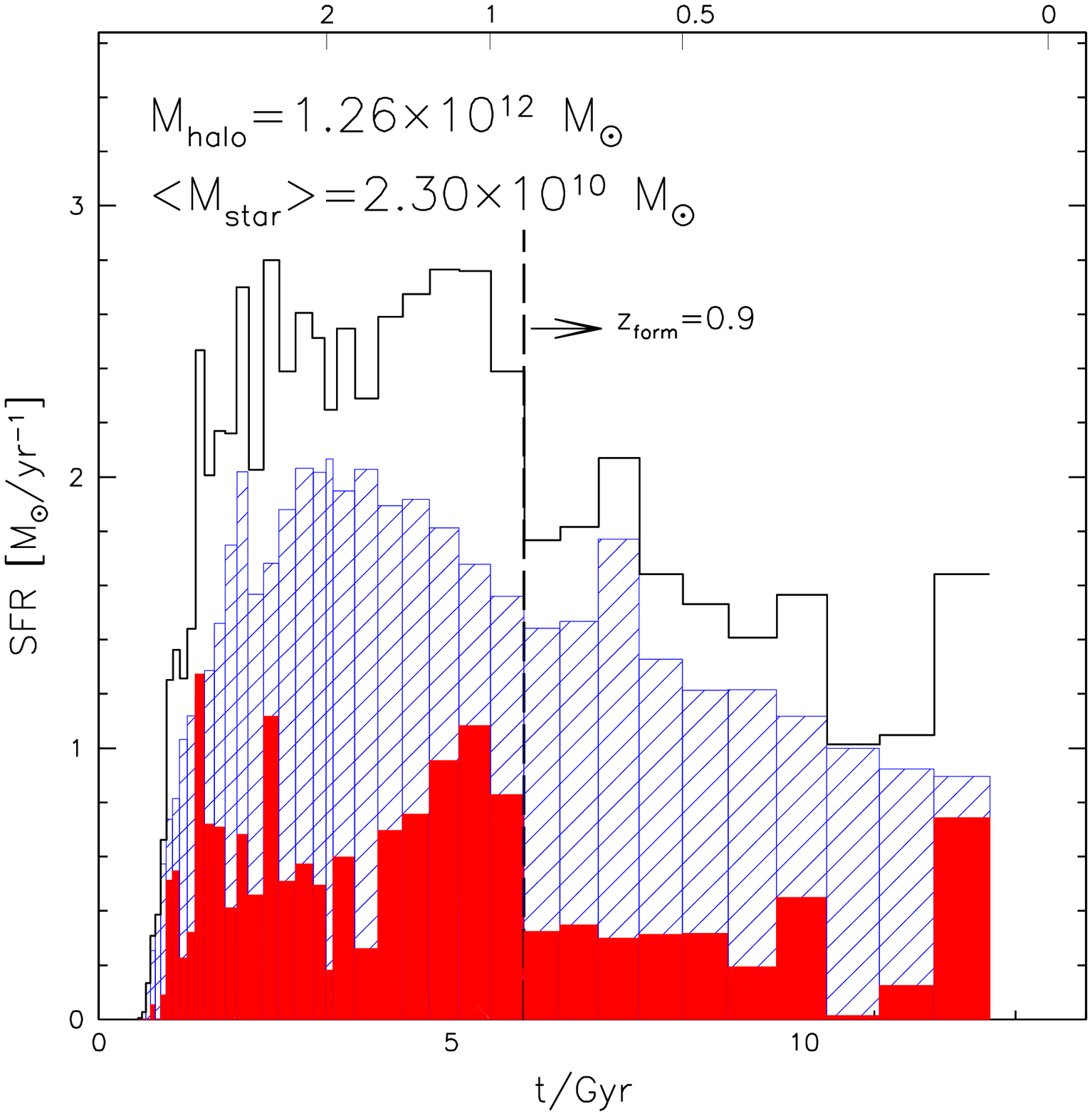}
\includegraphics[width=0.7\columnwidth]{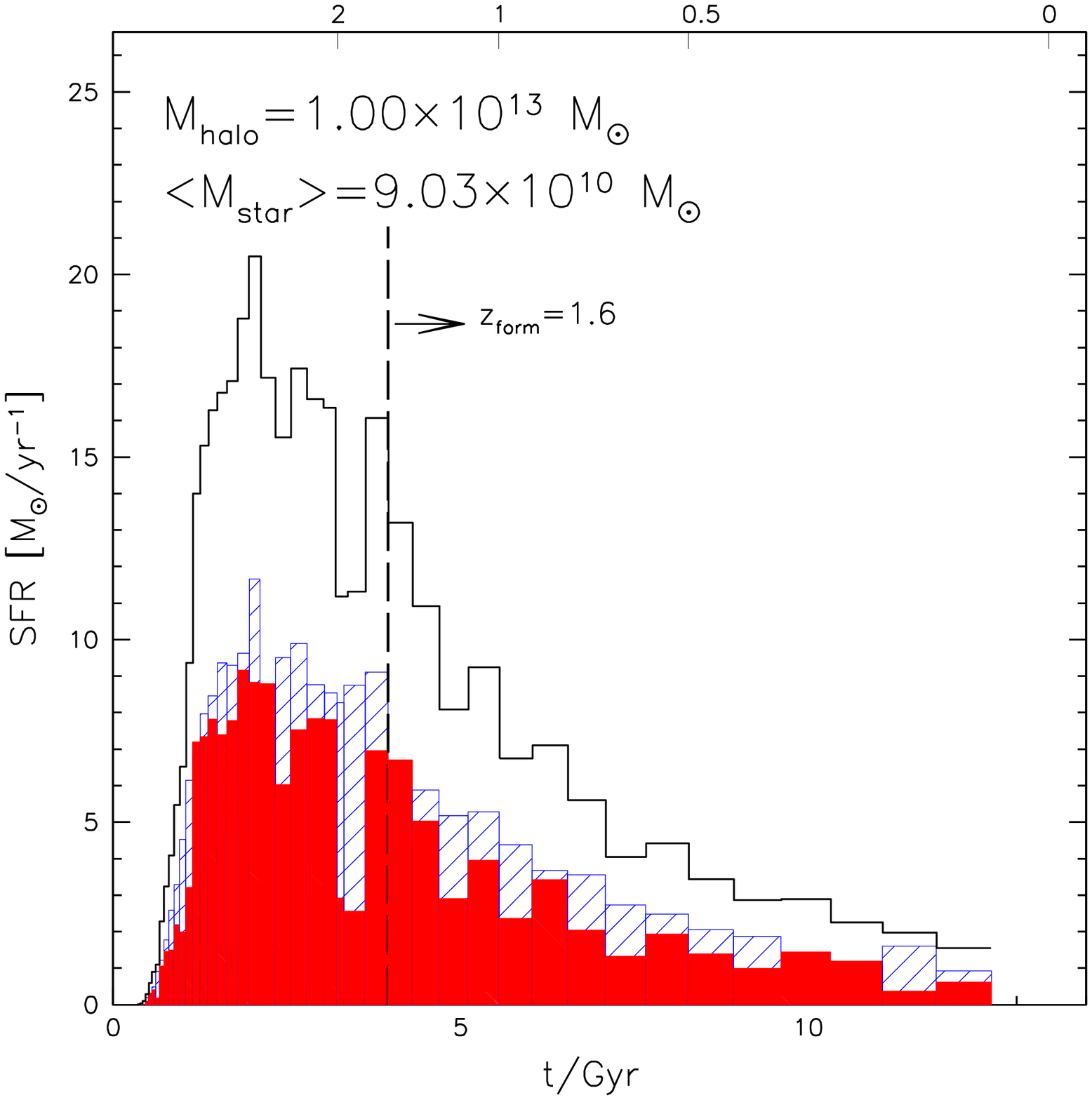}
\includegraphics[width=0.7\columnwidth]{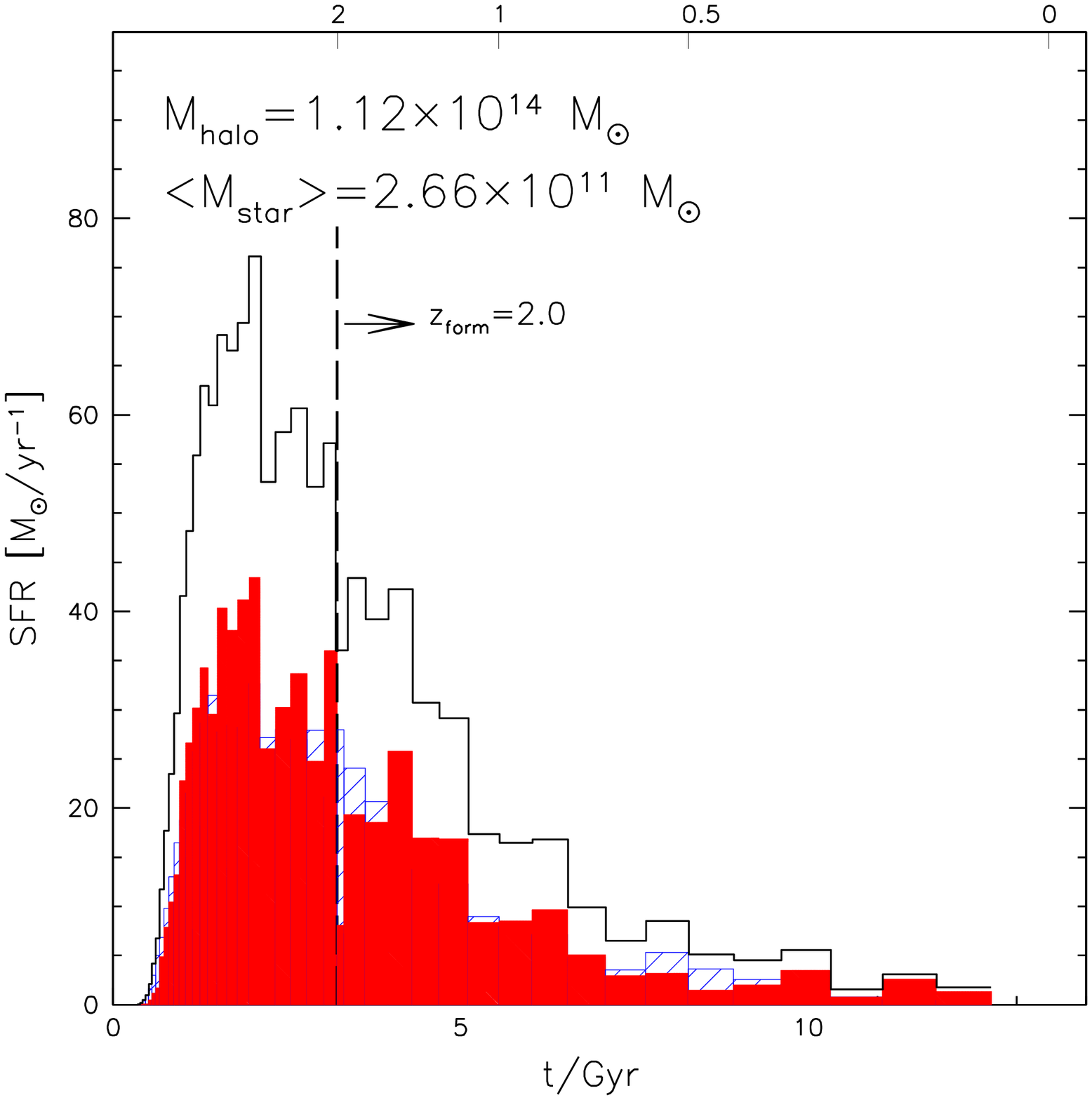}
\end{center}
\caption{Averaged SFHs for central galaxies of 4 different halos, with masses ranging from $\sim 10^{11}\Msun$
  to $\sim 10^{14}\Msun$. The average are computed  over 100 realizations. As in figure \ref{geco_sfh181}, blue shaded histograms show the   star formation occurring in quiescent mode, the red ones represent the bursty mode, while the black envelope is the total SFR.
The host halo mass and the averaged final stellar mass are indicated.}
\label{geco_sfhav}
\end{figure*}

\subsection{Formation Histories of the Stellar Populations}

\begin{figure*}
\includegraphics[width=\columnwidth]{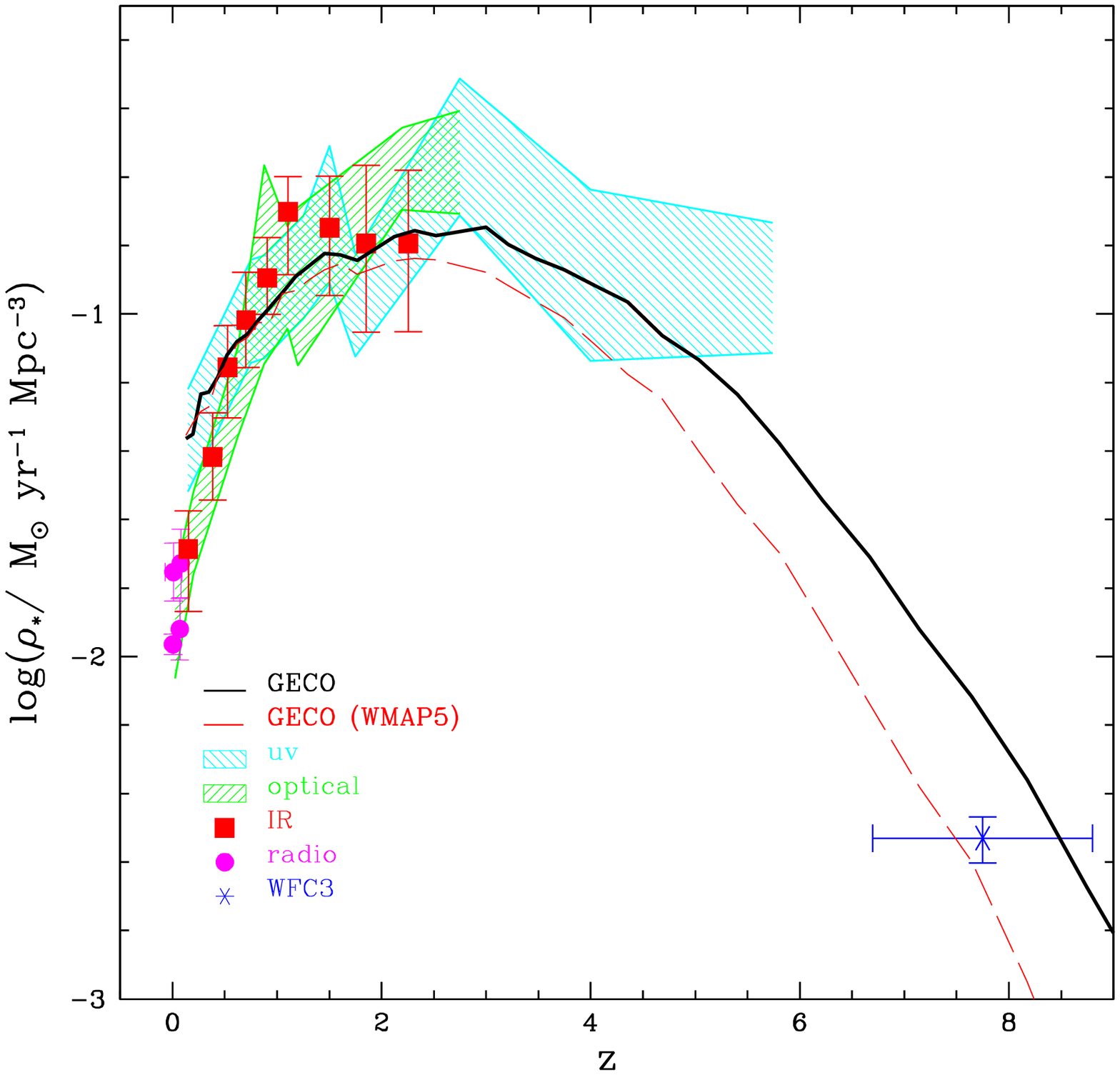}
\includegraphics[width=\columnwidth]{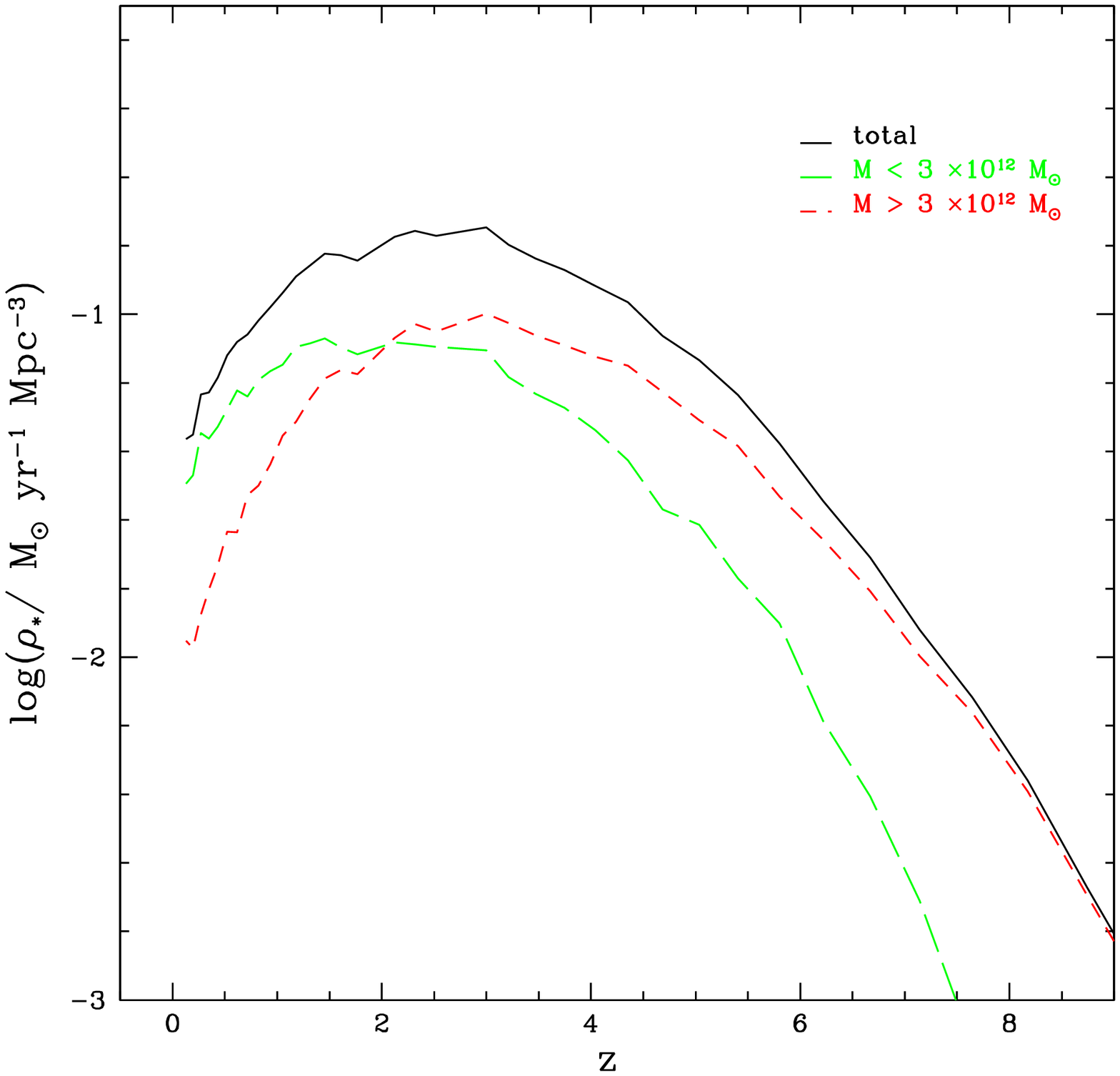}
\caption{Left panel: SFR density as a function of redshift. Solid
  black line shows the GECO predictions. The shaded area and coloured
  points are the SFR obtained from observations at different
  wavelengths. The cyan shaded region shows the SFR derived from  UV
  tracers, green region means that observations are in the optical,
  and magenta dots in the radio (from the compilation of
  \citealt{Hopk:04});
  red squares refer to infrared determinations, from \citet{Rod:09} and the blue star is the $z\simeq7$ determination from 
  HST-WFC3 data \citep{Bunker:09}.
  Right panel: SFR density splitted in the contribution by objects residing in
  different halo masses. Solid
  black line shows the total SFR density, red short-dashed shows the
  contribution from galaxies living in high-mass halos
  ($M_{halo}>3\times10^{12} M_{\odot}$), while the green long-dashed curve
  represents the contribution from galaxies in less massive halos
  ($M_{halo}<3\times10^{12} M_{\odot}$).}
\label{geco_madau}
\end{figure*}

In order to get a deeper insight into the process of the stellar build-up in our modelled galaxies, we investigate here their detailed star formation histories (SFH).
We derive the SFH {\it a-posteriori} for each present-day galaxies,
taking into account its baryonic tree and summing up all the stellar
populations formed inside its progenitors.  We compute the SFHs over
52 time-steps, our pre-defined grid of times used for the merger tree.
We show in Figure \ref{geco_sfh181}, as an example, the SFHs for 4 different realizations of the central galaxy of a halo with mass $5 \times 10^{12} \Msun$, i.e. very similar to the Milky-Way halo.
We indicate the stars formed in the quiescent mode with a blue shaded histogram and with the red one the stars formed during the starburst phase.
The black envelope is the sum of both contributions. In each panels, the host halo mass and stellar mass
are indicated, together with the parameter $r$, that is the ratio between the bulge and the total stellar mass.

In the case of such a Milky-Way-like halo, the majority of the central galaxies are spirals ($\sim 60\%$).
However, the mix of galaxy properties turned out to be rather heterogeneous.
In the lower-right panel we see an example of quite smooth star formation history with a steep rise of SFR at early times followed by a smooth decrease.
In the two upper panels we see that more than one peak of quiescent star formation are present, indicating that star formation has occurred in different progenitors of the final galaxy. In other cases, the starburst can occur at more recent times ($z\la 0.5$ in the lower-left panel) and is more efficient in consuming all the gas, hence the galaxy evolves passively until present. In all the cases represented starburst events are very frequent at early times, even in galaxies that at the present time have a late-type morphology.

In order to give an indication of the age of a galaxy, we define its formation redshift as the redshift when the galaxy has formed half of its present-day stellar mass. We indicate it in Fig. \ref{geco_sfh181} as a
vertical dashed line.
Galaxies with a smooth, quiescent star formation tend to form their stars later, while galaxies having experienced a more bursty star formation history, thus having an elliptical morphology, tend to have higher formation redshifts.

In Figure \ref{geco_sfhav} we show the SFH of the central galaxy of
halos with different mass, averaged over 100 realizations of the same
parent halo tree. 
Halo masses range from $\sim 10^{11} \Msun$ up to $\sim 10^{14} \Msun$. Note that, while in Fig. \ref{geco_sfh181} the SFHs refer to single realizations of one halo, each panel here refers to the average over 100 realizations of the same halo, hence SFHs result to be smoother.
Two trends with the halo mass can be noticed. The first is that, passing from low-mass to high-mass halos, the
starbursts become the dominant mechanism of star formation. The second
is that the star formation rate for galaxies living in high-mass
halos peaks at higher redshift compared to that in lower mass
objects. Therefore, the associated formation redshift for massive
objects is higher. In fact, the increasing importance of the starburst
mode, together with the AGN feedback connected to it, lead to a fast
consumption of gas at early times, by preventing the gas to cool and
inhibiting further star formation. 
The consequence is a lower rate of star formation at later times,
hence higher average formation redshifts for galaxies living in
high-mass halos. 

Since high-mass galaxies live in high-mass halos, these results indicate that stars in massive galaxies are on average older than in their less massive counterparts.
Nevertheless even in massive galaxies the star formation is not completely quenched, but a long tail of star formation at low redshift still occurs.
This behaviour of our SFHs appears to be in general agreement with the evidence for a {\it downsizing} pattern of galaxy evolution, an effect discussed since long time \citep{Cowie:96, Gavazzi:96, Fr:98, Thomas:05}. We will discuss further the physical reasons for this in Sect. \S\ref{concl}.
\begin{figure*}
\begin{center}
\includegraphics[width=0.65\columnwidth]{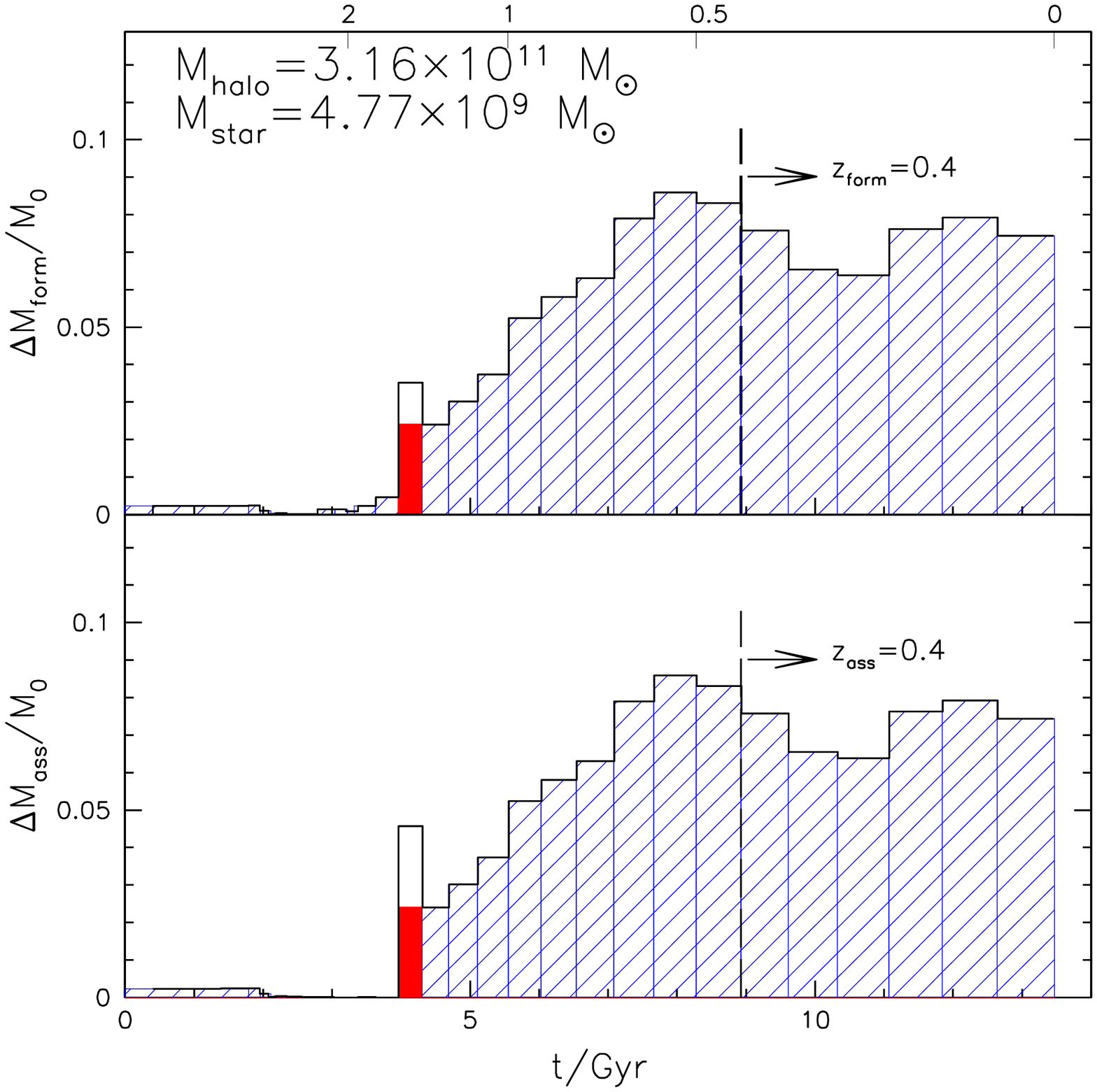}
\includegraphics[width=0.65\columnwidth]{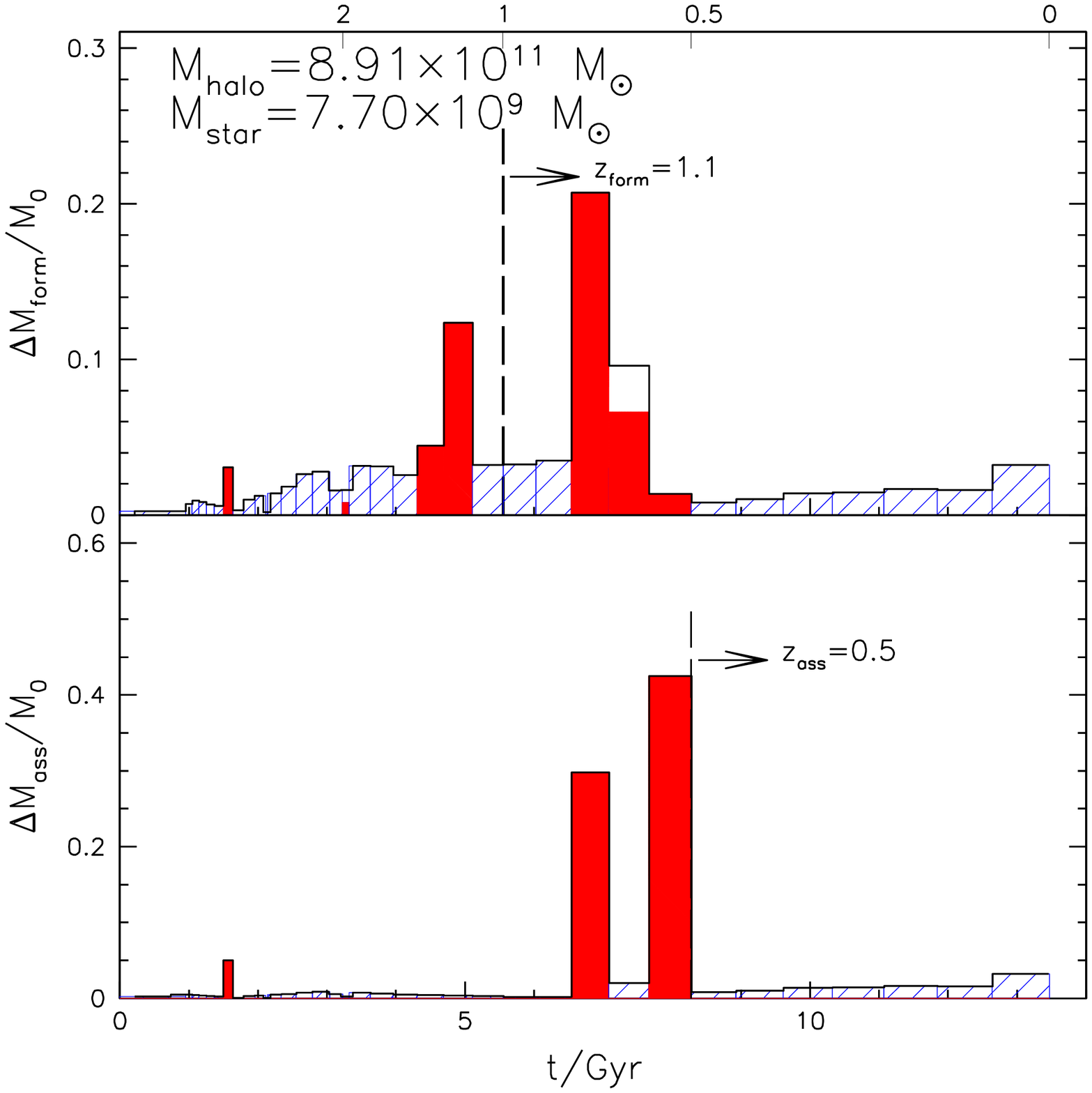}
\includegraphics[width=0.65\columnwidth]{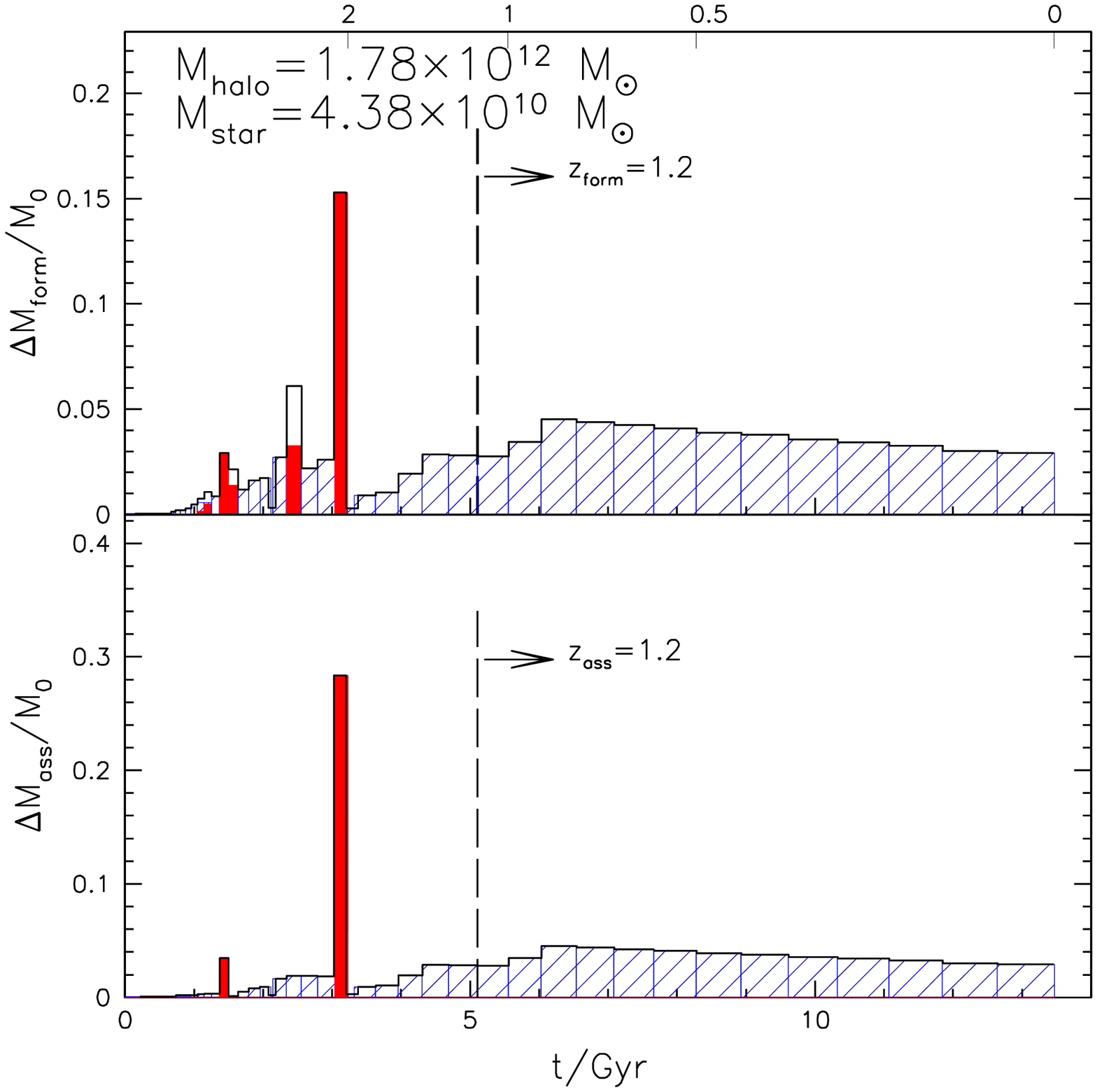}
\includegraphics[width=0.65\columnwidth]{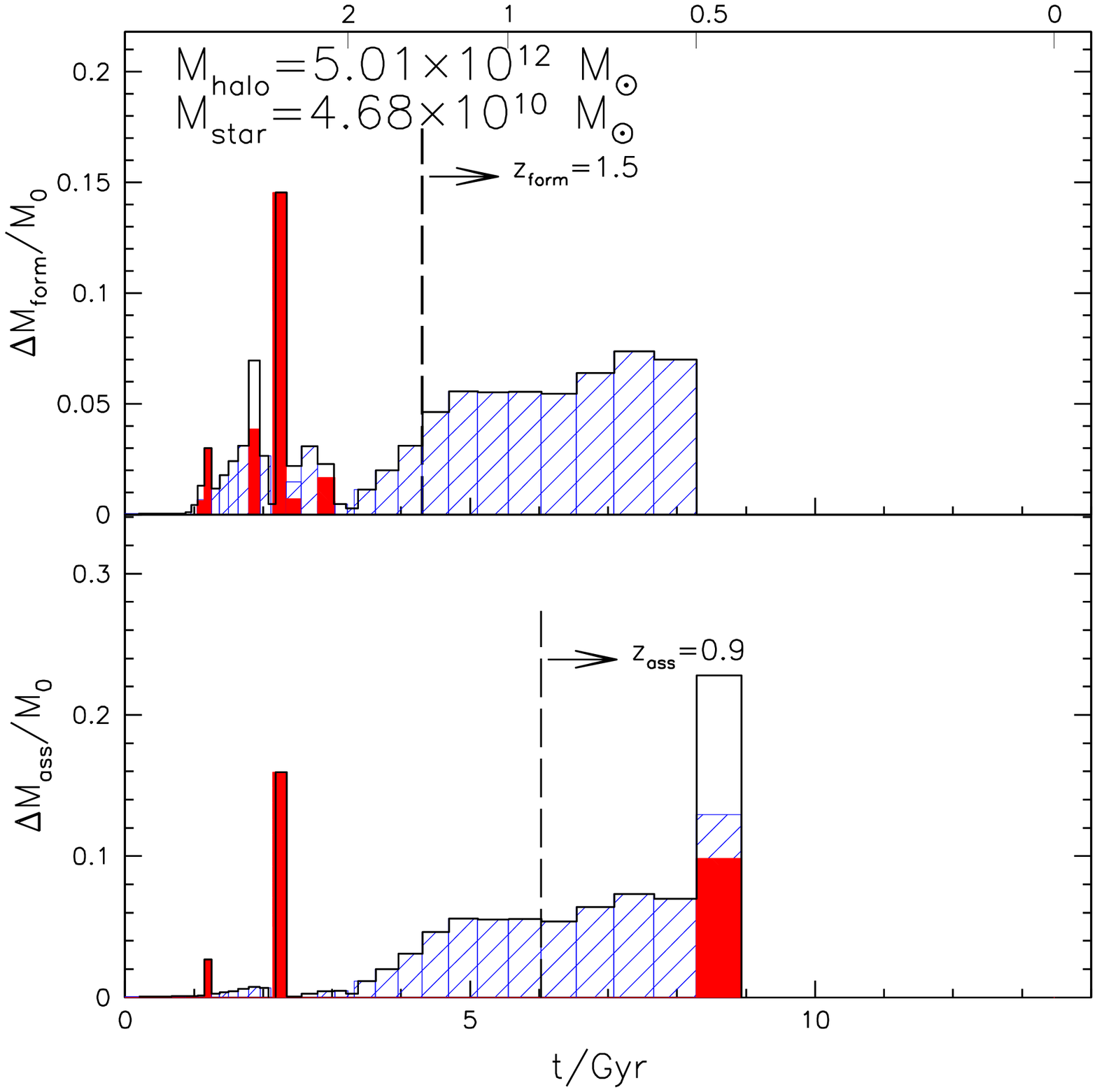}
\includegraphics[width=0.65\columnwidth]{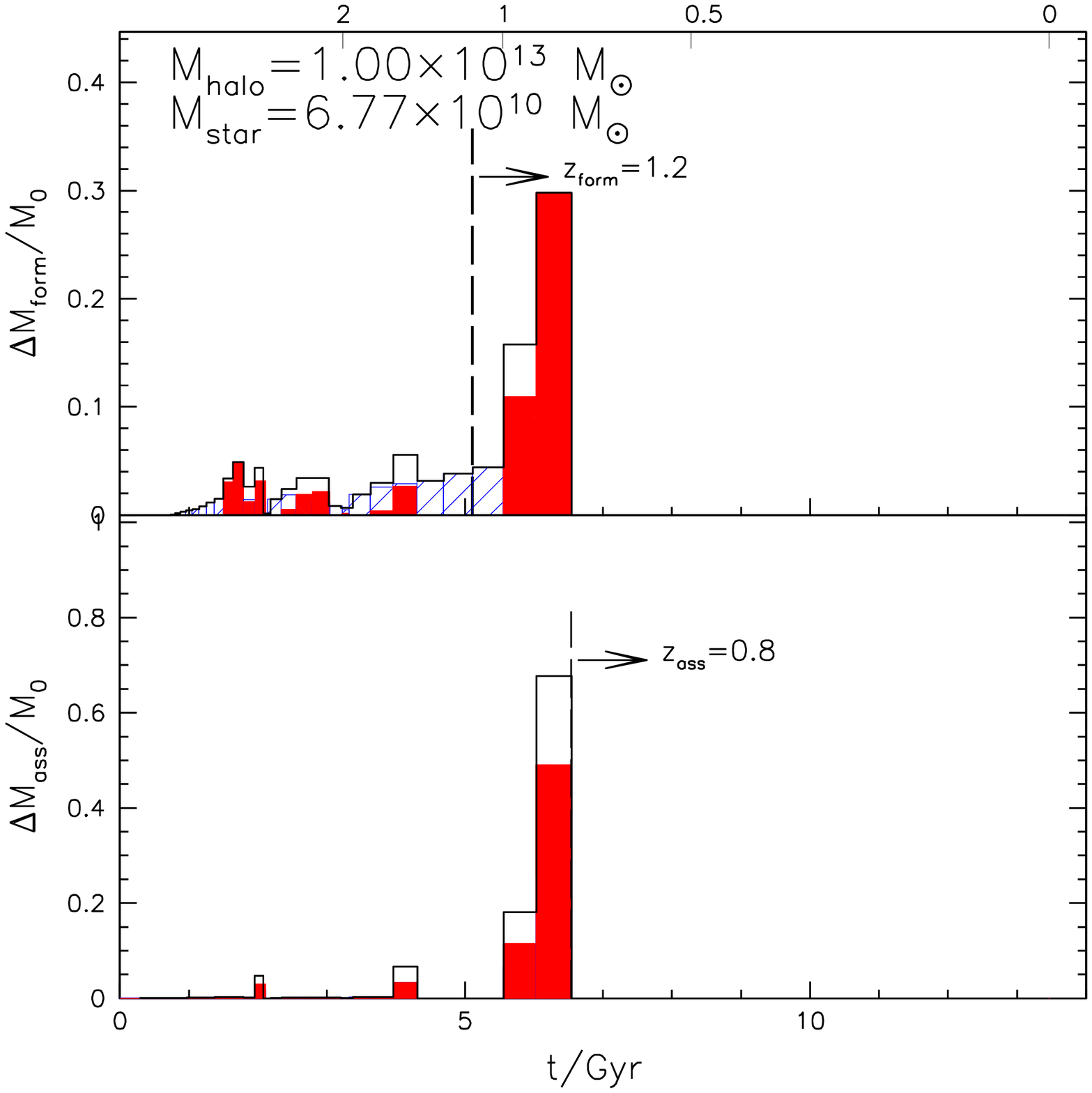}
\includegraphics[width=0.65\columnwidth]{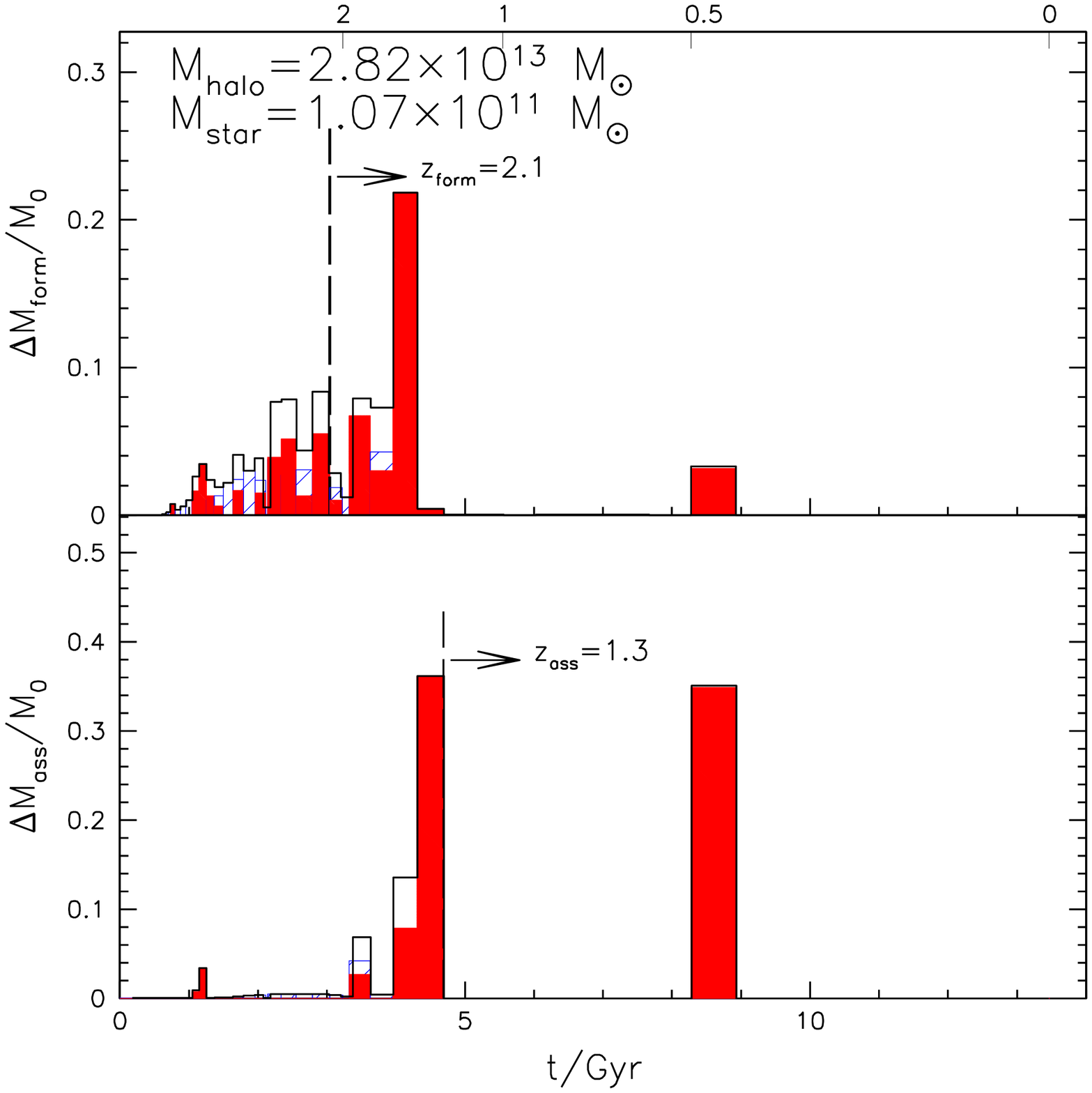}
\includegraphics[width=0.65\columnwidth]{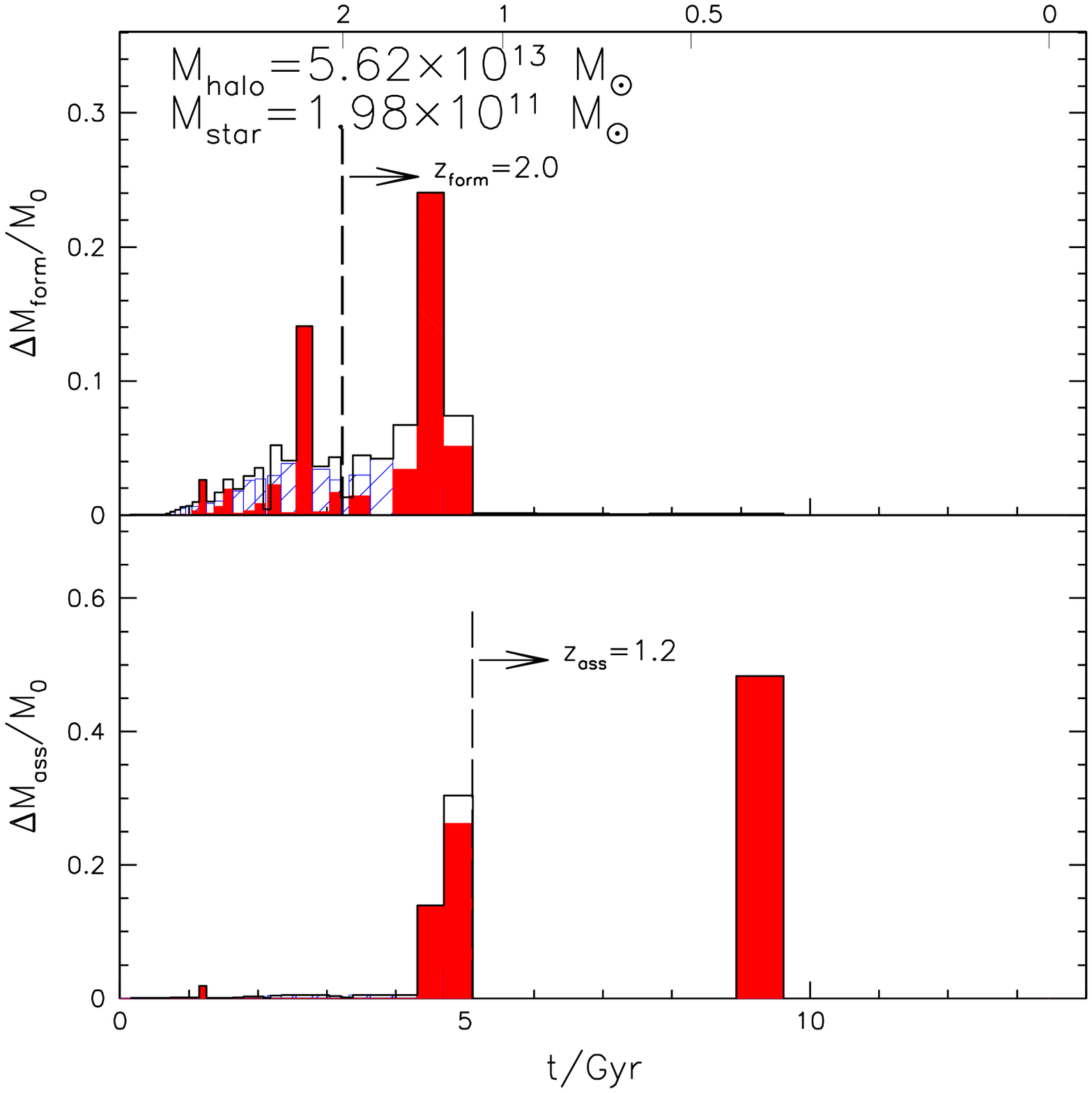}
\includegraphics[width=0.65\columnwidth]{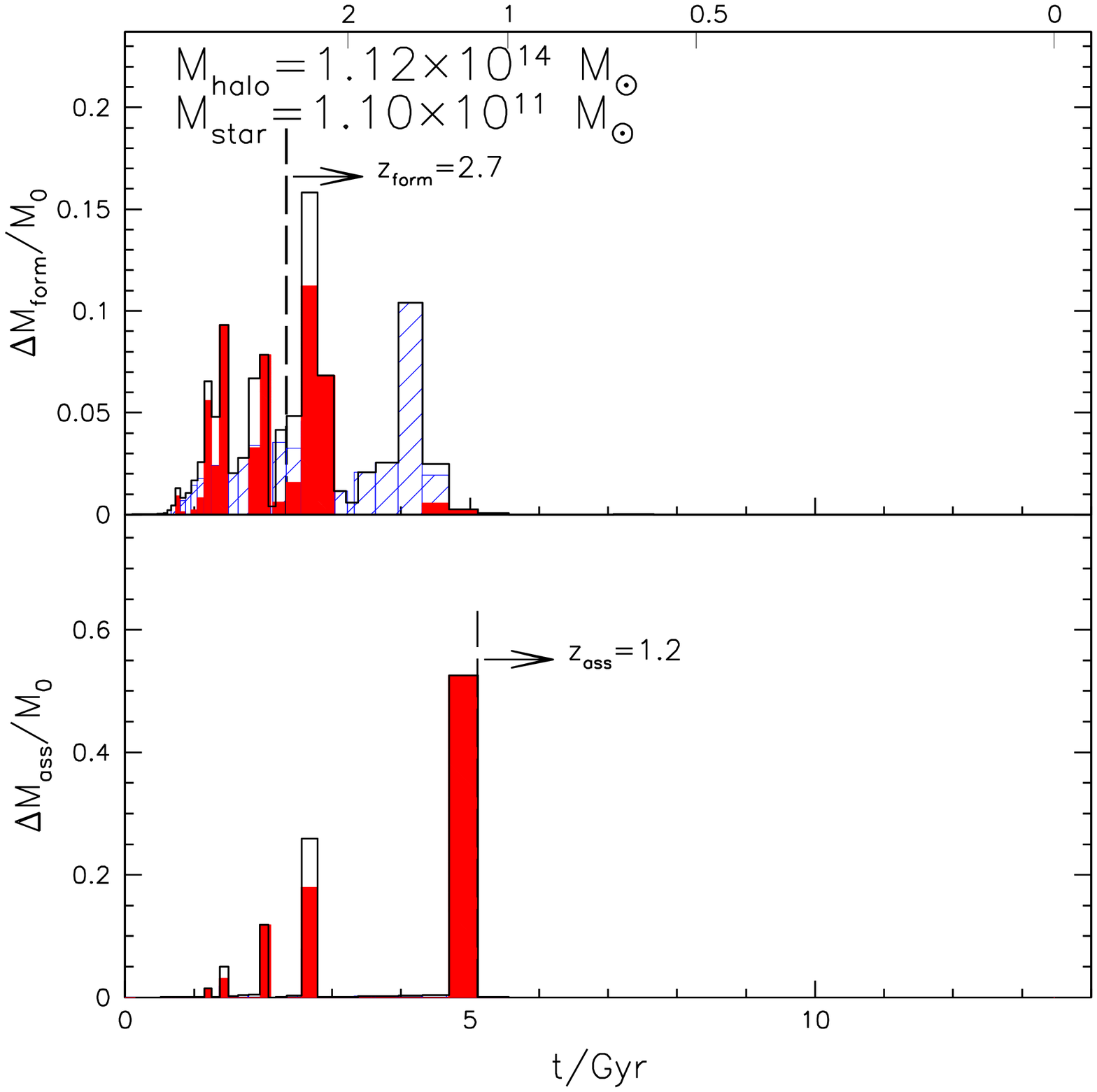}
\includegraphics[width=0.65\columnwidth]{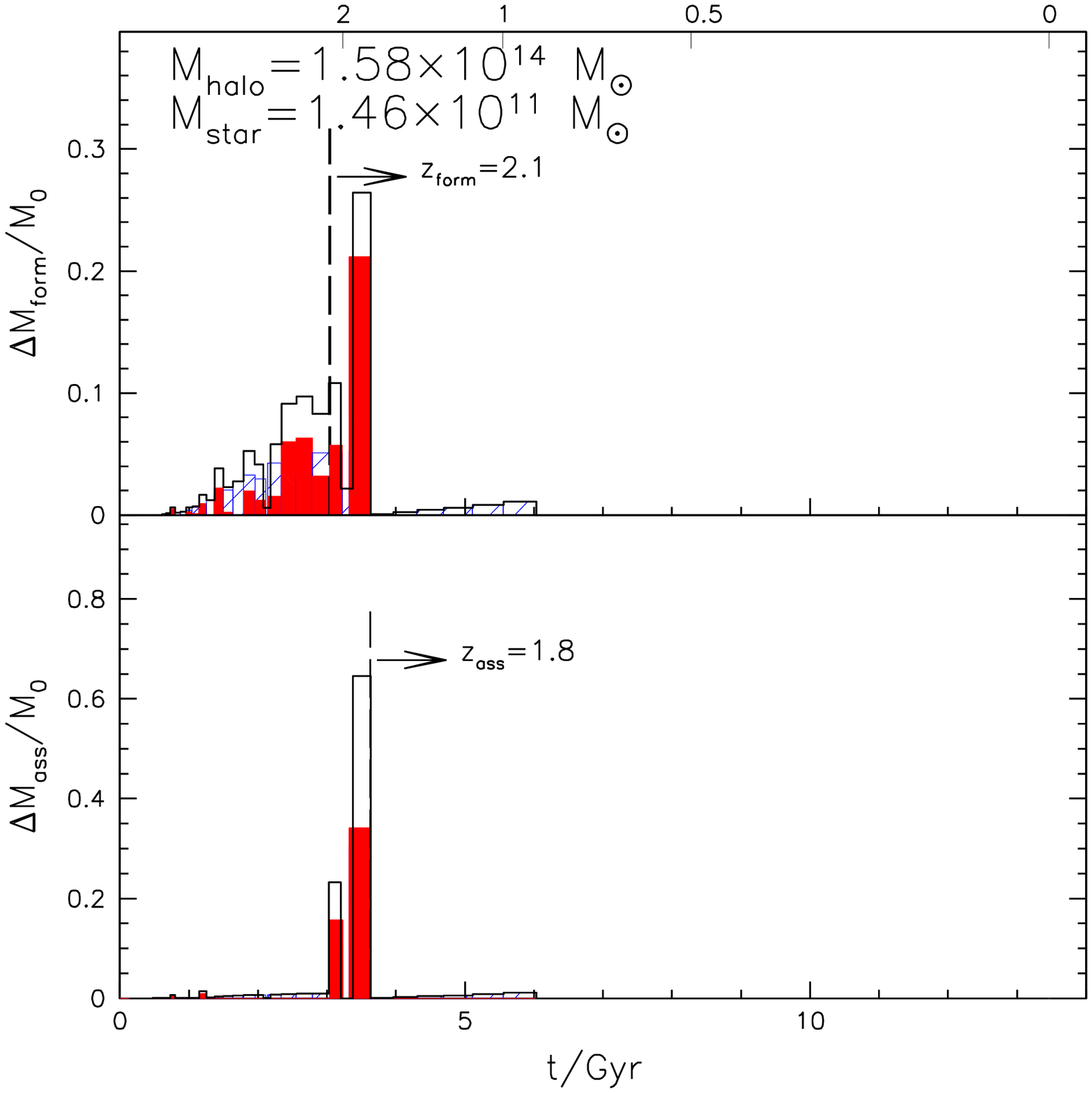}
\end{center}
\caption{Comparison between stellar mass formed (upper panels) and
  assembled into the main progenitor
  (lower panels) for
  different model galaxies, living in different halos. In the upper
  panel red and blue histograms represent,
  respectively, bursty and quiescent SF, while in the assembly panels
  red and blue stay for the mass assembly of bulges and discs respectively. }
\label{sfh_ass1}
\end{figure*}

\subsection{The Cosmic Star Formation Rate Density}

The evolution of the comoving star formation rate density for the whole galaxy population is shown
in Figure \ref{geco_madau} (left panel). The model prediction is shown by the black line and is compared with data from the compilation by \citet{Hopk:04}, for UV, optical and radio tracers, and from \citet{Rod:09} 
 for the IR tracers.
Different colours encode different tracers for the SFR:
cyan for star formation derived from the UV continuum,
green for optical tracers ({\OII},  {\Ha} and {\Hb}),
magenta for data using radio tracers and
red for SFR estimates derived from the IR luminosity.
The latter are believed to offer particularly reliable and robust
estimates of the star-formation rate because they trace rather
directly the emission by young stellar populations and are free of
dust extinction effects
\citep{Ken:98, Fr:01, Elbaz:02}.
 The blue star at z$\simeq$7 is the determination of the SFR density of \citet{Bunker:09} through HST-WFC3 data with the z'-drops technique, effective over the redshift range 6.7$<$z$<$8.8, as indicated by the x-axis bar. 

The model provides a remarkably good fit to the data over most of the redshift range.
It should be stressed, in particular, the nice agreement of the model with IR data, showing a fast rise from $z=0$ to $z=1$ and then a wide plateau up to $z\sim 4$.
Locally ($z\lesssim 0.2$) there is an indication that our model produces too many stars.
It appears consistent with UV determinations, but it keeps  higher with respect to the determination of \citet{Rod:09}  using the Spitzer $24\mu m$ luminosity function.
Even at very high redshift (z$\simeq7$), the model is marginally consistent with the observations, especially if we take into account that the  HST-WFC3 data can be considered as a lower limit due to incompleteness effects. 
We note also that the WMAP5 model predicts a much lower SFR at high redshift, ($z \gtrsim 2$) leading to a better agreement with the z$\simeq$ 7 data.  
Contrary to what found in \citet{Som:08}, with the new cosmology we find a better agreement with data in the SFR density, but a worst one in the stellar mass assembly, as noted above.

The high level of star formation rate at redshifts $z\simeq 3 - 5$ found in the model was already mentioned in the previous Sect. On one hand, the quiescent mode of star formation is very effective, due to the high
cooling efficiency in denser environment. Indeed, the peak in the quiescent mode of star formation (blue shaded histograms in Figs. \ref{geco_sfhav}) is always at $z\sim2$. On the other hand at high redshift
the star formation mainly occurs in starbursts, triggered by the high merging rate, which enhances the SFR at early cosmic times. The exhaustion of cold gas available explains the sharp later decline at lower redshifts.

In the right panel of Figure \ref{geco_madau} the contribution to the total star formation rate density (black solid line) is splitted according to the mass of the hosting halo:
the short-dashed red line shows the contribution from galaxies living in high-mass halos ($M>3 \times 10^{12} \Msun$), while the green long-dashed line shows the one from galaxies living in less massive halos.
The time dependence of the star formation rate in the two cases has a similar behaviour: a slow increase at high redshifts, a phase of peak activity, and finally a quick decline to the present time.
However, in addition to this general pattern, we notice a systematic shift in the SFR history between the two halo populations, with the high-mass halo activity being shifted at higher redshifts, $z\sim 2-4$, while it is more concentrated around $z=1-2$ in the less massive population.

We then confirm our previously found results: galaxies residing in high-mass halos form their stars at earlier
epochs, and then their star formation rapidly slows-down, while galaxies in less massive systems form stars at an enhanced rate even at recent times ($z\lesssim 1$).

\begin{figure*}
\centering
\includegraphics[width=1.5\columnwidth]{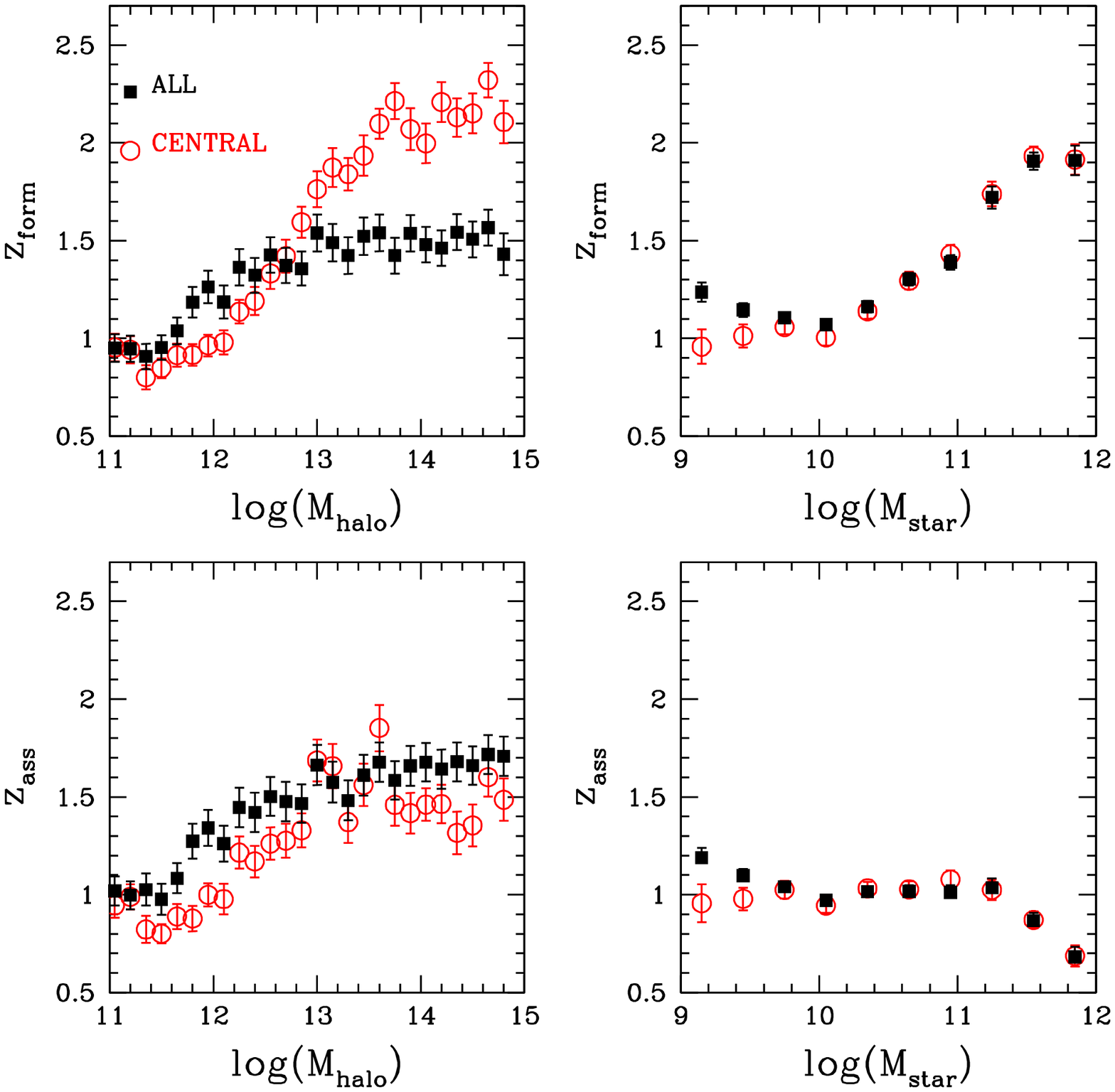}
\caption{Mean formation redshift (upper panels) and assembly redshift (lower
panels) of galaxies as a function of their host halo mass (left-hand
panels) and stellar mass (right-hand panels).
Black squares indicate the global population of galaxies, including
central and satellites, while red open circles stay for central
galaxies. Error-bars indicate the standard error on the mean over 100 MC
realizations of the same halo.
}
\label{zform}
\end{figure*}

\subsection{Star Formation and Mass Assembly}

We have discussed in the previous Sect. an important aspect of our model, its prediction that massive systems form their stars at earlier times then less massive objects. This however concerns the birth times of the galaxy's stellar populations, while the times of mass assembly may be very different when stars are formed at high redshift in a number of distinct progenitors and are assembled at more recent times.

In Figure \ref{sfh_ass1} we compare the star formation history with the ``assembly history'' for
single realizations of various parent halos.
In the upper boxes of each panel we show the SFH of all stellar populations contributing to the final galaxy, in analogy with those reported in Fig. \ref{geco_sfh181}.
The SFH is expressed here in terms of the fraction of the final stellar mass formed in each time-step (instead of the SFR), for an easier quantitative comparison with the assembly history. In the bottom boxes we show the mass assembled into the main progenitor at each time-step, where the main
progenitor identifies at each redshift the progenitor of the galaxy which survives
until the present day.
Here, red histograms indicate the amount of stars added to the bulge in a given time-step, while blue histograms show the stars added to the disc.  Although in a few cases the history of stellar assembly is very
similar to that of star formation, indicating that star formation took place mainly in one single object,
in general the two paths may be significantly different.

In most of the cases the assembly of bulges, due to mergers, is related to starbursts, although in some systems, mainly in merging episodes at low redshifts,   the mass assembly is not associated to any event of star formation because the gas content is already consumed (\emph{dry merging}).
Substantial dissipationless merging may lead to very different formation times for the process of star formation and mass build-up.
Although dry mergers in the local universe are observed, their frequency and their role in galaxy evolution are still poorly constrained by observations \citep{vanD:05, Bell:06}. Anyway galaxy formation models  predict that they are a crucial process in the formation of local ellipticals (see \citealt{Cattaneo:10}).

We define the redshift of assembly as that corresponding to when half
of the final stellar mass is assembled in one single object, and we
show it in Fig. \ref{sfh_ass1}. By definition, the assembly redshift
is always smaller than the formation one, they equal each other when
almost all the stars are formed in the main progenitor.

We report in Figure \ref{zform} the dependencies of the mean star formation redshift (upper panels) and the redshift of mass assembly (lower panels) as a function of the hosting halo mass (left-hand panels) and of the galactic stellar mass (right-hand panels). Error-bars indicate the standard error of the mean over
100 MC realizations of the same host halo. Note that in order to derive
the redshift of formation (or assembly) as a function of stellar  mass
we weight each galaxy mass by the number density of its host halo,
because any stellar mass may belong to a certain range of halo masses.

The mean formation redshift shows a clear increase with both halo and stellar mass. This effect is more pronounced for the central galaxies than for the whole population including satellites.
We have already shown in the previous Sects. how galaxies living in high-mass halos formed their stars earlier and at a faster rate than in low-mass halos. Here, we demonstrate that the trend is true even with the galactic stellar mass, although it is weaker. Note that the most massive halos ($M_{halo} > 10^{14} \Msun$) host galaxies with the higher formation redshift ($z_{form} \sim 2.5$, top left panel of the figure), but since these halos are very rare, the mean formation redshift of the more massive galaxies ($M_{star} \sim 10^{12} \Msun$), which belong to a certain range of halo masses, keeps somewhat lower ($ \sim 2$, top right panel).

The behaviour of the redshift of mass assembly is more difficult to interpret. We found that, on average, galaxies living in high mass halos assemble their stars earlier. Indeed, the merger time-scales due to both dynamical friction and random collisions among satellites increase with the halo mass. We already mentioned in \S\ref{dynfric} that the dynamical friction time increases with the ratio between the mass of the halo and that of the satellite (eq. \ref{df_epsilon}). Moreover, since the merger probability is strongly reduced in high-velocity encounters, the merger rate for satellites decreases with the ratio $V_c/v_{gal}$ (eq. \ref{satcoll}), and becomes very low for massive cluster-size halos.
Nevertheless, when the redshift of mass assembly is expressed in terms of the galactic stellar mass, the downsizing behaviour is no more recovered: the assembly time keeps approximately constant around $z_{ass} \sim 1$ for low and intermediate masses and decreases in the highest mass end, with $z_{ass} \sim 0.7$ for
$M_{star} \sim 10 ^{12} \Msun$. These stellar mass values correspond to those of the CD galaxies in local rich clusters or groups.
The combined effects of the formation redshift and the assembly redshift suggest that the only way to form the most massive galaxies in the local universe is in the form of dry mergers at recent times. Indeed, massive galaxies cease to form stars quite early and, without such events, they are not able to grow their mass above $M_{star} \simeq 10^{11} \Msun$.

In conclusion, our model expectation is that, although mergers and collisions are rare in massive halos, they do happen, and their effect is crucial in the formation of the today most massive galaxies.
Refined observational estimates of the galaxy mass functions at the high mass-end all over the z=0 to 4 redshift interval (on top of those already apparent in Fig. \ref{geco_ev_all}) will clarify how far this expectation of a substantial \emph{dry merging} process is born out by the data.

\section{ Discussion and Conclusions}\label{concl}

We have presented a new semi-analytical model of galaxy formation, the Galaxy Evolution Code, GECO which appears to reproduce several key statistical properties of local and high-redshift galaxies.

GECO  is based on a state-of-the-art Monte Carlo algorithm for the representation of the dark matter halo merging history, based on the Extended Press-Schechter formalism.
GECO includes detailed implementations for gas cooling, star formation, feedback from SN and galaxy mergers, due to both dynamical friction and random collisions.
Moreover, the parallel growth of BHs is followed in time and the subsequent AGN feedback is modelled.

We specifically tested our results directly on the observables involving the stellar mass and star-formation rate more then luminosities, as usually done by other published models. This is motivated on one hand by the fact that the stellar mass functions are the most direct outcome of the model.
On the other hand, stellar masses in galaxies have recently become a rather straightforward observable thanks to rest-frame near-infrared data by the Spitzer Space Telescope, directly probing the stellar mass content in high-redshift galaxies.
At the same time, Spitzer is also probing with deep far-infrared photometric imaging the rate of stellar formation in distant objects. Therefore, we can compare the outcomes of our model with the most reliable set of observables and a minimal number of free parameters.
We thus believe that the basic physics of the $\Lambda CDM$ hierarchical clustering concept of galaxy formation can be tested by us in a very effective way.

The main results obtained in the present work are summarised in the following.

\begin{enumerate}

\item
   The local stellar mass function results in a remarkably good
   agreement with the determination of \citet{Cole:01} and  \citet{Bell:03} (Fig. \ref{geco_smfz0}).
   At the high-mass end the total mass function is dominated by the contribution
   of bulges, while discs dominate at the faint-end.
   When the total stellar mass is splitted into the contributions of early-type
   and late-type galaxies, the former populate the bright-end side,
   while the latter mainly contribute at low masses.
  The number densities of the two morphological types cross each other at 
  $M_{star}\sim 3 \times 10^{10} \Msunh$, as observed.
   Although we reproduce the general trend of the morphological mass functions, our model fails in matching the low-mass end of spheroids, showing an excess of low-mass systems. 
Likely, this is due to an oversimplification of the satellite population, since satellite galaxies loose  their hot gas reservoir as soon as they are incorporated in a more massive halo
and the star formation is  quenched soon after. 

\item
   The co-evolution of galaxies and BHs is modelled following the
   prescriptions of \citet{Cr:06}. A first mode of accretion onto
   BHs considered is the so called 'radio-mode', that inhibits the quiescent star
   formation, while the second one is the `QSO-mode', that is triggered only during galaxy
   mergers and constitutes a major channel of BH accretion. As a consequence of
   mergers, a starburst is induced as well, feeding the galactic bulge
   component (and destroying the disc in the case of a major merger).
   This leads to a parallel growth of BH and bulges with
   the two masses very well correlated, in agreement with
   observational data \citep{HR:04} and to the local black-hole mass
   function in a remarkable agreement with the observations of \citep{Shankar:04}.

\item
   We compare the stellar mass functions resulting from the model with
   various observational determinations up to $z\sim 3.5$ (Fig. \ref{geco_ev_all}) and found a
   reasonably fair agreement over the whole redshift range considered.
   Nevertheless, the observed ratio between the evolution
   of the faint- and the bright-end of the stellar mass function is
   not very accurately reproduced: there is too much evolution in the model at the bright-end
   and too little at the faint-end compared to observations.
   Various sets of observables indicate a large increase in the number density of low-mass objects
   between $z\sim 2$ and the present-day and a lower rate of evolution for massive objects.
   However, we mentioned that the completeness and robustness
   of the observational mass function are to be proven there, before claiming more definite
   conclusions.
   In the case of  WMAP5 cosmology we observe a delay in the formation of cosmic structure, that lead to a further reduction of high-mass systems.

\item
   The bolometric quasar luminosity function is compared with \citet{Hopk:07a},
   showing a good level of agreement at low and intermediate redshift,
   but a tendency to underpredict the number of bright quasars at high redshift.
   A mechanism for enhancing the cooling rate at high redshift might
   simultaneously increase the fuel for BHs and enhance the SF at
   high redshift, as seems to be required to improve the match with
   the stellar mass functions.

\item
   The integrated star formation rate density (Fig. \ref{geco_madau}) shows an high level of star formation at high redshifts,
   a peak at $z\sim 1.5-3$ and then a sharp decline below $z\sim 1$. When
   compared with the determination of the SFH derived using various
   tracers (UV, optical, radio, IR), our predictions are in very close agreement with these observations. At very high redshift ($z \simeq 7$) our model is able to correctly reproduce the recent determination of the SFR density  by HST-WFC3 data.

\item
  We analysed in detail the SFH in simulated sets of galaxies, to gain insights into how the model treats
   star formation and how it depends on the galaxy or halo mass. We computed the averaged
   SFH for the central galaxies living in halos of different sizes
   (Fig. \ref{geco_sfhav}). We identified two main trends with halo mass.
   First, going to high-mass systems, the contribution of the starburst
   mode to the total SF becomes increasingly important, and indeed predominant in very high mass objects.
   Second, the formation redshift, defined as that when half of the present-day
   stellar mass is formed, increases, leading to older stellar populations in massive systems.
   Hence galaxies in our model form their stars following a \emph{downsizing} pattern,
   consistent, for instance, with the dating of stellar populations in local galaxies \citep{Thomas:05}.
   The naive expectation of early versions of hierarchical galaxy formation models was that,
   since massive halos are assembled later than their lower-mass counterparts, the most massive galaxies,
   hosted in the largest halos, should form their stellar content at the same late cosmic time.
   Actually, as shown in Fig. \ref{zform}, this is clearly not the case in our refined model.
   According to it, \emph{downsizing} in star-formation is an intrinsic feature of semi-analytical models
   (see also \citealt{Nei:06}).
   The present-day massive galaxies were formed through the assembly of a
   number of smaller progenitors, that collapsed at high redshifts
   from the highest density peaks of the primordial density field.
   According to this scenario, also named biased galaxy formation \citep{Dekel:86},
   bright and massive systems started to form stars early on.
   This feature is a natural outcome of the merger tree formalism:
   progenitors of high-mass halos fall below the resolution mass imposed to the merger tree
   after several time-steps back in cosmic time, so the leaves
   of the tree are found at high redshift. On the contrary, smaller systems, closer
   to the resolution mass, take only a few time-steps back to reach this minimum mass.
   Note that this not merely a computational artifact, but indeed corresponds
   to the fact that we expect quite negligible star formation
   to have occurred below such threshold mass, on consideration of the
   SF quenching by the UV photoionizing background (Sect. \ref{UV}).
   Since baryons are put into halos starting from the leaves, in
   high mass halos star formation took place at early times.
   Moreover, at high redshift both mechanisms of star formation were more effective.
   Thanks to the efficient cooling of the gas,
   the quiescent mode of star formation occurs at enhanced rate.
   Also, the frequency of mergers at these times is high,
   allowing an efficient conversion of gas in stars via a starburst.
   The exhaustion of cold gas is, therefore, very rapid,
   leaving the galaxy devoid of fuel, and preventing further star formation to occur.

   A further reason for the star formation quenching in massive systems may be ascribed to AGN feedback.
   In order to check the importance of such mechanism, we run a test simulation with the AGN emission
   switched off. The comparison between the two versions of the model
   are presented in Appendix A. 
   We found that AGN feedback has some effect in increasing the average age of stellar populations
   in the most massive galaxies, as it was point out by previous works \citep{Cr:06,DeL:06,Cattaneo:08}.
   However, in our model the \emph{downsizing} trend is still
   obtained, although slightly 
   weaker, even in the absence of an AGN. 
   
   Therefore, in GECO, the AGN feedback has not a dramatic effect in producing the local galaxy properties (see also \citealt{Menci:06} and  \citealt{Monaco:07})
and it is not 
   the main reason for the \emph{downsizing} pattern of galaxy evolution. In our modelling, the latter is 
   rather due to a higher efficiency of SF back in time for the most massive galaxies in
   rich environments, as explained above. 
   Although we have shown that the \emph{downsizing} is quite a natural feature of  our model, it is not the case for all the published models and some of them fail in reproducing such trend with mass,
   as it is shown in \citet{Fontanot:09}.

\item
   Finally, we compared the star formation history of simulated galaxies
   with the mass assembly history, that is the history of the mass assembled into the
   main progenitor at each time-steps (Fig. \ref{sfh_ass1}). As expected, the two
   processes may occur on very different time-scales, especially in
   high-mass systems, where the star formation took place at high
   redshift in many distinct progenitors, which assemble at low
   redshift. The late assembly of these systems occurs in
   the majority of the cases through \emph{dry mergers}, i.e. mergers between
   spheroidal systems with little or no gas. In these cases, the merger is not accompanied
   by any event of star-formation, neither quiescent nor bursting.
   An intriguing question is then whether galaxies in the model assemble their
   stellar mass in a \emph{downsizing} way, as it occurs for the star formation.
   Possible downsizing effects in the mass assembly were indicated by several authors
   \citep{Bundy:06, Cimatti:06, Hopk:07} as inferred from the lack of evolution in the high-mass
   end of the mass function and by the evolution of the blue-to-red galaxy crossover
   mass (the mass for which the early-type mass function intersects that of late-types).
   In our model, we found that although the assembly time shows a
   shallow dependence on the host halo mass, on average galaxies living in
   massive halos assemble their stars before galaxies in less massive
   hosts (Fig. \ref{zform}). Instead, the assembly times are almost constant with stellar
   mass, and decrease for very-high mass systems ($M_{star} > 10^{11}
   \Msun$), hence leading to an \emph{upsizing} trend with time in the high-mass end.
   In our modelling, dry mergers are the main reason for this late
   assembly of massive galaxies. These findings agree with other
   previously published semi-analytical models \citep{DeL:06,
   Cattaneo:08, Cattaneo:10}.
   The importance of dry mergers in the formation of the most massive
   galaxies that we observe in the nearby universe, is implied also by
   the recent observations of the size evolution of massive spheroids
   \citep{Trujillo:07, Cimatti:08}. 
   Indeed, if such compact galaxies are absent in the local universe as suggested by \citet{Trujillo:09}, the expected mechanisms that move the high-redshift compact galaxies to the local relation are 
    dissipationless mergers. Anyway, other works \citep{Valentinuzzi:09} suggest that such superdense galaxies in the local universe are not as rare as previously claimed.

   How far this might be incongruous with the
   observational indication of downsizing in mass assembly mentioned above
   will become clear with further observational confrontation. 
   For example,
   \citet{Cattaneo:08} argue that the upsizing in mass
   assembly can coexist with a downwards trend in the transition mass,
   which then turns out to be a poor indicator of downsizing.

\end{enumerate}

In conclusion, GECO presents an encouraging level of agreement with a wide range of observational data, at low and high-redshifts.
We focus on comparing GECO with data on the two main phases of the galaxy formation process, that are the star formation and the mass assembly.
On one hand, we confirmed that the observed \emph{downsizing} in star formation is natural part of our scheme of  hierarchical growth of structures.
On the other hand, the stellar mass assembly process remains more difficult to understand from both a theoretical and an observational point of view.
The times of galaxy assembly in our model, related with both the
galaxy merger time-scales and the star formation efficiency, strongly
depend on the details of the implementations of galaxy dynamics
(dynamical friction, satellite collisions and tidal stripping):
further work in this sense remains to be done in order to have a
deeper insight into the galaxy assembly process, and certainly a
comparison with N-body simulations will be helpful. 

The most striking conclusion is that, despite the simplicity of the
prescriptions adopted, and the small number of free parameters used, the
main features of the evolving galaxy population are
reproduced. In particular, the AGN feedback is needed only to improve
the match of the local stellar mass function, but its effect on
stellar ages is not determinant.

We have to keep in mind in any case that this paper includes just a
preliminary and partial confrontation with the data, showing at least
no obvious clash. 
Much more extensive analyses and tighter constraints will be obtained as soon as refined data on the evolutionary mass functions and stellar birthrates will become available.

\section*{Acknowledgements}

We thank the referee for helpful suggestion which improved the paper.
We are grateful to Nicola Menci for various illuminating discussions, criticisms and suggestions about the present model.
We also thank Alfonso Cavaliere and Pierluigi Monaco for discussions
and Antonio Cava, Carlo Giocoli and Ignacio Trujillo for helpful
suggestions.
We acknowledge financial support by the Padova University for the PhD studentship and the Italian Space Agency for funding a post-doc fellowship, which made this paper possible.

\newpage

\appendix

\section{Effect of AGN feedback}

\begin{figure}
\begin{center}
\includegraphics[width=\columnwidth]{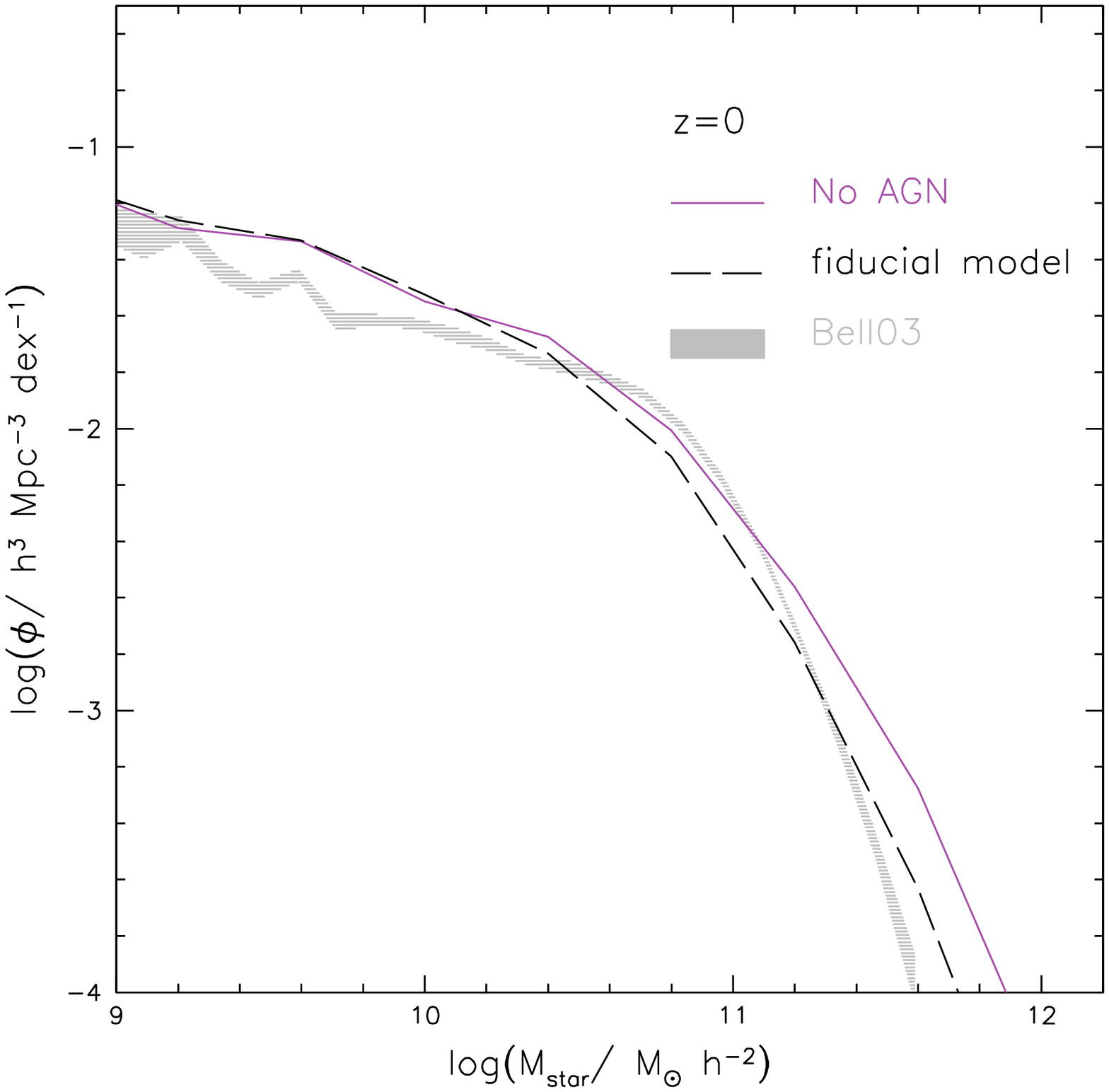}
\includegraphics[width=\columnwidth]{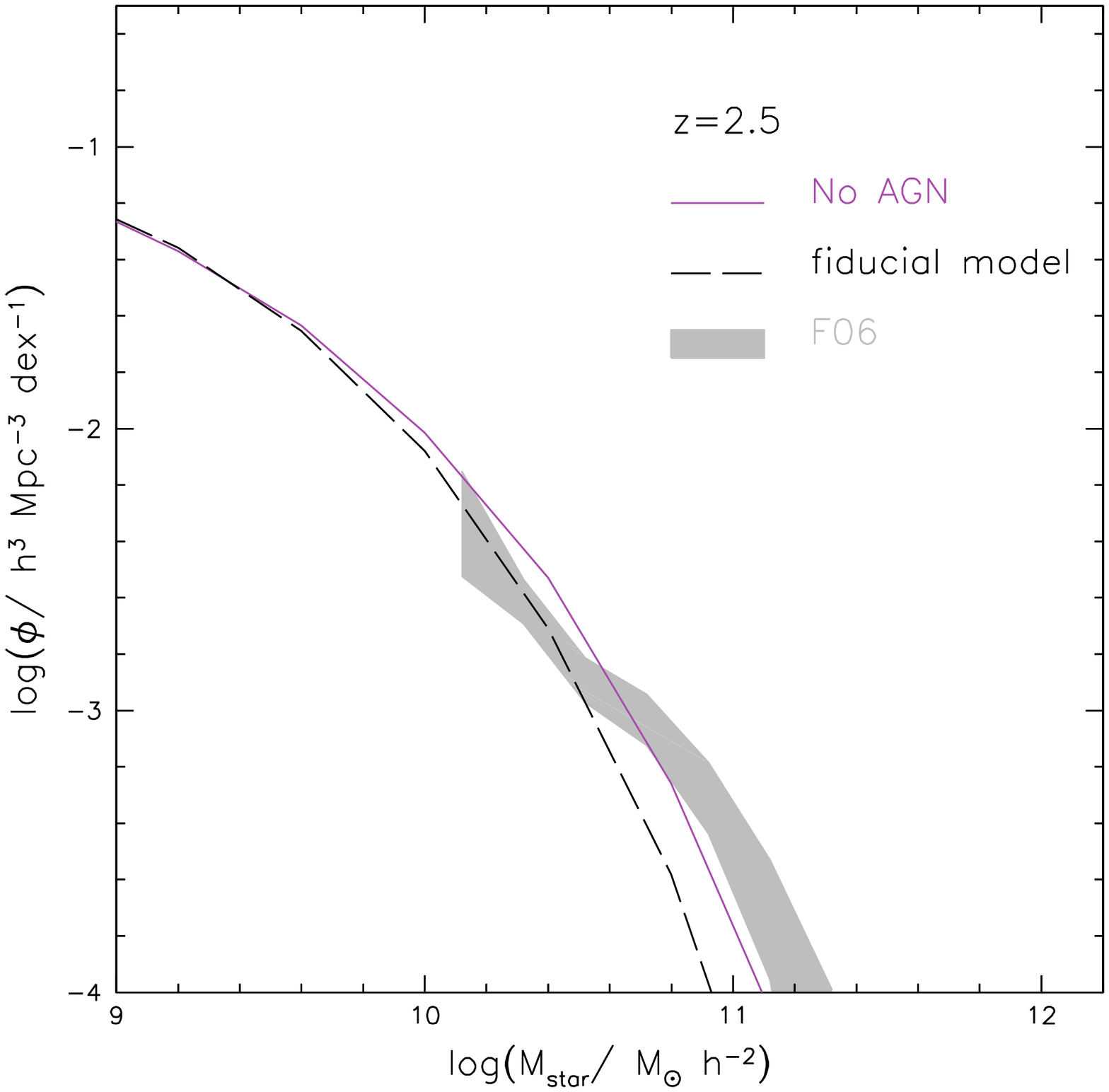}
\end{center}
\caption{Upper panel: comparison between the stellar mass function at
  z=0 of our fiducial model (dashed black line) with the model without
  AGN feedback (solid purple line). 
  Lower panel: the same as the upper panel but for z=2.5. Observational data 
  at the given redshift are also indicated by the shaded region.
 }
\label{cfr_smf}
\end{figure}

\begin{figure}
\begin{center}
\includegraphics[width=\columnwidth]{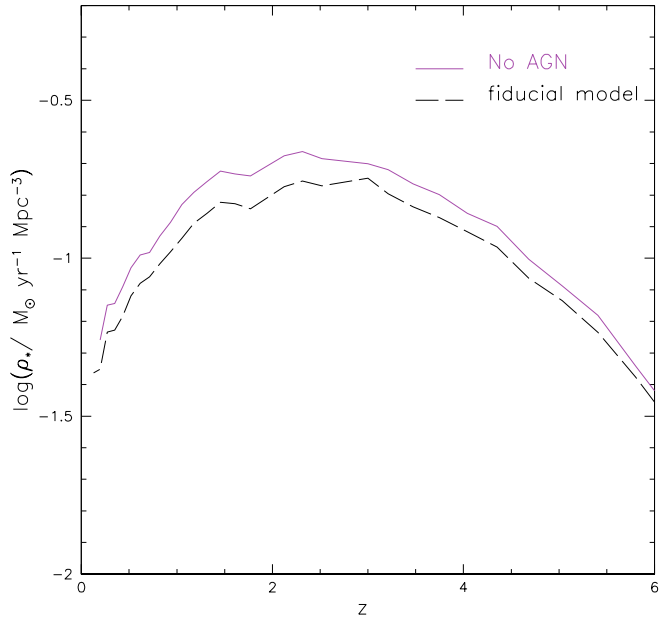}
\end{center}
\caption{Comparison between the SFR density as a function of redshift 
  in the model without AGN feedback (solid purple line)
  and the fiducial case (dashed black line). }
\label{cfr_madau}
\end{figure}

Several recent works \citep{Cr:06, Malbon:07, Sij:07}  advocated the AGN feedback as the only
mechanism able to quench star formation in order to match the local
abundance of massive galaxies as well as their old ages and red colours.
Here we check the relevance of such feedback in reproducing the
observations in GECO, comparing 
our fiducial model with a version in which
the AGN feedback (both in quasar and radio modes) is switched off.
In Figure \ref{cfr_smf} we can see the effect of the AGN in the
stellar mass function. As expected, the presence of the AGN influences only the
bright-end, reducing it by $\simeq$ 0.15 dex, both in the local universe
(upper panel) and at high redshift (lower panel), hence
leading to the same amount of evolution in the stellar mass function as
in the fiducial case. We note that given the overabundance
of massive galaxies by the same amount at all the redshift,
 in the model 
without AGN the comparison with the observational mass function 
at high redshift results in a closer agreement with respect to the
fiducial model. 
Figure \ref{cfr_madau} shows that, obviously, in the model without AGN
the star formation keeps lower at all redshifts, 
 with an indication that at very
high redshift ($z>4$) the difference between the two models tends to
disappear, meaning that the importance of AGN feedback increases at
late cosmic times.

\begin{figure}
\begin{center}
\includegraphics[width=\columnwidth]{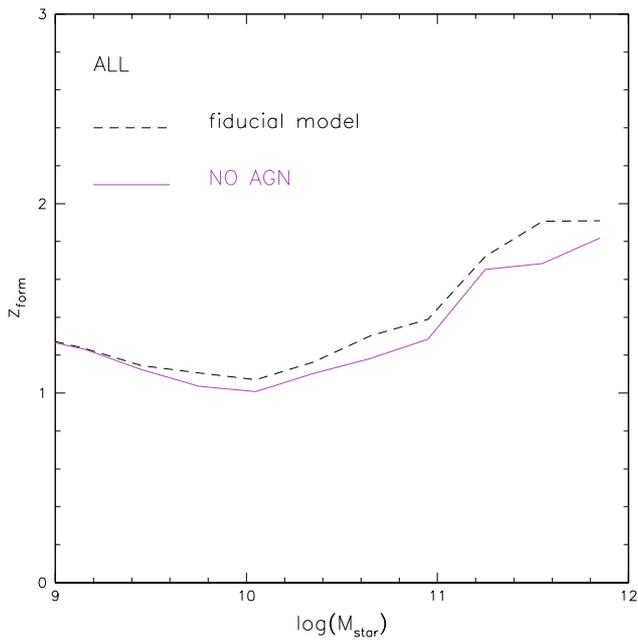}
\end{center}
\caption{Comparison between the formation redshifts as a function of
  stellar mass in the model without AGN feedback (solid purple line)
  and the fiducial case (dashed black line).  }
\label{cfr_zform}
\end{figure}

A further issue to be handled is the relevance of the AGN in producing
the downsizing trend in the stellar ages. In Figure \ref{cfr_zform} 
the trend of the formation redshift with mass is shown for both
models. Although the AGN is effective in increasing stellar ages,
the net effect is actually very
modest.
At variance with other models (\citealt{Cr:06} for instance) the
downsizing trend is still recovered, 
leading to the conclusion that the presence of AGN is not the main
reason for the onset of the downsizing pattern for galaxy evolution in
our model.

Therefore, in our model the AGN has the effect of reducing the star
formation activity in the high-mass objects, but the amount of
reduction is perhaps less striking than previously claimed
\citep{Cr:06, Malbon:07, Marulli:08, Lagos:08}. Indeed, the values for the efficiencies of black hole accretion
adopted here, are somehow smaller than those adopted by the models
mentioned above, but we found that they are suitable to obtain the
very good match of the local relations.\\

\clearpage

\section{Effect of cosmic reionization}

\begin{figure}
\begin{center}
\includegraphics[width=\columnwidth]{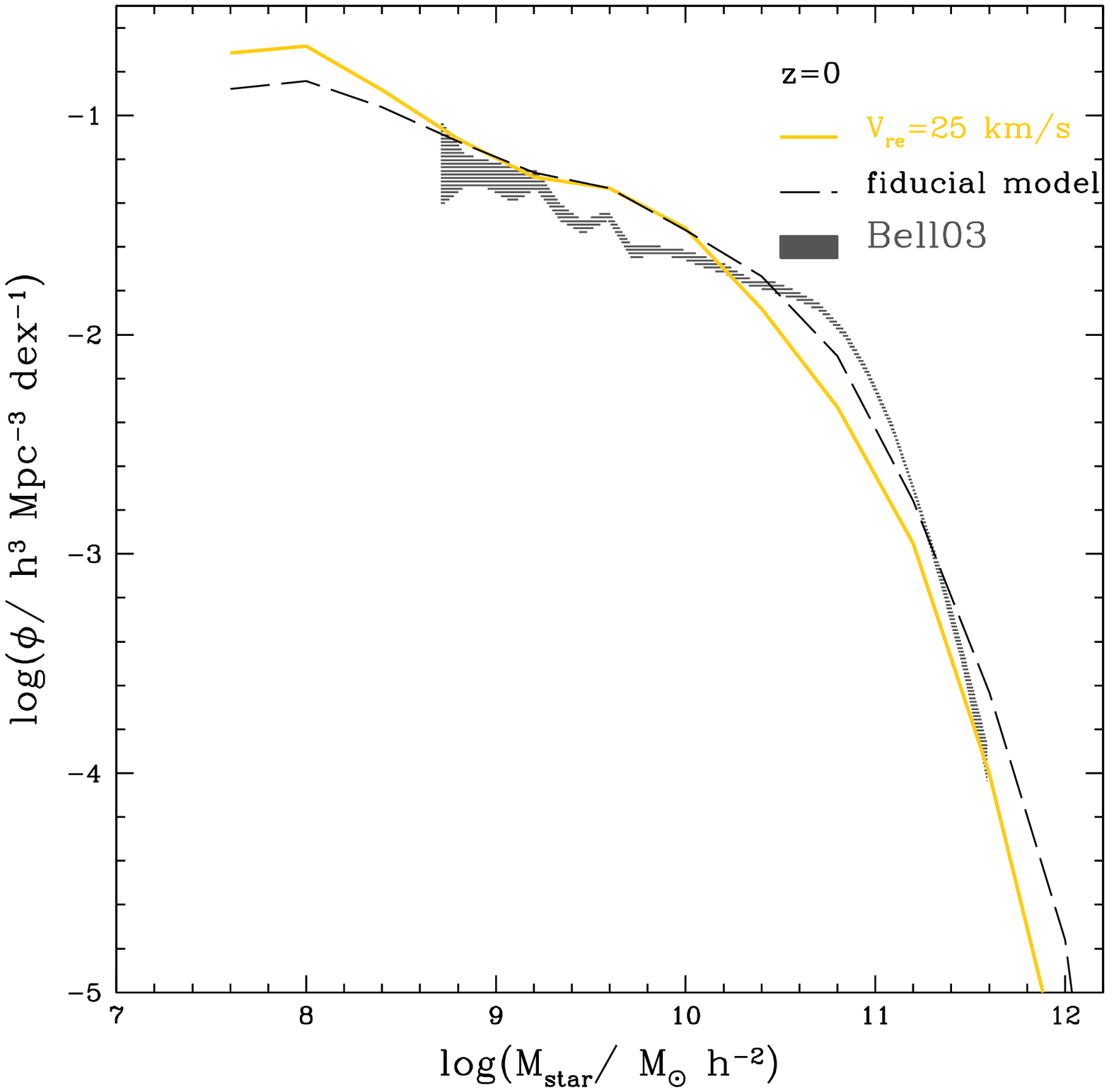}
\includegraphics[width=\columnwidth]{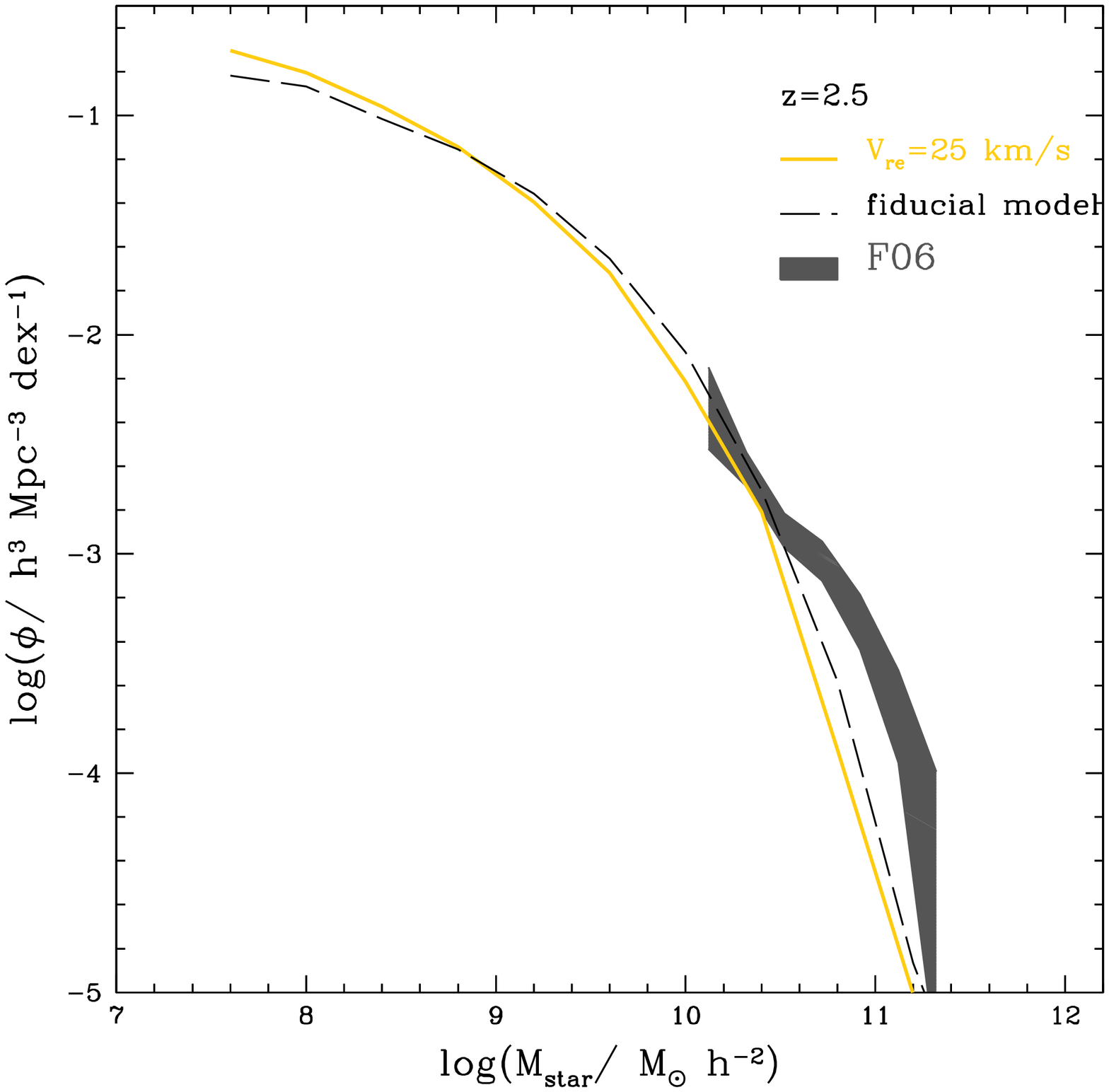}
\end{center}
\caption{Upper panel: comparison between the stellar mass function at
  z=0 of our fiducial model (dashed black line) with a model with a reduced 
  effect of the cosmic  reionization (solid yellow line). 
  Lower panel: the same as the upper panel but for z=2.5.
  Observational data 
  at the given are also indicated by the shaded region.
 }
\label{cfr_smf_re}
\end{figure}

As explained in Sect. \ref{UV},  our treatment for the cosmic reionization is based on the results previously found by \citet{Bens:03}. Anyway, 
it has been claimed, on the basis of recent hydrodynamical simulations, that the effect of the photoionizing background on structure formation is indeed weaker than previously assumed \citep{Hoeft:06, Okamoto:08}.
Taking into account such results, we test the effect of our assumption on the limit circular velocity for the reionization. 
In Figure \ref{cfr_smf_re} we show a comparison between our fiducial model with a model in which the circular velocity limit for baryon cooling has been lowered down to 25 km/s. 
To allow the formation of low-mass halos, we used for this model a merger tree
 with a resolution of $10^{9} \Msun$.  
We see that a lower limit on the circular velocity has the effect of slightly increasing the number of low mass galaxies. At the same time, the number of  massive objects is reduced, since  a greater amount of gas has been locked within satellites and is not available for star formation inside the brightest systems. In any case, the effect is rather moderate and the prediction of the fiducial model could be accommodated, for instance,  
by increasing the effect of SN.
 
Therefore, we conclude that although we might have overestimated the effect of reionization in our fiducial model, the main results found in the previous sections keep unchanged if we allow a milder dependence of galaxy formation on the cosmic reionization.

\label{lastpage}


\begin{thebibliography}{99}


\bibitem[\protect\citeauthoryear{Baldry et al.}{2004}]{Baldry:04}
Baldry I.~K., Glazebrook K., Brinkmann J., Ivezi{\'c} {\v Z}., Lupton
R.~H., Nichol R.~C., Szalay A.~S., 2004, ApJ, 600, 681

\bibitem[\protect\citeauthoryear{Baugh, Cole,
\& Frenk}{1996}]{Baugh:96} Baugh C.~M., Cole S., Frenk C.~S., 1996, MNRAS, 283, 1361

\bibitem[\protect\citeauthoryear{Bell et al.}{2003}]{Bell:03}
Bell E.~F., McIntosh D.~H., Katz N., Weinberg M.~D., 2003, ApJS, 149, 289

\bibitem[\protect\citeauthoryear{Bell et al.}{2006}]{Bell:06}
Bell E.~F., Phleps S., Somerville R.~S., Wolf C., Borch A., Meisenheimer
K., 2006, ApJ, 652, 270

\bibitem[\protect\citeauthoryear{Benson et al.}{2002}]{Bens:02}
Benson A.~J., Lacey C.~G., Baugh C.~M., Cole S., Frenk C.~S., 2002, MNRAS,
333, 156

\bibitem[\protect\citeauthoryear{Benson et al.}{2003}]{Bens:03}
Benson A.~J., Frenk C.~S., Baugh C.~M., Cole S., Lacey C.~G., 2003, MNRAS,
343, 679

\bibitem[\protect\citeauthoryear{Berta et
al.}{2007}]{Berta:07} Berta S., et al., 2007, A\&A, 476, 151

\bibitem[\protect\citeauthoryear{Bertone, De Lucia,
\& Thomas}{2007}]{Bertone:07} Bertone S., De Lucia G., Thomas P.~A., 2007, MNRAS, 379, 1143

\bibitem[\protect\citeauthoryear{Blumenthal et
al.}{1984}]{Blum:84} Blumenthal G.~R., Faber S.~M., Primack
J.~R., Rees M.~J., 1984, Nat, 311, 517

\bibitem[\protect\citeauthoryear{Bond et al.}{1991}]{Bond:91}
Bond J.~R., Cole S., Efstathiou G., Kaiser N., 1991, ApJ, 379, 440

\bibitem[\protect\citeauthoryear{Bond
\& Myers}{1996}]{Bond:96} Bond J.~R., Myers S.~T., 1996, ApJS, 103, 1

\bibitem[\protect\citeauthoryear{Bower et al.}{2006}]{Bower:06}
Bower R.~G., Benson A.~J., Malbon R., Helly J.~C., Frenk C.~S., Baugh
C.~M., Cole S., Lacey C.~G., 2006, MNRAS, 370, 645

\bibitem[\protect\citeauthoryear{Boylan-Kolchin, Ma, 
\& Quataert}{2008}]{BK:08} Boylan-Kolchin M., Ma C.-P., Quataert E., 2008, MNRAS, 383, 93 

\bibitem[\protect\citeauthoryear{Bundy, Ellis,
\& Conselice}{2005}]{Bundy:05} Bundy K., Ellis R.~S., Conselice C.~J., 2005, ApJ, 625, 621

\bibitem[\protect\citeauthoryear{Bundy et al.}{2006}]{Bundy:06}
Bundy K., et al., 2006, ApJ, 651, 120

\bibitem[\protect\citeauthoryear{Bunker et al.}{2009}]{Bunker:09} 
Bunker A., et al., 2009, arXiv, arXiv:0909.2255 

\bibitem[\protect\citeauthoryear{Cattaneo et
al.}{2006}]{Cattaneo:06} Cattaneo A., Dekel A., Devriendt J.,
Guiderdoni B., Blaizot J., 2006, MNRAS, 370, 1651

\bibitem[\protect\citeauthoryear{Cattaneo et
al.}{2008}]{Cattaneo:08} Cattaneo A., Dekel A., Faber S.~M.,
Guiderdoni B., 2008, MNRAS, 858

\bibitem[\protect\citeauthoryear{Cattaneo et 
al.}{2010}]{Cattaneo:10} Cattaneo A., Mamon G.~A., Warnick K., 
Knebe A., 2010, arXiv, arXiv:1002.3257 

\bibitem[\protect\citeauthoryear{Chandrasekhar}{1943}]{Ch:43}
Chandrasekhar S., 1943, ApJ, 97, 255

\bibitem[\protect\citeauthoryear{Cimatti, Daddi,
\& Renzini}{2006}]{Cimatti:06} Cimatti A., Daddi E., Renzini A., 2006, A\&A, 453, L29

\bibitem[\protect\citeauthoryear{Cimatti et
al.}{2008}]{Cimatti:08} Cimatti A., et al., 2008, A\&A, 482, 21

\bibitem[\protect\citeauthoryear{Croton et al.}{2006}]{Cr:06}
Croton D.~J., et al., 2006, MNRAS, 365, 11

\bibitem[\protect\citeauthoryear{Cole
\& Lacey}{1996}]{CL:96} Cole S., Lacey C., 1996, MNRAS, 281, 716

\bibitem[\protect\citeauthoryear{Cole et al.}{1994}]{Cole:94}
Cole S., Aragon-Salamanca A., Frenk C.~S., Navarro J.~F., Zepf S.~E., 1994,
MNRAS, 271, 781

\bibitem[\protect\citeauthoryear{Cole et al.}{2000}]{Cole:00}
Cole S., Benson A., Baugh C., Lacey C., Frenk C., 2000, ASPC, 200, 109

\bibitem[\protect\citeauthoryear{Cole et al.}{2001}]{Cole:01}
Cole S., et al., 2001, MNRAS, 326, 255

\bibitem[\protect\citeauthoryear{Cowie et al.}{1996}]{Cowie:96}
Cowie L.~L., Songaila A., Hu E.~M., Cohen J.~G., 1996, AJ, 112, 839

\bibitem[\protect\citeauthoryear{Cox et al.}{2008}]{Cox:08}
Cox T.~J., Jonsson P., Somerville R.~S., Primack J.~R., Dekel A., 2008,
MNRAS, 384, 386

\bibitem[\protect\citeauthoryear{De Lucia et
al.}{2004}]{DeL:04} De Lucia G., Kauffmann G., Springel V.,
White S.~D.~M., Lanzoni B., Stoehr F., Tormen G., Yoshida N., 2004, MNRAS,
348, 333

\bibitem[\protect\citeauthoryear{De Lucia et
al.}{2006}]{DeL:06} De Lucia G., Springel V., White S.~D.~M.,
Croton D., Kauffmann G., 2006, MNRAS, 366, 499

\bibitem[\protect\citeauthoryear{Dekel
\& Birnboim}{2006}]{Dekel:06} Dekel A., Birnboim Y., 2006,
  MNRAS, 368, 2

\bibitem[\protect\citeauthoryear{Dekel
\& Silk}{1986}]{Dekel:86} Dekel A., Silk J., 1986, ApJ,
  303, 39

\bibitem[\protect\citeauthoryear{Dunkley et 
al.}{2009}]{Dunkley:09} Dunkley J., et al., 2009, ApJS, 180, 306 


\bibitem[\protect\citeauthoryear{Elbaz et
al.}{2002}]{Elbaz:02} Elbaz D., Cesarsky C.~J., Chanial P.,
  Aussel H., Franceschini A., Fadda D., Chary R.~R., 2002, A\&A, 384,
  848

\bibitem[\protect\citeauthoryear{Fan et al.}{2000}]{Fan:00}
Fan X., et al., 2000, AJ, 120, 1167

\bibitem[\protect\citeauthoryear{Ferrarese
\& Merritt}{2000}]{Fer:00} Ferrarese L., Merritt D., 2000, ApJ, 539, L9

\bibitem[\protect\citeauthoryear{Fontana et
al.}{2006}]{Font:06} Fontana A., et al., 2006, A\&A, 459, 745

\bibitem[\protect\citeauthoryear{Fontanot et 
al.}{2009}]{Fontanot:09} Fontanot F., De Lucia G., Monaco P., 
Somerville R.~S., Santini P., 2009, MNRAS, 397, 1776 

\bibitem[\protect\citeauthoryear{Franceschini et
al.}{1998}]{Fr:98} Franceschini A., Silva L., Fasano G.,
Granato G.~L., Bressan A., Arnouts S., Danese L., 1998, ApJ, 506, 600

\bibitem[\protect\citeauthoryear{Franceschini et
al.}{2001}]{Fr:01} Franceschini A., Aussel H., Cesarsky C.~J., Elbaz D., Fadda D., 2001, A\&A, 378, 1

\bibitem[\protect\citeauthoryear{Franceschini et
al.}{2006}]{Fr:06} Franceschini A., et al., 2006, A\&A, 453, 397

\bibitem[\protect\citeauthoryear{Gavazzi
\& Scodeggio}{1996}]{Gavazzi:96} Gavazzi G., Scodeggio M.,
  1996, A\&A, 312, L29

\bibitem[\protect\citeauthoryear{Gebhardt et
al.}{2000}]{Geb:00} Gebhardt K., et al., 2000, ApJ, 539, L13

\bibitem[\protect\citeauthoryear{Giocoli et
al.}{2007}]{Giocoli:07} Giocoli C., Moreno J., Sheth R.~K., Tormen
G., 2007, MNRAS, 376, 977

\bibitem[\protect\citeauthoryear{Gnedin}{2000}]{Gnedin:00} Gnedin 
N.~Y., 2000, ApJ, 542, 535 


\bibitem[\protect\citeauthoryear{Heckman}{2001}]{Heckman:01}
Heckman T.~M., 2001, ASPC, 240, 345

\bibitem[\protect\citeauthoryear{H{\"a}ring
\& Rix}{2004}]{HR:04} H{\"a}ring N., Rix H.-W., 2004, ApJ, 604, L89

\bibitem[\protect\citeauthoryear{Hatton et al.}{2003}]{Hatton:03}
Hatton S., Devriendt J.~E.~G., Ninin S., Bouchet F.~R., Guiderdoni B.,
Vibert D., 2003, MNRAS, 343, 75

\bibitem[\protect\citeauthoryear{Hoeft et al.}{2006}]{Hoeft:06} 
Hoeft M., Yepes G., Gottl{\"o}ber S., Springel V., 2006, MNRAS, 371, 401 

\bibitem[\protect\citeauthoryear{Hopkins}{2004}]{Hopk:04}
Hopkins A.~M., 2004, ApJ, 615, 209

\bibitem[\protect\citeauthoryear{Hopkins, Richards, 
\& Hernquist}{2007}]{Hopk:07a} Hopkins P.~F., Richards G.~T., Hernquist L., 2007, ApJ, 654, 731 

\bibitem[\protect\citeauthoryear{Hopkins et
al.}{2007}]{Hopk:07} Hopkins P.~F., Bundy K., Hernquist L.,
Ellis R.~S., 2007, ApJ, 659, 976

\bibitem[\protect\citeauthoryear{Jiang et al.}{2008}]{Jiang:08} 
Jiang C.~Y., Jing Y.~P., Faltenbacher A., Lin W.~P., Li C., 2008, ApJ, 675, 
1095 

\bibitem[\protect\citeauthoryear{Kang 
\& van den Bosch}{2008}]{Kang:08} Kang X., van den Bosch F.~C., 2008, ApJ, 676, L101 

\bibitem[\protect\citeauthoryear{Kauffmann
\& Haehnelt}{2000}]{KH:00} Kauffmann G., Haehnelt M., 2000, MNRAS,
  311, 576

\bibitem[\protect\citeauthoryear{Kauffmann
\& White}{1993}]{kauff:93a} Kauffmann G., White S.~D.~M., 1993, MNRAS, 261, 921 

\bibitem[\protect\citeauthoryear{Kauffmann, White,
\& Guiderdoni}{1993}]{kauff:93b} Kauffmann G., White S.~D.~M., Guiderdoni B., 1993, MNRAS, 264, 201 

\bibitem[\protect\citeauthoryear{Kauffmann et
al.}{1999}]{kauff:99} Kauffmann G., Colberg J.~M., Diaferio A.,
White S.~D.~M., 1999, MNRAS, 307, 529

\bibitem[\protect\citeauthoryear{Kauffmann et
al.}{2003}]{Kauff:03} Kauffmann G., et al., 2003, MNRAS, 341, 54

\bibitem[\protect\citeauthoryear{Kennicutt}{1998}]{Ken:98}
Kennicutt R.~C., Jr., 1998, ApJ, 498, 541

\bibitem[\protect\citeauthoryear{Kere{\v s} et
al.}{2005}]{Keres:05} Kere{\v s} D., Katz N., Weinberg D.~H.,
Dav{\'e} R., 2005, MNRAS, 363, 2

\bibitem[\protect\citeauthoryear{Kimm et al.}{2009}]{Kimm:09} 
Kimm T., et al., 2009, MNRAS, 394, 1131 

\bibitem[\protect\citeauthoryear{Kitzbichler
\& White}{2006}]{Kit:06} Kitzbichler M.~G., White S.~D.~M., 2006, MNRAS, 366, 858

\bibitem[\protect\citeauthoryear{Kravtsov, Gnedin, 
\& Klypin}{2004}]{Krav:04} Kravtsov A.~V., Gnedin O.~Y., Klypin A.~A., 2004, ApJ, 609, 482 

\bibitem[\protect\citeauthoryear{Lacey
\& Cole}{1993}]{LC:93} Lacey C., Cole S., 1993, MNRAS, 262, 627

\bibitem[\protect\citeauthoryear{Lagos, Cora, 
\& Padilla}{2008}]{Lagos:08} Lagos C.~D.~P., Cora S.~A.,
  Padilla N.~D., 2008, MNRAS, 388, 587 

\bibitem[\protect\citeauthoryear{Lanzoni et 
al.}{2005}]{Lanzoni:05} Lanzoni B., Guiderdoni B., Mamon G.~A., 
Devriendt J., Hatton S., 2005, MNRAS, 361, 369 


\bibitem[\protect\citeauthoryear{Lapi et al.}{2006}]{Lapi:06} 
Lapi A., Shankar F., Mao J., Granato G.~L., Silva L., De Zotti G., Danese 
L., 2006, ApJ, 650, 42 

\bibitem[\protect\citeauthoryear{Macci{\`o} et 
al.}{2009}]{Maccio:09} Macci{\`o} A.~V., Kang X., Fontanot F., 
Somerville R.~S., Koposov S., Monaco P., 2009, MNRAS, 1907 

\bibitem[\protect\citeauthoryear{Malbon et al.}{2007}]{Malbon:07} 
Malbon R.~K., Baugh C.~M., Frenk C.~S., Lacey C.~G., 2007, MNRAS, 382, 1394 

\bibitem[\protect\citeauthoryear{Mamon}{1992}]{Mamon:92} Mamon
G.~A., 1992, ApJ, 401, L3

\bibitem[\protect\citeauthoryear{Marchesini et 
al.}{2009}]{Marchesini:08} Marchesini D., van Dokkum P.~G., 
F{\"o}rster Schreiber N.~M., Franx M., Labb{\'e} I., Wuyts S., 2009, ApJ, 
701, 1765 

\bibitem[\protect\citeauthoryear{Martin}{1999}]{Mar:99} Martin
C.~L., 1999, ApJ, 513, 156

\bibitem[\protect\citeauthoryear{Marulli et 
al.}{2008}]{Marulli:08} Marulli F., Bonoli S., Branchini E., 
Moscardini L., Springel V., 2008, MNRAS, 385, 1846 

\bibitem[\protect\citeauthoryear{McCarthy et 
al.}{2008}]{McCarthy:08} McCarthy I.~G., Frenk C.~S., Font A.~S., 
Lacey C.~G., Bower R.~G., Mitchell N.~L., Balogh M.~L., Theuns T., 2008, 
MNRAS, 383, 593 

\bibitem[\protect\citeauthoryear{Menci
\& Cavaliere}{2002}]{Menci:02} Menci N., Cavaliere A., 2002, ASPC, 253, 429

\bibitem[\protect\citeauthoryear{Menci et al.}{2004}]{Menci:04}
Menci N., Cavaliere A., Fontana A., Giallongo E., Poli F., Vittorini V.,
2004, ApJ, 604, 12

\bibitem[\protect\citeauthoryear{Menci et al.}{2006}]{Menci:06}
Menci N., Fontana A., Giallongo E., Grazian A., Salimbeni S., 2006, ApJ,
647, 753

\bibitem[\protect\citeauthoryear{Mihos
\& Hernquist}{1994}]{MH:94} Mihos J.~C., Hernquist L., 1994, ApJ, 425, L13

\bibitem[\protect\citeauthoryear{Mihos
\& Hernquist}{1996}]{MH:96} Mihos J.~C., Hernquist L., 1996, ApJ, 464, 641

\bibitem[\protect\citeauthoryear{Mo, Mao,
\& White}{1998}]{Mo:98} Mo H.~J., Mao S., White S.~D.~M., 1998, MNRAS, 295, 319

\bibitem[\protect\citeauthoryear{Monaco, Fontanot, 
\& Taffoni}{2007}]{Monaco:07} Monaco P., Fontanot F.,
  Taffoni G., 2007, MNRAS, 375, 1189 

\bibitem[\protect\citeauthoryear{Monaco et al.}{2009}]{Monaco:09} 
Monaco P., et al., 2009, AIPC, 1111, 48 


\bibitem[\protect\citeauthoryear{Navarro
\& White}{1994}]{NW:94} Navarro J.~F., White S.~D.~M., 1994, MNRAS, 267, 401

\bibitem[\protect\citeauthoryear{Navarro, Frenk, 
\& White}{1995}]{Navarro:95} Navarro J.~F., Frenk C.~S., White S.~D.~M., 1995, MNRAS, 275, 56 
\bibitem[\protect\citeauthoryear{Navarro, Frenk,
\& White}{1997}]{NFW:97} Navarro J.~F., Frenk C.~S.,
  White S.~D.~M., 1997, ApJ, 490, 493

\bibitem[\protect\citeauthoryear{Neistein, van den Bosch,
\& Dekel}{2006}]{Nei:06} Neistein E., van den Bosch F.~C., Dekel A., 2006, MNRAS, 372, 933

\bibitem[\protect\citeauthoryear{Okamoto, Gao, 
\& Theuns}{2008}]{Okamoto:08} Okamoto T., Gao L., Theuns T., 2008, MNRAS, 390, 920 

\bibitem[\protect\citeauthoryear{Percival et
al.}{2002}]{Perc:02} Percival W.~J., et al., 2002, MNRAS, 337,
1068

\bibitem[\protect\citeauthoryear{P{\'e}rez-Gonz{\'a}lez et 
al.}{2008}]{PG:07} P{\'e}rez-Gonz{\'a}lez P.~G., et al., 
2008, ApJ, 675, 234 

\bibitem[\protect\citeauthoryear{Perlmutter, Turner,
\& White}{1999}]{Perl:99} Perlmutter S., Turner M.~S., White M., 1999, PhRvL, 83, 670

\bibitem[\protect\citeauthoryear{Press
\& Schechter}{1974}]{PS:74} Press W.~H., Schechter P., 1974, ApJ, 187, 425

\bibitem[\protect\citeauthoryear{Riess et al.}{1998}]{Riess:98}
Riess A.~G., et al., 1998, AJ, 116, 1009

\bibitem[\protect\citeauthoryear{Rodighiero et 
al.}{2009}]{Rod:09} Rodighiero G., et al., 2009, arXiv, 
arXiv:0910.5649

\bibitem[\protect\citeauthoryear{Roos
\& Norman}{1979}]{Roos:79} Roos N., Norman C.~A., 1979, A\&A, 76, 75

\bibitem[\protect\citeauthoryear{Sanders
\& Mirabel}{1996}]{Sand:96} Sanders D.~B., Mirabel I.~F., 1996, ARA\&A, 34, 749

\bibitem[\protect\citeauthoryear{Shankar et 
al.}{2004}]{Shankar:04} Shankar F., Salucci P., Granato G.~L., De 
Zotti G., Danese L., 2004, MNRAS, 354, 1020 

\bibitem[\protect\citeauthoryear{Sheth
\& Lemson}{1999}]{SL:99} Sheth R.~K., Lemson G., 1999, MNRAS, 305, 946

\bibitem[\protect\citeauthoryear{Sheth
\& Tormen}{2002}]{ST:02} Sheth R.~K., Tormen G., 2002, MNRAS, 329, 61

\bibitem[\protect\citeauthoryear{Sijacki et 
al.}{2007}]{Sij:07} Sijacki D., Springel V., di Matteo T., 
Hernquist L., 2007, MNRAS, 380, 877 

\bibitem[\protect\citeauthoryear{Somerville}{2002}]{S:02} 
Somerville R.~S., 2002, ApJ, 572, L23 

\bibitem[\protect\citeauthoryear{Somerville
\& Kolatt}{1999}]{SK:99} Somerville R.~S., Kolatt T.~S., 1999, MNRAS, 305, 1

\bibitem[\protect\citeauthoryear{Somerville
\& Primack}{1999}]{SP:99} Somerville R.~S., Primack J.~R., 1999,
  MNRAS, 310, 1087

\bibitem[\protect\citeauthoryear{Somerville, Primack,
\& Faber}{2001}]{SPF:01} Somerville R.~S., Primack J.~R.,
  Faber S.~M., 2001, MNRAS, 320, 504

\bibitem[\protect\citeauthoryear{Somerville et 
al.}{2008}]{Som:08} Somerville R.~S., Hopkins P.~F., Cox 
T.~J., Robertson B.~E., Hernquist L., 2008, MNRAS, 391, 481 

\bibitem[\protect\citeauthoryear{Spergel et
al.}{2003}]{Spergel:03} Spergel D.~N., et al., 2003, ApJS, 148,
175

\bibitem[\protect\citeauthoryear{Spergel et
al.}{2007}]{Spergel:07} Spergel D.~N., et al., 2007, ApJS, 170,
377

\bibitem[\protect\citeauthoryear{Springel, Yoshida,
\& White}{2001}]{Spring:01} Springel V., Yoshida N., White S.~D.~M., 2001, NewA, 6, 79

\bibitem[\protect\citeauthoryear{Strickland et
al.}{2000}]{Strick:00} Strickland D.~K., Heckman T.~M., Weaver
K.~A., Dahlem M., 2000, AJ, 120, 2965

\bibitem[\protect\citeauthoryear{Stringer et 
al.}{2009}]{Stringer:08} Stringer M.~J., Benson A.~J., Bundy K., 
Ellis R.~S., Quetin E.~L., 2009, MNRAS, 393, 1127 

\bibitem[\protect\citeauthoryear{Sutherland
\& Dopita}{1993}]{SD:93} Sutherland R.~S., Dopita M.~A., 1993, ApJS, 88, 253

\bibitem[\protect\citeauthoryear{Taffoni et 
al.}{2003}]{Taffoni:03} Taffoni G., Mayer L., Colpi M., Governato 
F., 2003, MNRAS, 341, 434 

\bibitem[\protect\citeauthoryear{Tegmark et
al.}{2004}]{Tegm:04} Tegmark M., et al., 2004, ApJ, 606,
  702

\bibitem[\protect\citeauthoryear{Thomas et al.}{2005}]{Thomas:05}
Thomas D., Maraston C., Bender R., Mendes de Oliveira C., 2005, ApJ, 621,
673

\bibitem[\protect\citeauthoryear{Thoul 
\& Weinberg}{1996}]{TW:96} Thoul A.~A., Weinberg D.~H., 1996, ApJ, 465, 608 

\bibitem[\protect\citeauthoryear{Tormen}{1997}]{Tormen:97} Tormen
G., 1997, MNRAS, 290, 411

\bibitem[\protect\citeauthoryear{Trujillo et 
al.}{2009}]{Trujillo:09} Trujillo I., Cenarro A.~J., de 
Lorenzo-C{\'a}ceres A., Vazdekis A., de la Rosa I.~G., Cava A., 2009, ApJ, 
692, L118

\bibitem[\protect\citeauthoryear{Trujillo et
al.}{2007}]{Trujillo:07} Trujillo I., Conselice C.~J., Bundy K.,
Cooper M.~C., Eisenhardt P., Ellis R.~S., 2007, MNRAS, 382, 109

\bibitem[\protect\citeauthoryear{Valentinuzzi et 
al.}{2010}]{Valentinuzzi:09} Valentinuzzi T., et al., 2010, ApJ, 712, 
226 

\bibitem[\protect\citeauthoryear{van Dokkum}{2005}]{vanD:05}
van Dokkum P.~G., 2005, AJ, 130, 2647

\bibitem[\protect\citeauthoryear{Viola et al.}{2008}]{Viola:08} 
Viola M., Monaco P., Borgani S., Murante G., Tornatore L., 2008, MNRAS, 
383, 777 

\bibitem[\protect\citeauthoryear{Volonteri, Lodato, 
\& Natarajan}{2008}]{Volonteri:08} Volonteri M., Lodato G.,
  Natarajan P., 2008, MNRAS, 383, 1079 


\bibitem[\protect\citeauthoryear{Warren et al.}{1992}]{Warren:92}
Warren M.~S., Quinn P.~J., Salmon J.~K., Zurek W.~H., 1992, ApJ, 399, 405

\bibitem[\protect\citeauthoryear{Weinmann et 
al.}{2006}]{Weinmann:06} Weinmann S.~M., van den Bosch F.~C., Yang 
X., Mo H.~J., Croton D.~J., Moore B., 2006, MNRAS, 372, 1161 

\bibitem[\protect\citeauthoryear{White
\& Frenk}{1991}]{WF:91} White S.~D.~M., Frenk C.~S., 1991, ApJ, 379, 52

\bibitem[\protect\citeauthoryear{White
\& Rees}{1978}]{WR:78} White S.~D.~M., Rees M.~J., 1978, MNRAS, 183, 341

\bibitem[\protect\citeauthoryear{White et al.}{1993}]{White:93}
White S.~D.~M., Navarro J.~F., Evrard A.~E., Frenk C.~S., 1993, Nat, 366,
429

\bibitem[\protect\citeauthoryear{Zentner}{2007}]{Zen:07}
Zentner A.~R., 2007, IJMPD, 16, 763

\bibitem[\protect\citeauthoryear{Zhang, Ma,
\& Fakhouri}{2008}]{Zhang:08} Zhang J., Ma C.-P., Fakhouri O., 2008, MNRAS, 387, L13


\end{thebibliography}
\end{document}